\documentclass[12pt,preprint]{aastex}

\newcommand{\cn}{{\rm cn}}
\newcommand{\sn}{{\rm sn}}

\shorttitle{Blackbody Spectrum of a Thin Accretion Disk}
\shortauthors{Li et al.}

\begin{document}

\title{Multi-Temperature Blackbody Spectrum of a Thin Accretion Disk
around a Kerr Black Hole: Model Computations and Comparison with
Observations}

\author{Li-Xin Li\altaffilmark{1,2}, Erik R. Zimmerman, Ramesh Narayan,
Jeffrey E. McClintock}
\affil{Harvard-Smithsonian Center for Astrophysics, 60 Garden Street, 
Cambridge, MA 02138}
\email{lli,ezimmerman,rnarayan,jmcclintock@cfa.harvard.edu}

\altaffiltext{1}{Chandra Fellow}
\altaffiltext{2}{Present address: Max-Planck-Institut f\"ur Astrophysik,
Karl-Schwarzschild-Str. 1, Postfach 1317, 85741 Garching, Germany}

\begin{abstract}
We use a ray-tracing technique to compute the observed spectrum of a
thin accretion disk around a Kerr black hole.  We include all
relativistic effects such as frame-dragging, Doppler boost,
gravitational redshift, and bending of light by the gravity of the
black hole.  We also include self-irradiation of the disk as a result
of light deflection.  Assuming that the disk emission is locally
blackbody, we show how the observed spectrum depends on the spin of
the black hole, the inclination of the disk, and the torque at the
inner edge of the disk.  We find that the effect of a nonzero torque
on the spectrum can, to a good approximation, be absorbed into a
zero-torque model by adjusting the mass accretion rate and the
normalization.  We describe a computer model, called KERRBB, which we
have developed for fitting the spectra of black hole X-ray binaries.
Using KERRBB within the X-ray data reduction package XSPEC, and
assuming a spectral hardening factor $f_{\rm col}=1.7$, we analyze the
spectra of three black hole X-ray binaries: 4U1543-47, XTE J1550-564,
and GRO J1655-40. We estimate the spin parameters of the black  
holes in 4U1543-47 and GRO J1655-40 to be $a/M \sim 0.6$ and
$\sim0.6-0.7$, respectively.  If $f_{\rm col}\sim1.5-1.6$, as in a
recent study, then we find $a/M \sim0.7-0.8$ and $\sim0.8-0.9$,
respectively.  These estimates are subject to additional uncertainties
in the assumed black hole masses, distances and disk inclinations.
\end{abstract}


\section{Introduction}
\label{intro}

Although the standard theory of thin accretion disks around black
holes was developed over thirty years ago
\citep{pri72,sha73,nov73,lyn74}, a straightforward confrontation of
the model with observations is still not possible because of the
challenging task of computing the observed spectrum of an accretion
disk around a Kerr black hole.  When fitting the soft X-ray spectra of
black hole binaries, the multi-temperature disk model DISKBB is often
used, which describes an approximate Newtonian model of a thin disk
\citep{mit84,mak86}.  This model has been criticized by \citet{gie01}
for not applying the proper boundary condition at the inner edge of
the disk.  More seriously, the model does not include relativistic
effects.
 
Various efforts have been made to include relativistic effects and the
appropriate boundary condition in calculating the spectra of black
hole accretion disks
\citep{cun75,cun76,han89,ebi91,zha97,ago00,gie01,ebi03,dav04}. However,
accurate quantitative analysis of observational data, e.g., trying to
determine the radius of the disk inner edge from spectral data with a
view to estimating the spin of the black hole, requires a
sophisticated model that treats all relativistic effects including the
self-irradiation of the disk caused by light deflection by the central
black hole.  At present, no such complete model exists that is
suitable for inclusion with standard data reduction software, e.g.,
the X-ray spectral fitting package XSPEC \citep{arn96}. As
high-quality observational data on black-hole accretion disks become
increasingly available (e.g., McClintock \& Remillard 2004), the need
for such a model has become urgent.  In this paper we present the
tools for a general relativistic model of an accretion disk around a
Kerr black hole, and describe a numerical code for computing the
observed blackbody spectrum of the disk.  The code is efficient, and
we have developed the appropriate software for interfacing it with
XSPEC for analyzing spectral data.

Two basic approaches have been described in the literature for
calculating the observed spectrum of an accretion disk around a Kerr
black hole.  The first method makes use of ``transfer functions''
\citep{cun75,cun76,lao91,spe95,ago00,dov04}. In this approach, all the
information about the Doppler boost due to the disk rotation as well
as the relativistic light deflection in the vicinity of the black hole
is contained in a transfer function, which operates as an integration
kernel for calculating the overall disk spectrum.  The transfer
function has been calculated and discussed in great detail by
\citet{spe95}, who have provided a fast and easy-to-use computer
program to do the calculations.  The second method uses the
technique of ``ray-tracing'' \citep{rau94,fan97,cad98,mul04,sch04}.
In this method one divides the image of the disk on the observer's sky
into a number of small elements.  For each image element, the orbit of
a photon is traced backward from the observer by following the
geodesics in a Kerr spacetime, until the orbit crosses the plane of
the disk. The flux density of the radiation emitted by the disk at
that point, as well as the redshift factor of the photon and the angle
between the wavevector of the photon and the normal to the disk
surface, are calculated. The observed flux density contributed by each
image element is thus obtained, and summing over all the elements
gives the total observed flux density of the disk (see
eq.~[\ref{feo3}] below).

In this paper we use the ray-tracing approach, which we find more
straightforward for numerical computations. We follow the procedures
described by \citet{fan97} and \citet{cad98}, who used elliptic
integrals to simplify the calculation of the orbit of a photon in the
background of a Kerr black hole (see also Rauch \& Blandford 1994).  
Fanton et al.'s code was written for
computing line emission from disks, whereas our code is designed for
the continuum blackbody emission.  We have made several improvements
to Fanton et al.'s calculations; in particular, we include the effect
of returning radiation, and we allow a nonzero torque to be set
at the inner edge of the disk.

Using our code, we have calculated spectra corresponding to a
three-dimensional grid of models spanning different values of the
black hole spin parameter $a_* \equiv a/M$, disk inclination angle
$\vartheta_{\rm obs}$, and dimensionless torque parameter $\eta$
(defined in eq.~[\ref{eta}]).  We have also developed associated
software which goes by the model name KERRBB for use with XSPEC.
KERRBB reads in the above table of spectra and uses it to fit spectral
data.  In addition to the three parameters, $a_*$, $\vartheta_{\rm
obs}$, and $\eta$ spanned by the table, other parameters such as the
mass of the black hole $M$, the distance to the source $D$, the mass
accretion rate of the disk $\dot M$, and the spectral hardening factor
$f_{\rm col}$ (see eqs.~[\ref{f_col}] and [\ref{iem}]), are also
included. The user can also decide whether to assume isotropic
emission from the disk surface or to include a standard limb-darkening
law (eq.~[\ref{iout}]).  The model works efficiently within XSPEC and
some sample results are presented.

The paper is organized as follows. In \S\ref{assump} we summarize the
assumptions behind our model, and in \S\ref{math} we present the basic
mathematical formalism.  We discuss the effect of returning radiation
in \S3.1 and show examples of calculated spectra in \S3.2, where we
explain how the spectrum is affected by parameters such as the spin of
the black hole, the inclination of the disk, and the torque on the
inner edge of the disk.  In \S\ref{novae} we apply KERRBB to spectral
data on three black hole X-ray binaries, 4U1543-47, XTE J1550-564, and
GRO J1655-40, and compare the results to those obtained with other
models. We conclude with a summary in \S\ref{sum}.

The ray-tracing technique involves complicated mathematics to describe
the orbits of photons in Kerr spacetime. Although many of these
formulae can be found in the literature (e.g., Chandrasekhar 1983;
Rauch \& Blandford 1994; Fanton et al. 1997; \v{C}ade\v{z} et al.
1998), we feel it is important to present the relevant mathematics in
full detail to make the paper more useful to future workers in this
field.  To avoid distracting the reader, we give the technical details
in Appendixes~\ref{integral}--\ref{spectra_obs}.

In Appendix~\ref{grad} we compare KERRBB with GRAD---a subroutine in
XSPEC for calculating the blackbody spectrum of a Keplerian disk
around a nonrotating ($a_*=0$) black hole.  We show that KERRBB and
GRAD give consistent results, once some errors in GRAD are corrected.

\section{Basic Assumptions}
\label{assump}

Throughout the paper we use units in which $G = c = h = 1$, where $G$
is the Newtonian constant, $c$ is the speed of light, and $h$ is the
Planck constant. We use cylindrical coordinates ($t,r,z,\varphi$) as
described by \citet{pag74}. In these coordinates, the rotation axis of
the black hole is along the $z$-axis, and the equatorial plane
corresponds to $z=0$.

We consider a geometrically thin and optically thick Keplerian
accretion disk around a Kerr black hole, with the spin axis of the
black hole perpendicular to the disk plane. The black hole has a mass
$M$ and a specific angular momentum $a$, where $-M \le a \le M$. The
disk has an inner boundary of radius $r_{\rm in}$, and an outer
boundary of radius $r_{\rm out}\gg r_{\rm in}$ (KERRBB assumes $r_{\rm
out} = 10^6 M$).

In the standard theory of accretion disks, it is usually assumed that the 
torque at the inner boundary of the disk is zero. However, this assumption
has only been justified for nonmagnetized or weakly magnetized flows 
\citep{muc82,abr89} and for very thin disks \citep{afs03}. Recent 
theoretical works on accretion disks have suggested that 
a nonzero torque at the inner boundary can arise from a magnetic field which 
either connects a disk to a central black hole, or couples a disk to the 
material in the plunging region 
\citep{kro99,gam99,ago00,li00,li02a,li02b,li04,wan02,wan03,uzd04a,uzd04b}.
Although this issue is still under debate 
\citep{pac00,arm01,haw02,afs03,li03a,li03b}, to make our model as general
as possible we assume that the torque at the inner boundary of the disk can
have any nonnegative value. 

In the presence of a torque $g_{\rm in}\geq0$ at the inner edge, the
total power of the disk, i.e. the net amount of energy flowing out of
the disk per unit time as measured by an observer at infinity, is
\citep{li02a}\footnote{When the effect of returning radiation is
considered, the total power of the disk corresponds to the
energy-at-infinity carried away from the disk per unit coordinate time
by the photons that permanently leave the disk---they either escape to
infinity or fall into the black hole. See \S\ref{return} for details.}
\begin{eqnarray}
    {\cal L}_{\rm total} = g_{\rm in} \Omega_{\rm in} + \epsilon_{\rm in} 
        \dot{M}\;, \label{power_t}
\end{eqnarray}
where $\Omega_{\rm in}$ is the angular velocity of the disk inner
boundary, $\dot{M}$ is the mass accretion rate of the disk, and
$\epsilon_{\rm in}= 1-E^\dagger_{\rm in}$ is the specific
gravitational binding energy at the inner boundary, where
$E^\dagger_{\rm in}$ is the specific energy of disk particles at the
inner boundary.

Equation~(\ref{power_t}) shows that the total power of the disk comes
from two sources: the gravitational binding energy between the disk
and the black hole, and a contribution from the torque at the inner
boundary of the disk (whenever the torque is nonzero). If the torque
is produced by the black hole, then the power source for the second
component is the spin energy of the black hole.

Let us define a dimensionless parameter
\begin{eqnarray}
    \eta\equiv\frac{g_{\rm in} \Omega_{\rm in}}{\epsilon_{\rm in} \dot{M}} \;,
    \label{eta}
\end{eqnarray}
which measures the ratio of the power from the torque to the power
from the gravitational binding energy of the accreting gas.  Then we
have
\begin{eqnarray}
    {\cal L}_{\rm total} = (1+\eta)\epsilon_{\rm in} \dot{M} \;. 
        \label{power_t2}
\end{eqnarray}
In this paper we treat $\eta$ as a free parameter that can have any
nonnegative value.  It is useful to define an effective mass accretion
rate
\begin{eqnarray}
    \dot{M}_{\rm eff} \equiv (1+\eta)\dot{M} \;, \label{meff}
\end{eqnarray}
such that the total power of the disk is simply
\begin{eqnarray}
    {\cal L}_{\rm total} = \epsilon_{\rm in} \dot{M}_{\rm eff} \;.  
        \label{power_t3}
\end{eqnarray}

With our conventions, $\eta = 0$ corresponds to the case when the
torque at the inner boundary of the disk is zero (we call this {\em
the standard case}), and $\eta=\infty$ corresponds to $\dot{M} = 0$
(we call this {\em the nonaccreting case}). In these two extreme
cases, all the power of the disk comes from either accretion ($\eta =
0$) or from the torque at the inner boundary ($\eta=\infty$). 

The nonaccreting case is similar to the case of a ``dead disk'' around a
magnetized spinning neutron star \citep{sun77}, in the sense that both
correspond to a zero mass accretion rate. However, the nonaccreting
case considered in this paper is not necessarily in a ``dead''
state. If the black hole rotates fast and the inner boundary of a
nonaccreting disk is located at the marginally stable orbit close to
the horizon of the black hole, then a disk powered by the spin energy
of the black hole can, in fact, be very bright \citep{li04}.

The model and the computer code described in this paper apply to a
geometrically thin Keplerian disk with its inner edge located at any
$r_{\rm in}\ge r_{\rm ms}$, where $r_{\rm ms}$ is the radius of the
marginally stable orbit.  For simplicity, in all plots presented in
the paper, and in the model which we have developed for use with
XSPEC, we assume that $r_{\rm in} = r_{\rm ms}$.

We assume that the disk radiates like a blackbody. However, due to the
complicated scattering processes in the disk atmosphere (predominantly
electron scattering and Comptonization), the color temperature $T_{\rm
col}$ of the emitted radiation is generally higher than the effective
temperature $T_{\rm eff}$ of the disk
\citep{ros92,shi93,shi95,dav04}. To take this effect into
account, we follow \citet{shi95} and \citet{ebi03} and assume that
the ratio $f_{\rm col}=T_{\rm col}/T_{\rm eff}$ is a constant (see
eqs. [\ref{teff}]-[\ref{iem}]).  We take $f_{\rm col}= 1.7$, the mean
value recommended by \citet{shi95}, when we model observations
in \S4.  Undoubtedly, it is an over simplification to model the
spectral modification due to Comptonization and electron scattering
with a single scaling parameter, but this is a standard approach in
the literature and is the best that we can do at this time.

\section{Mathematical Formalism for the Calculation of the Observed Spectrum 
of Disk Radiation}
\label{math}

\subsection{Effects of the returning radiation}
\label{return}

Because of the gravity of the central black hole, not all of the
radiation emitted by the disk escapes to infinity: a part of it, which
we call ``returning radiation,'' returns to the disk or is permanently
captured by the black hole. The returning radiation that strikes the
disk will interact with the disk particles and eventually be scattered
or absorbed. For simplicity, we make the following assumption: All the
radiation returning to the disk in the region beyond the inner
boundary is absorbed by the disk, then reprocessed and reradiated. All
the radiation returning in the region inside the inner boundary (i.e.,
the plunging region) is advected or scattered inward by the infalling
gas (which has a large inward velocity), and is captured by the black
hole \citep{ago00}. In this subsection we study the effects of the
returning radiation on the emission of the disk and the evolution of
the central black hole.

The mathematics for studying the effect of the returning radiation is 
presented
in Appendix~\ref{flux}. In brief, considering the effect of returning
radiation, at each point in the disk the net flux density $F$ is
composed of two nonnegative components: an outgoing component $F_{\rm
out}$ which represents the flux density of the energy emitted by the
disk {\it in situ}; and an ingoing component $F_{\rm in}$, which is
the flux density in the returning radiation, i.e. radiation emitted
elsewhere and focused back onto the disk by the gravity of the black
hole. In the steady state, the net flux density $F$ is determined by
the balance of the energy and angular momentum in the disk:
$F=F_0+F_S$, where $F_0$ is the standard solution for the flux density
when the effect of returning radiation is ignored (eq.~[\ref{f0}]),
and $F_S$ represents the work done by the returning radiation on the
disk (eq.~[\ref{fs}]). The total outgoing flux density is then $F_{\rm
out} = F+ F_{\rm in} = F_0+F_{\rm in}+F_S$ (eq.~[\ref{sol_fout}]).
The inclusion of $F_S$ in our energy balance equation is the main
difference between our calculations and those of \cite{ago00}.  Note,
however, that this has very little effect on the flux or spectrum of a
standard disk.

The self-irradiation of the disk arising from returning radiation is
essentially a nonlocal process. Both $F_{\rm in}$ and $F_S$ of the
incoming radiation---and also $F_{\rm out}$ at the point on the disk
where the incoming photons cross the disk---are functionals of $F_{\rm
out}$ at the point on the disk where the incoming photons were
emitted.  Therefore, $F_{\rm out}$ (and thus $F_{\rm in}$ and $F_S$)
must be obtained by solving the following functional equation
\begin{eqnarray}
    F_{\rm out} = F_0 + F_{\rm in}[F_{\rm out}] + F_S[F_{\rm out}] \;.
	\nonumber
\end{eqnarray}
Self-consistent solutions for $F_{\rm out}$, $F_{\rm in}$, and $F_S$
can be obtained by an iterative method using ray-tracing (see
Appendix~\ref{flux}).

In Figure~\ref{ret_flux_0999} we show the solutions of $F_{\rm in}$
and $F_S$ for the returning radiation as functions of disk radius, for
the case of a Kerr black hole of $a=0.999M$ with $\eta=0$ (upper
panel) and $\eta= \infty$ (lower panel). The radiation emitted by the
disk is assumed to be isotropic in the disk frame, rather than
limb-darkened (see eq.~[\ref{iout}] in Appendix~\ref{flux}).  As
mentioned earlier, in all figures in this paper we assume that the
inner boundary of the disk is at the marginally stable orbit, $r_{\rm
in} = r_{\rm ms}$. For comparison, the outgoing flux density when the
returning radiation is ignored ($F_0$, eq.~[\ref{f0}]) is also
shown. The flux densities $F_{\rm in}$ and $F_S$ always have the
following asymptotic behaviors: $F_{\rm in}\propto r^{-3}$,
$F_S\propto r^{-7/2}$ for $r\gg r_{\rm in}$. Thus, asymptotically,
$F_{\rm in}$ behaves as if it is produced by a ``lamp'' on the axis of
the black hole, while $F_S$ behaves like it is produced by a torque at
the inner boundary of the disk. We remind the reader of the asymptotic
behavior of $F_0$: $F_0\propto r^{-3}$ when $\eta =0$, $F_0\propto
r^{-7/2}$ when $\eta=\infty$. This asymptotic behavior is the same as
that in the Newtonian case (see, e.g., Syunyaev \& Shakura 1977),
which is not surprising since at large radii the effect of relativity
is unimportant.

From Figure~\ref{ret_flux_0999} we see that, for the case $\eta =0$
(upper panel), the flux density of the returning radiation is always
dominated by the original outgoing flux density $F_0$, except near the
inner boundary of the disk where $F_0$ and $F_S$ approaches zero but
$F_{\rm in}$ remains finite.  When $\eta$ is large (lower panel), at
small radii the flux density is dominated by the original outgoing
flux density $F_0$, while at large radii the flux density is dominated
by that of the returning radiation ($F_{\rm in}$). Since at large
radii $F_{\rm in}$ is always $\propto r^{-3}$, which is the same as
the asymptotic behavior of $F_0(\eta=0)$, the lower panel of
Figure~\ref{ret_flux_0999} indicates that at large disk radii the
signature of a large torque at the inner boundary is smeared out by
the returning radiation. This makes it hard to detect the torque at
the disk inner boundary by observing the spectrum of the disk (see
\S\ref{spectra} for more discussion).

From Figure~\ref{ret_flux_0999} we also see that $F_S$ is always less
important than $F_{\rm in}$.

A photon emitted by the disk has three possible fates: it may be
captured by the black hole, return to the disk, or escape to
infinity. Thus, when the returning radiation is considered, we have
three different definitions for the ``power'' of the disk: 1) The {\em
total emission} of the disk, which is the total energy-at-infinity
emitted by the disk per unit coordinate time (i.e., the total
energy-at-infinity carried away from the disk per unit coordinate time
by all photons emitted by the disk, whether the photons escape to
infinity, fall into the black hole, or return to the disk)
\begin{eqnarray}
    P_{\rm emit} = 4\pi\int_{r_{\rm in}}^\infty E^\dagger F_{\rm out}
        \, r dr \;,
\end{eqnarray}
where $E^\dagger$ is the specific energy of a disk particle on a circular
orbit of radius $r$ \citep{pag74,tho74}.
2) The {\em total power} of the disk, defined by
equations~(\ref{power}) and (\ref{power_t}), which is the total
energy-at-infinity carried away from the disk per unit coordinate time
by the ``net'' radiation, i.e., the photons that permanently leave the
disk---they either escape to infinity or fall into the black hole.  3)
The {\em total luminosity} of the disk, which is the total
energy-at-infinity carried away from the disk per unit coordinate time
by the photons that escape to infinity.

By definition, the total emission of the disk, $P_{\rm emit}$, can be
decomposed into three components: $P_{\rm emit,BH}$, for the radiation
that falls into the black hole; $P_{\rm emit,ret}$, for the radiation
that returns to the disk; and $P_{\rm emit,esc}$, for the radiation
that escapes to infinity.  Therefore, we can define three fractions
\begin{eqnarray}
    \iota_{\rm BH} \equiv \frac{P_{\rm emit,BH}}{P_{\rm emit}} \;, 
        \hspace{1cm}
    \iota_{\rm ret} \equiv \frac{P_{\rm emit,ret}}{P_{\rm emit}} \;, 
        \hspace{1cm}
    \iota_{\rm esc} \equiv \frac{P_{\rm emit,esc}}{P_{\rm emit}} \;.
    \label{frac1}
\end{eqnarray}
Clearly, the three ratios satisfy $\iota_{\rm BH}+\iota_{\rm ret}+\iota_{\rm 
esc} = 1$. 

By the conservation of energy, we must have
\begin{eqnarray}
    P_{\rm emit,esc} + P_{\rm emit,BH} = {\cal L}_{\rm total} \;,
    \label{conserv}
\end{eqnarray}
where ${\cal L}_{\rm total}$ is the total power of the disk. So we can
define two other ratios
\begin{eqnarray}
    f_{\rm BH} \equiv \frac{P_{\rm emit,BH}}{{\cal L}_{\rm total}} \;,
        \hspace{1cm}
    f_{\rm esc} \equiv \frac{P_{\rm emit,esc}}{{\cal L}_{\rm total}} \;,
    \label{frac2}
\end{eqnarray}
which respectively represent the fraction of the energy going into the
black hole and the fraction of the energy escaping to infinity in the 
``net'' energy radiated by the disk. The two ratios must satisfy $f_{\rm 
BH}+f_{\rm esc} = 1$.

The fractions with respect to the ``total'' radiation, $\iota_{\rm
ret}$, $\iota_{\rm BH}$, and $\iota_{\rm esc}$, are shown as functions
of the spin of the black hole in Figure~\ref{fraction}. Again, we assume
that the radiation emitted by the disk is isotropic in the disk frame.
Two extreme cases are shown: a standard Keplerian disk, where $g_{\rm in} 
= 0$ (equivalent to $\eta=0$, thin lines); and a nonaccreting disk, where
$\dot{M}=0$ but $g_{\rm in}\neq 0$ (equivalent to $\eta=\infty$, thick
lines).  For the $\eta=0$ case, we show the spin of
the black hole from $a=0$ to $a=0.9999M$. By considering the
thermodynamics of black holes, \citet{ago00} argued that a
nonaccreting disk ($\eta=\infty$) can exist only for $a>0.3594M$. A
physical explanation for this is that a black hole rotates faster than
the inner boundary of the disk (at the marginally stable orbit) only
if $a>0.3594M$.  Therefore, if $a\le 0.3594M$ a black hole cannot
exert a positive torque on the disk \citep{li00,li02a}. Hence, for the
case of a nonaccreting disk, we show the spin of the black hole from
$a=0.3594M$ to $a=0.9999M$.

The corresponding fractions with respect to the ``net'' radiation, $f_{\rm BH}$ 
and $f_{\rm esc}$, are shown in Figure~\ref{fraction2} (upper panel). To check
the conservation of energy, i.e. equation~(\ref{conserv}), the difference
between the computed $f_{\rm BH}+f_{\rm esc}$ and $1$ is also shown (lower
panel). Within the errors of the computation the conservation of energy is
confirmed. 

From Figures~\ref{fraction} and \ref{fraction2} we see that the
effect of the returning radiation crucially depends on the spin of the
black hole and the torque at the inner boundary of the disk. For a
standard accretion disk ($\eta = 0$) around a Schwarzschild black hole
($a=0$), $1.7\%$ of the total radiation emitted by the disk returns to the 
disk, $0.66\%$ is captured by the black hole, and the
remaining $97.6\%$ escapes to infinity. For a standard accretion disk
($\eta = 0$) around a Kerr black hole of $a=0.9999M$, $27\%$ of the total
radiation emitted by the disk returns to the disk, $4\%$ is captured by the 
black hole, and $69\%$ escapes to infinity. For a nonaccreting disk ($\eta 
= \infty$) around a Kerr black hole of $a= 0.9999M$, $59\%$ of the total 
radiation emitted by the disk returns to the disk, $7\%$ is captured by the 
black hole, and $34\%$ escapes to infinity. 
Therefore, the effect of the returning radiation is most important for a fast 
spinning black hole and a directly rotating disk (rotating in the same
direction as the black hole) with a large torque at its inner
boundary. The obvious reason for this is that as $a$ goes up the inner
boundary of the disk shrinks, and when $g_{\rm in}> 0$ more energy is
dissipated in and radiated from the inner region of the disk, which is
close to the central black hole.

In terms of the ``net'' radiation of the disk, for a standard accretion 
disk around a Schwarzschild black hole, about $0.7\%$ of the ``net'' 
radiation is captured by the black hole, and the remaining $99.3\%$ escapes to 
infinity. For a standard accretion disk around a Kerr black hole of $a=0.9999M$, 
about $6\%$ of the ``net'' radiation is captured and $94\%$ escapes to infinity. 
For a nonaccreting disk around a Kerr black hole of $a=0.9999M$, about 
$17\%$ of the ``net'' radiation is captured and $83\%$ escapes to infinity. 

Similar results have been obtained by \citet{ago00} using the
``transfer function'' approach. However, in their calculations, Agol
and Krolik ignored the stress of the returning radiation (i.e., the
term $F_S$) and obtained a fraction for the radiation captured by the
black hole in the ``net'' radiation that is somewhat smaller than the
fraction that we have obtained. For example, for a nonaccreting disk
around a Kerr black hole of $a=0.9999M$, they find $f_{\rm BH} =
15\%$, while we obtain $f_{\rm BH} =17\%$.

Figures~\ref{ret_flux_0999}--\ref{fraction2} show that when the black
hole is rotating rapidly and there is a large torque at the inner
boundary of the disk the effect of returning radiation is extremely important.

In order to study the effect of the radiation that is captured by the
black hole on the spinup/spindown of the black hole, following
\citet{tho74} we define a capture function $C$ for each photon emitted
by the disk at radius $r$: $C=1$ if the photon is eventually captured
by the black hole, and $C=0$ if the photon escapes to infinity or
returns to the disk.  The calculation of the capture function is
described in Appendix~\ref{con_bh}.

Here we focus on the case when the stress at the inner boundary of the disk is 
zero, i.e., $g_{\rm in} = 0$. Then, the torque that spins up/down the black 
hole comes from two sources: gas accreted from the disk, and radiation captured 
by the black hole. Then, the resultant spinup/spindown of the black hole is 
governed by \citep{tho74}
\begin{eqnarray}
    \frac{d a_*}{d\ln M} = \frac{L_{\rm in}^\dagger + \zeta_L}{
	   M\left(E_{\rm in}^\dagger + \zeta_E\right)} - 2 a_* \;, 
        \hspace{1cm} a_*\equiv \frac{a}{M} \;,
    \label{daM}
\end{eqnarray}
where $L_{\rm in}^\dagger$ is the specific angular momentum of disk particles
at the inner boundary, and
\begin{eqnarray}
    \zeta_E &\equiv& \frac{3}{2\pi r_g^2} \int_{r_{\rm in}}^\infty \left[
        \int_{\Omega_+}C \Upsilon(-n_t) \cos\theta d\Omega\right]
	   \tilde{f}_{\rm out}\,rdr \;, \label{zeta_e}\\
    \zeta_L &\equiv& \frac{3}{2\pi r_g^2} \int_{r_{\rm in}}^\infty \left(
        \int_{\Omega_+}C \Upsilon n_\varphi \cos\theta d\Omega\right)
	   \tilde{f}_{\rm out}\,rdr \;, \label{zeta_l}
\end{eqnarray}
where $r_g=M$ is the gravitational radius of the black hole, $n_t$ and 
$n_\varphi$ are given by equation~(\ref{ntf}). See Appendix~\ref{flux}
for the meanings of $\theta$, $\Omega$, $\Omega_+$, and $\Omega_-$.
For convenience, in 
equations~(\ref{zeta_e}) and (\ref{zeta_l}) we have used a dimensionless 
outgoing flux function $\tilde{f}_{\rm out}$ defined by
\begin{eqnarray}
    F_{\rm out} = \frac{3 \dot{M}_{\rm eff}}{8\pi r_g^2}\, \tilde{f}_{\rm 
	   out} \;. \label{ffout}
\end{eqnarray}

The function $d a_*/d\ln M$, which is defined by equation~(\ref{daM})
and is a function of $a_*$ only when $r_{\rm in}=r_{\rm ms}$, is
plotted in Figure~\ref{spin_func} (upper panel). The value of $a$ at
which $d a_*/d\ln M =0$ gives the equilibrium spin of the black hole 
[which is called the ``canonical state'' by \citet{tho74}]: $a=
a_{\rm eq}$. When the black hole is in a state of $a< a_{\rm eq}$, the
effect of accretion from the disk dominates, which will spin up the
black hole, until the equilibrium state $a= a_{\rm eq}$ is reached.
On the other hand, when the black hole is in a state of $a> a_{\rm
eq}$, the effect of the capture of photons dominates, which will spin
down the black hole to the equilibrium state $a= a_{\rm eq}$. When the
disk emission is isotropic, we obtain $a_{\rm eq} = 0.9983M$, and when
the disk emission is limb-darkened, we obtain $a_{\rm eq} =
0.9986M$.\footnote{These numerical results are correct only if
accretion from a thin disk and capture of photons is the unique
process to spin up/down the black hole, and if the torque at the inner
boundary of the disk is zero.} These results show that the
limb-darkening effect does not significantly affect the gross disk
radiation.

The efficiency of an accretion disk in converting rest mass into outgoing
radiation is defined by the ratio of the total luminosity of the disk
to the mass-energy accretion rate as measured at infinity, which is given
by $1-E_{\rm in}^\dagger-\zeta_E\,$ \citep{tho74}. In the lower panel of 
Figure~\ref{spin_func} we show the efficiency of a standard accretion disk 
as a function of the spin of the black hole when the effect of the returning 
radiation is considered. In the canonical state, the total efficiency is 
$0.309$ when the disk emission is isotropic and $0.315$ when the emission is
limb-darkened. When the effect of the returning radiation is ignored,
the radiation efficiency of a standard disk is $0.326$ when
$a=0.9983M$, and $0.331$ when $a=0.9986M$.

For comparison, the results of \citet{tho74} are also shown in the
lower panel of Figure~\ref{spin_func} (the two plus signs). Without
considering the effect of the radiation returning to the disk, Thorne
obtained: $a_{\rm eq} = 0.9978M$ when the disk radiation is isotropic,
and $a_{\rm eq} = 0.9982$ when the disk radiation is
limb-darkened. The corresponding radiation efficiencies of the disk
are respectively $0.302$ and $0.308$.

\subsection{Blackbody radiation spectrum from a Keplerian accretion disk}
\label{spectra}

The energy flux density of the blackbody radiation emitted by an
accretion disk around a black hole as observed by a remote
observer---$F_{E_{\rm obs}}$, where $E_{\rm obs}$ is the photon
energy---is given by equation~(\ref{feo3}). However, for the purpose
of comparison with X-ray observations, it is more convenient to use
the photon number flux density rather than the energy flux density. So
we define $N_{\rm obs} \equiv F_{E_{\rm obs}}/E_{\rm obs}$.

For convenience, let us define 
\begin{eqnarray}
    d\tilde{\Omega}_{\rm obs} \equiv \left(\frac{D}{r_g}\right)^2
        d\Omega_{\rm obs} \;,
\end{eqnarray}
where $D$ is the distance from the observer to the black hole and
$d\Omega_{\rm obs}$ is the element of the solid angle subtended by the
image of the disk on the observer's sky. With the above definition,
$d\tilde{\Omega}_{\rm obs}$ defined above is independent of the
distance $D$. Then, by equations~(\ref{feo3}) and (\ref{teff}), the
formula for calculating $N_{\rm obs}$ is
\begin{eqnarray}
    N_{E_{\rm obs}} = N_0 \left(\frac{E_{\rm obs}}{\rm keV}\right)^2
        \int \frac{d\tilde{\Omega}_{\rm obs}}{\exp\left[\mu \left(
	   \frac{E_{\rm obs}}{{\rm keV}}\right) g^{-1} \tilde{f}_{\rm 
	   out}^{-1/4}\right]-1} \;,
\end{eqnarray}
where $g$ is the photon redshift (defined by eq.~[\ref{red_shift}]), 
$\tilde{f}_{\rm out}$ is defined by equation~(\ref{ffout}), and
\begin{eqnarray}
    N_0 &=& 0.07205\, f_{\rm col}^{-4}\left(\frac{M}{M_\odot}\right)^2 
        \left(\frac{D}{\rm kpc}\right)^{-2} {\rm photons~}
	   {\rm keV}^{-1}{\rm cm}^{-2}\, {\rm sec}^{-1} \;, \\[1mm]
    \mu &=& 0.1202\, f_{\rm col}^{-1}\left(\frac{\dot{M}_{\rm eff}}{10^{18} 
	   {\rm g}\, {\rm sec^{-1}}}\right)^{-1/4} \left(\frac{M}{M_\odot}
        \right)^{1/2} \;,
\end{eqnarray}
where $f_{\rm col}$ is the spectrum hardening factor defined by 
equation~(\ref{f_col}).

Some examples of the observed spectra calculated with our ray-tracing
code are shown in Figures~\ref{spec_spin_incl}--\ref{spec_limb}.  In
Figures~\ref{spec_spin_incl}--\ref{spectra_torque_0999} we assume the
disk radiation is isotropic in the frame corotating with the disk,
i.e., there is no limb-darkening, whereas in Figure~\ref{spec_limb} we
include limb-darkening according to equation (\ref{iout}).
 
In the upper panel of Figure~\ref{spec_spin_incl}, we show the
dependence of the observed spectrum on the spin of the black hole. The
disk has $\eta=0$ (zero-torque) and an inclination angle
$\vartheta_{\rm obs} =30^\circ$ (for other parameters see the caption
of the figure). Different lines correspond to different spins of the
black hole: $a/M = 0, 0.5, 0.9$, and $0.999$ (left to right).  We see
that the spectrum becomes harder as the spin of the black hole goes
up. Physically, this is caused by the fact that when $M$ is fixed and
$a/M$ increases, the radius of the inner edge of the disk (by
assumption located at the marginally stable orbit) decreases, so that
the disk has a higher radiation efficiency (see Fig.~\ref{spin_func},
lower panel) and a higher temperature.  We also see another effect: as
$a/M$ increases, the flux density at low energies increases even
though this radiation comes from large radii where the spin of the
black hole should have negligible effect. This increase in flux density 
is caused by the returning radiation. As $a/M$ goes up, more radiation 
emitted by the disk in the
inner region is focused back to the disk by the gravity of the black
hole (Figs.~\ref{fraction} and \ref{fraction2}).  This increases the
radiation of the disk even at large radii (corresponding to the low
energy end of the spectrum).

In the lower panel of Figure~\ref{spec_spin_incl}, we test the effect
on the observed spectrum of the inclination angle of the disk. The
case considered is for $\eta =0$ and $a/M = 0.9$.  Different lines
correspond to different inclination angles of the disk:
$\vartheta_{\rm obs} = 0^\circ, 40^\circ, 70^\circ$ and $85^\circ$. At
the low energy end, the flux density goes down as $\vartheta_{\rm
obs}$ increases. This is caused by the projection effect.  The low
energy radiation is primarily emitted by the disk at large radii,
where the effect of relativity is not important. The projection causes
the flux density of the disk radiation to be proportional to
$\cos\vartheta_{\rm obs}$. At the high energy end, the flux density
goes up as $\vartheta_{\rm obs}$ increases. This is caused by the 
effects of Doppler beaming and gravitational focusing. The high
energy radiation is primarily emitted by the disk in the region near
the black hole, where the orbital velocity of the disk is mildly
relativistic so that special relativistic beaming boosts the disk radiation
to higher energy.  In addition, near the black hole, the gravity of
the black hole is strong and focuses the disk radiation back to the
disk plane, thereby modifying the projection effect. The joint action of the
two effects leads to an enhancement at the high energy end of the
observed spectrum.

Figure~\ref{spectra_0999} shows the effect of the returning radiation
on the observed spectrum of the disk. The three panels correspond to
$\eta = 0$ (upper panel, the standard disk case), $\eta =1$ (middle
panel), and $\eta =\infty$ (lower panel, the nonaccreting case). The
black hole has a spin $a/M = 0.999$, and the disk has an inclination
angle $\vartheta_{\rm obs} =30^\circ$ (other parameters are given in
the caption of the figure).  The spectra when returning radiation is
included are shown with solid lines, and those without the returning
radiation are shown with dashed lines.  We see that the returning
radiation enhances the disk radiation (especially at the high energy
end), and the effect is more prominent for a disk with a larger torque
at its inner boundary. For the standard disk with $\eta = 0$ (upper panel), 
where the power of the disk comes purely from disk accretion, the effect 
of the returning radiation is almost indistinguishable from the effect of 
a change in the mass accretion rate: The dotted line (almost coincident 
with the solid line) represents a disk spectrum minus the returning 
radiation with the same parameters except that the mass accretion rate is 
larger by a factor of $1.23$. For the case $\eta = 1$ (middle panel), for 
which the power of the disk comes equally from disk accretion and a torque 
at the inner boundary of the disk, the effect of the returning radiation 
can again be well approximated by adjusting the effective mass accretion
rate. The dotted line represents the spectrum without the returning
radiation of a disk with the same parameters except that the effective
mass accretion rate is larger by a factor of $1.7$. In this case also, the 
dotted line agrees very well with the solid line. However, when $\eta$ is 
very large, e.g. for the nonaccreting case with $\eta =\infty$ (lower panel), 
where the power of the disk comes purely from the torque at the inner
boundary, the effect of the returning radiation is so prominent that
it cannot be fitted by simply modifying the effective mass accretion
rate.

Figure~\ref{spectra_torque_0000} shows the effect of a nonzero torque
at the inner boundary of the disk for the case of a Schwarzschild
black hole ($a = 0$). Each panel corresponds to a different value of
$\eta$: $0.1$, $0.3$, $1$ and $16.49$. The last value corresponds to
the case that the total efficiency of the disk is equal to unity---the
upper limit for the efficiency of a Keplerian disk around a
nonrotating black hole. In each panel, the solid line is the spectrum
when the disk has a nonzero torque at the inner boundary, and the
dashed line is the spectrum when the disk has a zero torque with other
parameters remaining the same.  In particular, for both the solid and
the dashed lines the effective mass accretion rate is $\dot{M}_{\rm
eff} = 10^{19}{\rm g\,sec}^{-1}$, which means that the two disks have
the same total power (but different mass accretion rates $\dot{M}$,
see eqs.~[\ref{meff}] and [\ref{power_t3}]). From the figure we see
that the effect of the torque at the inner boundary is to make the
spectrum harder.  When the torque at the inner boundary is positive,
more energy is dissipated and more radiation is emitted in the inner
region of the disk where the temperature is higher. 
Also, the disk rotates faster there, so the Lorentz boost is more
prominent.  Both effects cause a hardening of the spectrum.

In the case of the returning radiation, we showed earlier that its
effect can be modeled very well by adjusting the mass accretion rate
in a model without returning radiation.  Can the effect of a nonzero
torque be similarly absorbed by adjusting the parameters of a
zero-torque model?  The first panel of Figure
\ref{spectra_torque_0000} shows that for $\eta=0.1$, which corresponds
to a weak torque, a zero-torque model with the same power ($\dot
M_{\rm eff}$) as the finite-torque model gives an almost
indistinguishable spectrum.  So, in this case, the answer to our
question is a definite yes.  For the cases of $\eta = 0.3$, $1$, and
$16.49$, we see that just keeping $\dot M_{\rm eff}$ the same is not
enough, since the dashed lines are noticeably different from the solid
lines.  However, by adjusting both $\dot M_{\rm eff}$ and $f_{\rm
col}$ of a zero-torque model, we can get a very good fit to the 
finite-torque models with $\eta = 0.3$ and $1$, as shown by the dotted lines;
the corresponding values of these parameters are given in the caption
to the Figure.  Only in the extreme case $\eta=16.49$ are we unable to
fit the spectrum by adjusting $\dot M_{\rm eff}$ and $f_{\rm col}$.
Note that adjusting $f_{\rm col}$ is equivalent to adjusting the
normalization, which could be done equally well by adjusting $M$,
$D$, or $\cos\vartheta_{\rm obs}$.  Since none of these parameters is
known precisely in a real system, one always has some freedom in the
normalization.  These results indicate that for a Schwarzschild black hole 
the spectrum of a disk with a modest torque at the inner boundary can be 
fitted with a zero torque at the inner boundary by adjusting $\dot{M}$ and 
the normalization.

Similar results for the case of a Kerr black hole with $a/M =0.999$
are shown in Figure~\ref{spectra_torque_0999}.  Now we find that the
spectrum can always be fitted by a disk with a zero torque at the
inner boundary by adjusting $\dot{M}_{\rm eff}$ and $f_{\rm col}$; in
fact, we can do this even when $\eta=\infty$.  Interestingly, the
effect of a nonzero torque is more important for a non-rotating black
hole than for a rapidly rotating hole, e.g., note that the spectrum is
significantly hardened in Figure~\ref{spectra_torque_0000} whereas
there is almost no effect in Figure~\ref{spectra_torque_0999}. This is
caused by the effect of the returning radiation. For a nonrotating
black hole, the effect of the returning radiation is not important due
to the fact that the inner edge of the disk has a large radius, so the
torque at the inner boundary produces a radiation flux density ($\sim
r^{-7/2}$ at large radii) that is distinctly different from the
radiation flux density arising from accretion ($\sim r^{-3}$ at large
radii). For a fast rotating black hole, the effect of the returning
radiation is important, it makes the radiation flux density of the
disk more or less similar to that arising from accretion (both
going as $\sim r^{-3}$ at large radii) (see
Figs.~\ref{ret_flux_0999}--\ref{fraction} and the relevant discussions
in \S\ref{return}).

Figure~\ref{spectra_torque_0999} shows another effect of the torque
at the inner boundary for a rapidly spinning hole: when the torque
changes from zero to nonzero and other parameters (including
$\dot{M}_{\rm eff}$) are left unchanged, the flux at infinity decreases
(compare the solid and dashed lines). This is caused by the fact
that when the torque at the inner boundary is nonzero more energy is
dissipated in the inner region of the disk; for a rapidly rotating
black hole the inner boundary of the disk is closer to the horizon of
the black hole and so more radiation is focused to the equatorial plane.

Overall, the results shown in Figures
\ref{spectra_0999}--\ref{spectra_torque_0999} suggest that
when modeling observational data the effects of returning radiation
and nonzero torque can be ignored since they can be absorbed by
modifying the mass accretion rate and the spectral hardening factor of
a zero-torque model without returning radiation.

Finally, in Figure~\ref{spec_limb} we show the effect of
limb-darkening (eq.~[\ref{iout}], Appendix~\ref{flux}) on the observed
spectra of the disk. As can be expected from equation~(\ref{iout}),
compared to the case when the disk emission is isotropic, when the
disk emission is limb-darkened we see more radiation when the disk has
a low inclination angle, and we see less radiation when the disk has a
high inclination angle. The effect of limb-darkening is most important
when the disk is nearly edge-on: in this case the effect cannot be
absorbed by simply adjusting the mass accretion rate and the spectral
hardening factor.

\section{Modeling the Spectra of Black Hole X-ray Binaries}
\label{novae}

To test the performance of our model KERRBB and to compare it with other
models of disk spectra, we have analyzed spectral data on three black
hole X-ray binaries: 4U1543--47 (hereafter U1543), XTE J1550-564
(hereafter J1550), and GRO J1655--40 (hereafter J1655).  The data were
obtained using the Proportional Counter Array aboard the {\em Rossi
X-ray Timing Explorer} ({\em RXTE}).  The data themselves and the methods 
of data analysis are thoroughly described in Zimmerman et al. 2004
(hereafter Z04).  These three black hole binaries have all had recent
outbursts that have been analyzed in detail using {\em RXTE}
observations.  \citet{par04} studied the 2002 outburst of U1543;
\citet{sob00} analyzed the outburst of J1550 in 1998-1999; and
\citet{sob99} analyzed J1655's 1996-1997 outburst.  In fitting
the spectra of these sources, we followed as closely as possible the
procedures described in these papers.

In addition to KERRBB and the other disk models discussed below, we
used four other supplementary XSPEC models in our fits.  Primary among
these was the power-law model, which has two parameters: the photon
index, $\Gamma$, and a normalization constant, which we will call
$K_{\rm PL}$.  The remaining three spectral components, which model
the effects of interstellar absorption, a smeared Fe absorption edge,
and a Gaussian Fe line, are relatively unimportant.  Because our focus
was on the disk models, we froze a number of the extraneous parameters
in these other spectral components to minimize their influence on our
fits by fixing the parameters at average values obtained from the
previously cited papers (see Z04).  The spectra were fitted using
XSPEC over the energy range $\approx 3-20$ keV.  All of the data sets
correspond to the high/soft state, for which the accretion flow is
believed to be dominated by a geometrically thin, optically thick disk
component of emission that contributes $\gtrsim 90$\% of the flux
\citep{mcc04}.  For further details on the data and the analysis
techniques, see Z04.

In addition to (i) KERRBB, the other XSPEC models we considered were:
(ii) GRAD: a general relativistic code for non-spinning black holes
(Hanawa 1989); (iii) DISKPN: a disk model employing a pseudo-Newtonian
potential (Gierli\'nski et al. 1999); (iv) EZDISKBB: a Newtonian model
with a zero-torque inner boundary condition (Z04); and (v) DISKBB: a
Newtonian model (Mitsuda et al. 1984) which includes a finite torque
at the inner edge of the disk (Gierli\'nski et al. 1999; Z04).  Note
that KERRBB is the most complete model of these five since it includes
the effects of general relativity and can handle a black hole with any
spin and any nonnegative torque at the inner edge of the disk.  In
this section we assume that the torque is zero ($\eta=0$).  Likewise
the torque at the inner disk edge is zero for all of the other models
except DISKBB (see below).  For the specific case of a non-spinning
black hole ($a=0$), KERRBB should agree with GRAD, but there are in
fact some deviations because of an error in GRAD discussed in
Appendix~\ref{grad}.  In order to facilitate comparison between KERRBB
and GRAD, we did not include the returning radiation in KERRBB in the
calculations reported here.  As we have shown in \S\ref{spectra}, the
effect is in any event small for $\eta=0$, and can be absorbed in the
fitted value of $\dot M$.  Of the other three models, DISKPN should be
closest to KERRBB since it attempts to include some relativistic
effects through the pseudo-Newtonian potential of Paczy\'nski \& Wiita
(1980).  The main difference between the final two Newtonian models is
that EZDISKBB includes a zero-torque boundary condition at the inner
edge of the disk whereas DISKBB has a finite torque.  As argued in
Z04, the zero-torque condition is expected to be valid in a number of
situations.  Furthermore, even in those cases in which a finite torque
might be expected, it is unlikely that the specific magnitude of the
torque would be equal to the value assumed in DISKBB.

\subsection{Comparison of normalizations}

For our first comparison, we analyzed data from ten epochs each on
U1543 and J1550 (see Z04 for details of the particular observations).
We fixed the mass of the black hole and the inclination of the disk at
their respective estimated values of $9.4M_\odot$ and $20.7\degr$ for
U1543 \citep{oro98,oro03} and $10.6M_\odot$ and $73.5\degr$ for J1550
\citep{oro02}.  We also assumed that the two black holes are not
spinning ($a=0$) and fixed the inner edge of the disk at the
marginally stable orbit ($r_{\rm in} = r_{\rm ms} = 6M$).  The
constancy of this inner disk radius over a wide range of X-ray
luminosity has been established in synoptic studies of several black
hole binaries (Tanaka \& Lewin 1995; Sobczak et al. 2000).  Although
there are reasonable estimates of the distances to the sources, we
left the distance $D$ as an adjustable parameter; in effect, this
parameter played the role of a normalization in these calculations.
We also left the mass accretion rate $\dot M$ as a free parameter, and
used a spectral hardening factor $f_{\rm col}=1.7$ \citep{shi95}.

Figure \ref{U1543_norm} shows the results of fitting the data on U1543
with the five models.  We see a remarkably consistent pattern.  KERRBB
and GRAD agree almost perfectly in their estimates of both $D$ and
$\dot M$; DISKPN has modest deviations from these two models; EZDISKBB
deviates somewhat more; and DISKBB deviates in both parameters by an
enormous factor.  Table 1 gives the average ratio by which the
estimated parameters obtained with each of the models deviates from
the value obtained with KERRBB.  Here and in what follows, we assume
that the result from KERRBB is correct and view any deviation from it
as a measure of the error in a particular model.

To understand the patterns seen in Figure \ref{U1543_norm}, we note
that each model adjusts two parameters to the data.  These parameters
are obtained essentially by fitting the integrated flux from the
source and the position of the peak in the spectrum; the latter in
effect measures the peak temperature (as observed at infinity) of the
emission from the disk.  For a non-rotating black hole, the radiative
efficiency of KERRBB and GRAD is $\epsilon = 1-(8/9)^{1/2}=0.057$;
that is, for a given $\dot M$, the luminosity is equal to $0.057\dot
Mc^2$.  In contrast, DISKPN has $\epsilon = 0.0625$ and EZDISKBB has
$\epsilon = 0.0833$.  Thus, these models produce the same luminosity
with a smaller $\dot M$.  Moreover, for a given luminosity, the fully
relativistic models KERRBB and GRAD include Doppler and gravitational
redshift factors, whereas the other models do not.  These cause
additional deviations.  The net effect is that the estimate of $\dot
M$ obtained with DISKPN is lower than the correct relativistic result
by a factor of $\sim 0.6$, while $\dot M$ with EZDISKBB is only a
quarter of the correct value.  The variations in the derived values of
$D$ with the various models are straightforward to understand --- each
model adjusts $D$ so as to fit the observed flux $\propto {\rm
luminosity}/D^2$.

Compared to the other models, DISKBB makes an extraordinarily large
error: the estimate of $\dot M$ is reduced by more than a factor of 50
relative to KERRBB.  The reason for this is well understood (e.g.,
Kubota et al. 1998; Gierli\'nski et al. 2001; Z04), though not widely
appreciated.  DISKBB assumes a finite torque at the inner edge and so
its radiative efficiency is three times that of EZDISKBB: $\epsilon=
0.25$.  Furthermore, this model has its temperature maximum at the
inner edge of the disk, whereas all the other models have their maxima
at larger radii.  Therefore, it predicts a substantially larger value
of $T_{\rm max}$ for a given luminosity.  Since the spectral fit tries
to reproduce the position of the peak in the observed spectrum, both
effects act in the same direction and lead to a large decrease in the
estimated value of $\dot M$.  Despite this well-known deficiency in
DISKBB, the model is still widely used to model spectra of black hole
X-ray binaries.  

Some authors have used DISKBB but included correction factors to make
the results more consistent with other models. For example, \citet{zha97}
and \citet{gie04} apply correction factors to the temperature and the 
flux derived from DISKBB. Other authors have attempted to include these 
corrections in the spectral hardening factor $f_{\rm col}$ (see Davis et 
al. 2004 for a review of this topic).  While these approaches have
some merit, we feel it is better to use the correct model, viz.,
EZDISKBB if one wishes to assume a Newtonian model, DISKPN if one is
interested in a pseudo-Newtonian potential, and KERRBB if one would
like to include all relativistic effects including returning
radiation, and would like to consider different black hole spins and
inner torques.  (GRAD is valid only for a non-rotating hole.) Indeed,
if we apply the approach of \citet{zha97} and \citet{gie04} to DISKBB
or EZDISKBB for U1543, where we expect $\dot{M}/D^2 \propto 1/g_{\rm 
GR}$ where $g_{\rm GR}$ is the relativistic correction factor
introduced by \citet{zha97}, we get $\dot{M}/D^2 \approx 0.22$ for the
DISKBB results, and $\dot{M}/D^2 \approx 0.68$ for the EZDISKBB results
in Table~1. These corrections improve somewhat the results of the
models without relativistic corrections ($\dot{M}/D^2 \approx 0.18$
for DISKBB and $\approx 0.56$ for EZDISKBB), but are still far from the
correct relativistic results of KERRBB, which by definition imply 
$\dot{M}/D^2 = 1$.

In the case of J1550, which is nearly edge-on, we find that the
relativistic corrections introduced by \citet{zha97} do not improve
the results of DISKBB and EZDISKBB at all. Indeed, from Table~1 of
\citet{zha97}, by interpolation we have $g_{\rm GR} \approx 1.14$
for $\theta = 73.5^\circ$ and $a_* = 0$, which brings the results of
DISKBB and EZDISKBB farther from the results of KERRBB (see Table~2).

We turn now to a comparison of the results for J1550 obtained using the
five models; these results are summarized in Figure~\ref{J1550_norm} 
and Table~2. Similar patterns are seen as in the case of U1543,
but there are also differences.  The latter are all caused by the fact
that U1543 is a nearly face-on system ($\vartheta_{\rm
obs}=20.7\degr$) whereas J1550 is a nearly edge-on system
($\vartheta_{\rm obs}=73.5\degr$).  For the three nonrelativistic
models, viz., DISKPN, EZDISKBB, and DISKBB, the effect of inclination
is straightforward: the flux simply decreases by a factor of $\cos
\vartheta_{\rm obs}$, and there is no change in the spectral shape.
However, the two relativistic models KERRBB and GRAD have additional
effects.  First, they have stronger Doppler beaming for an edge-on
system like J1550 and consequently appear brighter and hotter for a
given $\dot M$.  Second, because of light deflection, the effective
projected area of the disk as viewed by the observer is larger than
one might expect with a simple $\cos \vartheta_{\rm obs}$ scaling,
especially for the hot inner regions of the disk.  For both reasons,
these two models are able to fit a given observed flux with a smaller
value of $\dot M$ than one might expect by naively scaling from a
face-on system like U1543.  As a result the ratios in Table 2
corresponding to DISKPN, EZDISKBB, and DISKBB are higher by a factor of
$\sim 2-3$ relative to those in Table 1.  In addition, we see that the
values obtained with GRAD are higher than those from KERRBB.  This is
the result of the error in GRAD pointed out in Appendix~\ref{grad}.
The error has almost no effect for a face-on system like U1543, but
becomes more serious for edge-on systems.

Finally, we note that the $\chi^2$ values of the fits vary somewhat
erratically from one epoch to the next.  Epochs 2 and 3 of U1543 are
relatively bad for all models, while epochs 6 and 9 of J1550 are bad
for three of the models (DISKPN, EZDISKBB, and DISKBB) but perfectly
reasonable for the other two models (KERRBB and GRAD).  In the case of
our third source, J1655, we obtained large $\chi^2$ values for all
except the first two epochs.  For this reason, we did not include this
source for the calculations discussed in this subsection.

\subsection{Estimating the black hole spin}

In the previous subsection, we treated the distance $D$ to each source
as a free parameter (equivalent to the normalization) and fitted $D$
and $\dot M$ from the data.  However, we do have independent estimates
of the distances: $7.5\pm1.0$ kpc for U1543 \citep{oro98,oro02}; 
$5.9_{-3.1}^{+1.7}$ kpc for J1550 \citep{oro02}; and
$3.2\pm0.2$ kpc for J1655 \citep{hje95}.  By
using this additional information we should in principle be able to
constrain the spin of the black hole (Zhang et al. 1997).  (Note
that we fixed $a=0$ in \S4.1.)  In the case of KERRBB, the calculation
is straightforward---we simply fix the value of $D$ and let $a_*\equiv
a/M$ and $\dot M$ be the free parameters.  For the other models,
however, $a_*$ is not an adjustable parameter.  In the case of the
three non-relativistic models, DISKPN, EZDISKBB, and DISKBB, we allow
the radius of the inner edge $r_{\rm in}$ and $\dot M$ to be the free
parameters.  Having fitted $r_{\rm in}$ from the data, we then use the
ratio $r_{\rm in}/M$ to estimate $a_*$, assuming that the inner edge
is at the marginally stable orbit corresponding to the particular spin
parameter.  This procedure is somewhat arbitrary, but is fairly
standard practice in the literature.  We have not attempted a similar
exercise with GRAD.

Figure~\ref{BH_spin} shows the estimated spin parameter $a_*$ for each
of the 10 observations of U1543 and J1550, and the first two observations 
of J1655 (with $M=7.0M_\odot$ and $\vartheta_{\rm obs} =69.5\degr$; Orosz 
\& Bailyn 1997).  We have ignored the remaining 8 observations of
J1655 because (for some unexplained reason) the $\chi^2$ values are
very large.  In the case of U1543, KERRBB gives a consistent estimate
of $a_* \sim 0.6$ for all the epochs.  The near-constancy of the
estimate is notable, especially since the luminosity and the mass
accretion rate do vary from one epoch to the next (see
Fig. \ref{U1543_norm}).  The models DISKPN and EZDISKBB give different
values of $a_*$, suggesting that it is dangerous to use these models
to estimate the black hole spin.  The model DISKBB is especially
poor---in the case of both U1543 and J1550, it gives estimates of the
disk inner edge that are too large ($>9M$) to be consistent with any
choice of the spin parameter.  The results are more variable in the
case of J1550.  We find with KERRBB values of $a_*$ ranging all the
way from $-0.8$ to $-0.1$.  In part this is because the spectra are
less sensitive to the value of $a_*$ when the parameter is negative,
but in part it might also reflect the quality of the data (e.g., the
very uncertain distance to the source; see below).  In the case of
J1655, the two epochs that we have analyzed give $a_*\sim 0.6-0.7$,
slightly larger than for U1543.

The ability to estimate $a_*$ depends on having independent estimates
of the disk inclination $\vartheta_{\rm obs}$, the black hole mass
$M$, the distance to the source $D$, and the spectral hardening factor
$f_{\rm col}$.  However, even for the best sources, there are
substantial uncertainties in these parameters.  Table 3 shows for one
epoch of U1543 how the fitted value of $a_*$ varies as each of the
four input parameters is allowed to range over its $1\sigma$ range of
uncertainty.  Leaving aside $f_{\rm col}$, we see that the
uncertainties in the other three parameters are not particularly
severe for this favorable case of U1543.  Even allowing for the
uncertainties it appears that one could infer that $a_* > 0.5$.  The
spectral hardening factor, however, introduces a large uncertainty.
The value of this parameter is not well constrained.  \citet{shi95}
suggested a value of 1.7, but said that the value could be anywhere in
the range from 1.5 to 1.9 (the range covered in Table 3).  A recent
comprehensive analysis by Davis et al. (2004), which includes better
opacities, comes down in favor of somewhat smaller values in the range
$f_{\rm col} \sim 1.5-1.6$, depending on the luminosity of the disk.
For such values, the spin estimates go up by a modest amount, to $a_*
\sim 0.7-0.8$ for U1543 and $\sim 0.8$ for J1655.  Apart from the
question of what is the correct value of $f_{\rm col}$ to use, a more
serious concern is that it is a rather severe simplification to model
the spectral modification due to Comptonization and opacity effects
with a single scaling parameter $f_{\rm col}$.  It is hard to quantify
the error from this approximation.  In our opinion, if one wishes to
estimate black hole spin via spectral fitting, one must first develop
more reliable spectral models of disk atmospheres.  The work of Davis
et al. (2004) is a first step in this direction.

In the case of J1550, the distance is highly uncertain: $D = 2.8-7.6$
kpc (Orosz et al. 2002).  Correspondingly, $a_*$ is poorly
constrained.  We analyzed the 10 observations of this source using the
two extreme $1\sigma$ values of $D$ and obtained values of $a_*$
ranging all the way from $-1$ to $+0.7$.  This emphasizes once again
that we cannot hope to constrain the spin of a black hole via spectral
fitting unless we have accurate input parameters.

\section{Summary}
\label{sum}

We have developed a ray-tracing computer code to calculate the spectrum 
of a thin accretion disk around a black hole, assuming that the emission 
from each point on the surface of the disk is locally blackbody-like with 
a constant spectral hardening factor $f_{\rm col}$.  The code includes
all relativistic effects.  It also includes the effect of
self-irradiation whereby radiation emitted at one point on the disk is
deflected by the gravity of the black hole and illuminates another
part of the disk.  The code can handle any value of the black hole
spin, the disk inclination, the disk inner radius ($r_{\rm in} \geq
r_{\rm ms}$), and the torque on the inner edge (the dimensionless
parameter $\eta$).  It also allows the user to choose between
isotropic emission and a Chandrasekhar (1950) limb-darkening law
(eq.~[\ref{iout}]). 

In terms of the scope of the calculations, the present work is not
markedly different from previous studies.  For instance, Fabian et
al. (1989) and Stella (1990) calculated line emission from a
relativistic disk around a non-rotating black hole, while Laor (1991)
repeated the calculation for a rapidly rotating black hole with a
specific value of $a/M = 0.998$ (i.e., a nearly maximally rotating
hole).  Our code calculates the continuum emission, rather than the line
emission, and is more comparable to the blackbody continuum models
calculated by \citet{han89} and \citet{ebi91} for a disk 
around a non-rotating black hole and by \citet{cun75,cun76} and 
\citet{gie01} for a disk around a rotating black hole.  Our model
supersedes Cunningham's model by including a torque at the inner
boundary of the disk, and that of Gierli\'nski et al. by including both
the torque and the effect of the returning radiation.  The calculations 
of Agol \& Krolik (2000) are very close
to the present work.  The only difference is that they did not include
the additional energy released as a result of the torque applied by
the returning radiation (the term $F_S$ in \S3.1 and Appendix~\ref{flux}).  
As we have shown, this term is quantitatively not important.

Using the ray-tracing code, we have calculated tables of spectra and 
developed a model called KERRBB for use with the X-ray spectral fitting 
package XSPEC.  This model can handle black hole spins in the range 
$a_* = a/M= -1$ to $+0.9999$, inclinations in the range $\vartheta_{\rm
obs}=0\degr$ to $85\degr$, and torque parameters in the range $\eta =
0$ to 1.  One can choose either isotropic emission or limb-darkened
emission from the disk (eq.~[\ref{iout}]).  In addition, as usual, one
can set (or fit) the black hole mass $M$, the distance to the source
$D$, the mass accretion rate $\dot M$, and the spectral hardening
factor $f_{\rm col}$.  KERRBB thus goes well beyond other models that
are presently available in XSPEC for fitting the continuum emission
from thin disks.  The other models are based on a Newtonian potential
(DISKBB, EZDISKBB) or a pseudo-Newtonian potential (DISKPN), or if
they solve the relativistic problem it is only for a specific case,
e.g., a non-rotating black hole (GRAD, see Appendix~\ref{grad}).
KERRBB is more general and complete than any of these models.  One
limitation of KERRBB is that the present implementation in XSPEC only
considers disks that extend down to the marginally stable orbit
($r_{\rm in} = r_{\rm ms}$).  Disks that evaporate into a corona/ADAF
at radii $r_{\rm in} >r_{\rm ms}$ are not yet modeled, but could
easily be incorporated in the future if there is a need.

The effects of some of the model parameters on the observed spectrum
are illustrated in Figures~\ref{spec_spin_incl}--\ref{spec_limb}.  The 
black hole spin parameter $a/M$ and the inclination angle $\vartheta_{\rm
obs}$ are seen to have a strong effect on the spectrum.  These model
parameters must clearly be included when fitting spectral data.  The
effect of the returning radiation appears to be less important.  At
least for torque parameters $\eta \lesssim 1$, its effect can be
almost completely absorbed by a rescaling of the mass accretion rate.
Likewise, the effect of limb-darkening, which is
important only when the disk is nearly edge-on, can otherwise be absorbed
by adjusting the mass accretion rate and the normalization 
(Fig.~\ref{spec_limb}).

\citet{afs03} have argued that a thin accretion disk will have a negligible 
torque at its inner edge.  Magnetohydrodynamic simulations of disks have
found nonzero torques \citep{haw02}, but the magnitude appears to be
relatively small ($\eta\sim 0.2$) even for relatively thick disks
($H/R\sim 0.1-0.2$).  For such values of $\eta$, and indeed even for
larger values up to $\eta\sim1$, Figures~\ref{spectra_torque_0000} and
\ref{spectra_torque_0999} show that the effect of a finite torque can
be modeled with a zero-torque disk by simply adjusting the mass
accretion rate and the normalization (however, see the caveat in the
following paragraph).  Since $\dot M$ is almost always fitted to the
data, and the normalization is rarely known accurately a priori
because of uncertainties in the black hole mass, source distance, disk
inclination, etc., it appears that there is no particular advantage to
including a finite torque in fitting spectral observations.
Nevertheless, we have retained $\eta$ as a parameter in KERRBB for the
occasions when it might be needed.

The following additional caveat must be noted with respect to models
with a finite torque at the disk inner edge.  Most such models have
continued viscous dissipation at radii inside $r_{\rm ms}$ (e.g.,
Krolik 1999; Gammie 1999).  KERRBB calculates the effect of the torque
and the radiation it produces only for radii $r\geq r_{\rm in}=r_{\rm
ms}$, and does not include the emission from smaller radii.  The model
is thus incomplete.  This is one more reason for not including a
finite torque in fitting spectral data with KERRBB.  When $\eta=0$, no
additional energy dissipation is expected for $r<r_{\rm in}$.  In this
case, KERRBB includes all the emission and is self-consistent.

We have compared KERRBB with the various other models mentioned above:
GRAD, DISKPN, EZDISKBB, DISKBB.  As \S\S3.1, 3.2,
Figures~\ref{U1543_norm}--\ref{BH_spin} and Tables~1 and 2 show, the
differences are quite large (except in the case of GRAD), indicating
that these models make significant errors in computing the spectra of
disks.  The errors are the result of the approximations made by the
models, e.g., Newtonian or pseudo-Newtonian dynamics, neglect of
Doppler and redshift effects, inappropriate torque at the inner edge,
etc. When the relativistic corrections of \citet{zha97} are applied
to DISKBB and EZDISKBB, the results are improved somewhat for the 
low inclination case, but the differences from the results of KERRBB
are still significant. For the high inclination case the relativistic 
corrections of \citet{zha97} do not improve the nonrelativistic 
results of DISKBB and EZDISKBB at all.

In \S3.2 we showed that, in principle, if we knew $M$, $\vartheta_{\rm
obs}$ and $D$, then we could determine the spin parameter $a_*$ of the
black hole by fitting spectral data.  This was first attempted by
Zhang et al. (1997) for the black hole X-ray binary J1655 (also, see
Sobczak et al. 1999).  For this approach to work, we need fairly
accurate estimates of $M$, $\vartheta_{\rm obs}$ and $D$ (see Table
3), which are not always available (e.g., the distance to J1550 is
very uncertain and so the estimates of $a_*$ obtained for this source
in Fig. 12 are highly suspect).  Moreover, even if we did have
accurate estimates of these parameters, the results would still be
uncertain because of the questionable blackbody assumption.
Non-blackbody effects are usually modeled by means of the spectral
hardening factor $f_{\rm col}$, and a value $f_{\rm col}=1.7$ is
conventionally used (Shimura \& Takahara 1995).  Using this value, and
allowing $\vartheta_{\rm obs}$, $M$ and $D$ to vary over their allowed
1$\sigma$ ranges, we estimate for the source U1543 that
$a_*\sim0.5-0.7$ (Table~3).  If instead we assume $f_{\rm col} \sim
1.5-1.6$, as in the recent paper of Davis et al. (2004), then we find
a somewhat more rapidly rotating black hole with $a_* \sim
0.7-0.8$.  On the other hand, if we allow the full range of
uncertainty as estimated by Shimura \& Takahara (1995), viz. $f_{\rm
col} \sim 1.5-1.9$, then $a_*$ could be as low as $\sim 0.3$. This
illustrates that the biggest obstacle to estimating black hole spin
via this method is the uncertainty in modeling the emitted spectrum
from the disk.  The crude treatment in terms of a simple hardening
factor $f_{\rm col}$ needs to be improved significantly before we can
make full use of spectral observations of black hole accretion disks.
More work along the lines of Davis et al. (2004) is highly desirable.

Other limitations include the fact that the model assumes that the disk is
perfectly flat with zero vertical thickness, and all the returning radiation
is absorbed by the disk and reemitted as blackbody radiation.  A real disk 
has a finite thickness which moreover varies with radius (i.e., the disk is
flared); it probably also has a substantial warp.  These complications
will modify the effect of returning radiation and will also change the
geometrical projection factors needed when computing the observed
spectrum.  Furthermore, at least a part of the returning radiation is 
scattered by the
disk, which adds a high-energy component to the spectrum \citep{cun76}.
Modeling such effects would involve too many additional parameters and we 
have not attempted it.

\acknowledgments

LXL acknowledges Massimo Calvani and Claudio Fanton for discussions on
ray-tracing and for sharing their computer code, Craig Gordon and
Keith Arnaud for helps on loading KERRBB to XSPEC.  LXL's research was
supported by NASA through Chandra Postdoctoral Fellowship grant number
PF1-20018 awarded by the Chandra X-ray Center, which is operated by
the Smithsonian Astrophysical Observatory for NASA under contract
NAS8-39073.  EZ, RN and JEM's research was supported in part by NASA
grants NAG5-9930 and NAG5-10780 and NSF grant AST 0307433.

\appendix

\section{Integration of Photon Orbits in a Kerr Spacetime I: The $E_\infty \neq 
0$ Case}
\label{integral}

The general orbit of a photon (indeed for any particle) in a Kerr spacetime
is described by three constants of motion \citep{car68}: the energy-at-infinity
$E_\infty$, the angular momentum about the axis of the black hole $L_z$, and
another constant ${\cal Q}$ (when $a=0$, ${\cal Q}+L_z^2$ is the square of the 
total angular momentum). Let us write $L_z = \lambda E_\infty$ and ${\cal Q} = 
Q E_\infty^2$.\footnote{The case of $E_\infty\approx 0$ is treated separately 
in Appendix~\ref{integral_p}.} Then, the equations governing the orbital 
trajectories are separable. Since the system is stationary and axisymmetric, 
only the motions in the $r$- and $\vartheta$-directions are required 
in the calculation of the radiation spectrum from the disk. The motion in the 
$r-\vartheta$ plane is governed by 
\citep{bar72,cha83}
\begin{eqnarray}
    \int_{r_e}^r \frac{dr}{\sqrt{R(r)}} = \pm 
        \int_{\vartheta_e}^\vartheta \frac{d\vartheta}{\sqrt{\Theta(
	   \vartheta)}} \;,
    \label{r-th}
\end{eqnarray}
where
\begin{eqnarray}
    R(r) &=& r^4 + \left(a^2-\lambda^2-Q\right)r^2 +2M\left[Q+(\lambda-a)^2
        \right]r - a^2 Q \;, \label{rr} \\[2mm]
    \Theta(\vartheta) &=& Q + a^2 \cos^2\vartheta - \lambda^2 \cot^2\vartheta\;,
        \label{thth}
\end{eqnarray}
and $r_e$ and $\vartheta_e$ are the starting values of $r$ and $\vartheta$. The 
$\pm$ signs in equation~(\ref{r-th}) must be carefully chosen according to the
sign of $dr/d\vartheta$ on the orbit of the photon.

Define $\mu = \cos\vartheta$, then equation~(\ref{r-th}) becomes
\begin{eqnarray}
    \int_{r_e}^r \frac{dr}{\sqrt{R(r)}} = \pm 
        \int_{\mu_e}^\mu \frac{d\mu}{\sqrt{\Theta_\mu(\mu)}} \;,
    \label{r-mu}
\end{eqnarray}
where $\mu_e = \cos\vartheta_e$ and
\begin{eqnarray}
    \Theta_\mu(\mu) = Q + (a^2-\lambda^2-Q)\mu^2 - a^2\mu^4 =
        a^2\left(\mu_-^2+\mu^2\right)\left(\mu_+^2-\mu^2\right)\;,
        \label{thmu}
\end{eqnarray}
and $\mu_\pm^2$ are defined by
\begin{eqnarray}
    \mu_\pm^2 = \frac{1}{2a^2}\left\{\left[\left(\lambda^2+Q-a^2\right)^2
	    +4a^2Q\right]^{1/2}\mp\left(\lambda^2+Q-a^2\right)\right\} \;.
    \label{mupm}
\end{eqnarray}
For a photon crossing the equatorial plane (which is the case that we are 
interested in), we have $Q>0$. Then, both $\mu_+^2$ and $\mu_-^2$ are 
nonnegative. Note, $\mu_+^2 \mu_-^2 = Q/a^2$.

For a photon emitted by the disk, we have $\vartheta_e=\pi/2$ and $\mu_e=0$, 
so $\mu^2$ can never exceed $\mu_+^2$ since otherwise $\sqrt{\Theta_\mu}$
and $\sqrt{\Theta}$ become imaginary. Then, the integral over $\mu$ can be 
worked out with the inverse Jacobian elliptic integral 
\citep[eq.~213.00]{byr54}
\begin{eqnarray}
    \int_\mu^{\mu_+}\frac{d\mu}{\sqrt{\Theta_\mu}} = 
        \int_\mu^{\mu_+}\frac{d\mu}{\sqrt{a^2\left(\mu_-^2+\mu^2\right)
	   \left(\mu_+^2-\mu^2\right)}} = \frac{1}{\sqrt{a^2\left(\mu_+^2+
        \mu_-^2\right)}}\,\cn^{-1} \left(\left.\frac{\mu}{\mu_+}\right|
        m_\mu\right) \;,
    \label{mu_int}
\end{eqnarray}
where $0\leq \mu < \mu_+$ and
\begin{eqnarray}
    m_\mu = \frac{\mu_+^2}{\mu_+^2+\mu_-^2}\;. \label{m_mu}
\end{eqnarray}
In this paper we use the convention for the modulus in an elliptic 
integral (e.g., $m_\mu$ in the above inverse Jacobian elliptic integral)
which is the same as that of \cite{abr72}, and differs from that of 
\cite{byr54} by a square.

The integral over $r$ can also be worked out with inverse Jacobian elliptic 
integrals. To do so, we need to find out the four roots of $R(r)=0$. Since
$R(r\rightarrow\pm\infty)=\infty$ and $R(r=0)=-a^2 Q\leq 0$, $R(r)=0$ has
at least two real roots. The remaining two roots can be either real or complex
(in the latter case the two must be complex conjugates of each other). We 
consider the two cases separately:

{\em Case A. $R(r)=0$ has four real roots~~} Let us denote the four roots by 
$r_1$, $r_2$, $r_3$, and $r_4$, with $r_1 \ge r_2 \ge r_3 \ge 
r_4$. By equation~(\ref{rr}), we have $r_1+r_2+r_3+r_4 = 0$ and $r_1 r_2 
r_3 r_4 = -a^2 Q \leq 0$; the latter implies that $r_4$ must be $\le 0$. 
Physically allowed regions for photons are given by $R\ge 0$, i.e., $r \ge r_1$ 
(region I) and $r_3 \le r \le r_2$ (region II). (Although $R \ge 0$ also for 
$r \le r_4$, this region is unphysical since $r_4 \le 0$.)

In region I ($r \ge r_1$), the integral over $r$ can be worked out by the
following integration \citep[eq.~258.00]{byr54}
\begin{eqnarray}
    \int_{r_1}^r \frac{dr}{\sqrt{R}} &=& \int_{r_1}^r \frac{dr}{\sqrt{
            (r-r_1)(r-r_2)(r-r_3)(r-r_4)}} \nonumber\\[2mm]
        &=& \frac{2}{\sqrt{(r_1-r_3)(r_2-r_4)}}\, \sn^{-1} \left[\left.
	       \sqrt{\frac{(r_2-r_4)(r-r_1)}{(r_1-r_4)(r-r_2)}}\right| m_4
	       \right] \;,
    \label{r4_int1}
\end{eqnarray}
where
\begin{eqnarray}
    m_4 = \frac{(r_1-r_4)(r_2-r_3)}{(r_1-r_3)(r_2-r_4)} \;.
    \label{m4}
\end{eqnarray}
It is easy to show that
\begin{eqnarray}
    0\le \sqrt{\frac{(r_2-r_4)(r-r_1)}{(r_1-r_4)(r-r_2)}} < 1 \;,
    \hspace{1cm} 0\le m_4 \le 1 \;, \nonumber
\end{eqnarray}
when $r_1 \neq r_2$. Thus, the inverse Jacobian elliptic function $\sn^{-1}$ in
equation~(\ref{r4_int1}) is well defined. 

When $r_1 = r_2$, the integral in equation~(\ref{r4_int1}) becomes 
$\propto\sn^{-1}(1|1) = \infty$. In this case the integral over $r$ needs 
special treatment. Indeed, when $r_1 = r_2$, the integral over $r$ can be 
expressed in terms of a logarithm,
\begin{eqnarray}
    \int \frac{dr}{\sqrt{R}} &=& \int \frac{dr}{(r-r_1)\sqrt{(r-r_3)
            (r-r_4)}} \;=\; -\frac{1}{\sqrt{(r_1-r_3)(r_1-r_4)}} 
            \nonumber\\[2mm]
        && \times\ln \left[\frac{\sqrt{(r-r_3)(r-r_4)}}{r-r_1}+ 
            \frac{2r_3 r_4-r_1(r_3+r_4)+(2r_1-r_3-r_4)r}{2(r-r_1)\sqrt{
            (r_1-r_3)(r_1-r_4)}}\right] \nonumber\\[2mm]
        &=& -\frac{1}{\sqrt{(r_1-r_3)(r_1-r_4)}} \nonumber\\[2mm]
        && \times \ln \left[\frac{\sqrt{(r-r_3)(r-r_4)}}{r-r_1}+
            \frac{r_1^2+r_3r_4+2 r_1 r}{(r-r_1)\sqrt{
            (r_1-r_3)(r_1-r_4)}}\right] \;,
    \label{r4a_int1}
\end{eqnarray}
where in the last step we have used $2r_1+r_3+r_4 = 0$.

In region II ($r_3 \le r \le r_2$), the integral over r can be worked out by the
following integration \citep[eq.~255.00]{byr54}
\begin{eqnarray}
    \int_r^{r_2} \frac{dr}{\sqrt{R}} &=& \int_r^{r_2} \frac{dr}{\sqrt{
            (r_1-r)(r_2-r)(r-r_3)(r-r_4)}} \nonumber\\
        &=& \frac{2}{\sqrt{(r_1-r_3)(r_2-r_4)}}\, \sn^{-1} \left[\left.
	       \sqrt{\frac{(r_1-r_3)(r_2-r)}{(r_2-r_3)(r_1-r)}}\right| m_4
	       \right] \;.
    \label{r4_int2}
\end{eqnarray}
It can be shown that
\begin{eqnarray}
    0\le \sqrt{\frac{(r_1-r_3)(r_2-r)}{(r_2-r_3)(r_1-r)}} \le 1 \;, \nonumber
\end{eqnarray}
so the inverse Jacobian elliptic function $\sn^{-1}$ in equation~(\ref{r4_int2}) 
is well defined when $r_1 \neq r_2\,$.

Again, we need special treatment for the case $r_1 = r_2\,$: the integral over 
$r$ is then given by
\begin{eqnarray}
    \int \frac{dr}{\sqrt{R}} &=& \int \frac{dr}{(r_1-r)\sqrt{(r-r_3)
            (r-r_4)}} \;=\; \frac{1}{\sqrt{(r_1-r_3)(r_1-r_4)}} 
            \nonumber\\[2mm]
        && \times \ln \left[\frac{\sqrt{(r-r_3)(r-r_4)}}{r_1-r}+
            \frac{r_1^2+r_3r_4+2 r_1 r}{(r_1-r)\sqrt{
            (r_1-r_3)(r_1-r_4)}}\right] \;.
    \label{r4a_int2}
\end{eqnarray}

{\em Case B. $R(r)=0$ has two complex roots and two real roots~~}  Let us 
assume that $r_1$ and $r_2$ are complex, $r_3$ and $r_4$ are real and $r_3 > 
r_4\,$. Then, we must have $r_1 = r_2^\star$, where $\star$ stands for complex 
conjugate. Since $r_1 r_2 r_3 r_4 = |r_1|^2 r_3 r_4 = -a^2 Q \le 0\,$, $r_4$ 
must be $\le 0$ and $r_3$ must be $\ge 0\,$. [Thus, it is impossible that 
$r_3 = r_4$, since otherwise it gives $r_3 = r_4 = 0$, which by 
eq.~(\ref{rr}) implies that $Q = \lambda -a = 0$ and $r_1 = r_2 = 0$ too.] 
Since $R\ge 0$ implies $r \ge r_3$ or $r \le r_4\,$, the physically allowed 
region for photons is given by $r \ge r_3\,$.

Define $u = \Re(r_1) = \Re(r_2)$ and $v = \Im(r_1) = - \Im(r_2)\,$, where
$\Re$ and $\Im$ denote the real and imaginary parts of a complex
number. Then, the integral over $r$ can be worked out with the following 
integration \citep[eq.~260.00]{byr54}
\begin{eqnarray}
    \int_{r_3}^r \frac{dr}{\sqrt{R}} &=& \int_{r_3}^r \frac{dr}{\sqrt{
            (r-r_1)(r-r_2)(r-r_3)(r-r_4)}} \nonumber\\[2mm]
        &=& \frac{1}{\sqrt{AB}}\, \cn^{-1} \left[\left.\frac{(A-B)r
	       +r_3 B -r_4 A}{(A+B)r -r_3 B -r_4 A}\right| m_2
	       \right] \;,
    \label{r2_int}
\end{eqnarray}
where
\begin{eqnarray}
    A^2 = (r_3-u)^2 + v^2 \;, \hspace{1cm} B^2 = (r_4-u)^2 + v^2  \;,
    \label{ab}
\end{eqnarray}
and
\begin{eqnarray}
    m_2 = \frac{(A+B)^2-(r_3-r_4)^2}{4AB} \;. \label{m2}
\end{eqnarray}

It is easy to verify that 
\begin{eqnarray}
    \left[\frac{(A-B)r+r_3 B -r_4 A}{(A+B)r -r_3 B -r_4 A}\right]^2
        = 1- \frac{4AB (r-r_3)(r-r_4)}{\left[(A+B)r -r_3 B -r_4 A\right]^2}
	   \le 1 \;, \nonumber
\end{eqnarray}
for $r\ge r_3$. Using the identity $(2AB)^2-\left[A^2+B^2-(r_3-r_4)^2\right]^2 
= 4(r_3-r_4)^2 v^2 > 0$, we have
\begin{eqnarray}
    -2AB < A^2+B^2-(r_3-r_4)^2 < 2AB \;. \nonumber
\end{eqnarray}
Then we can show that
\begin{eqnarray}
    0< m_2 <1 \;. \nonumber
\end{eqnarray}
Hence, the inverse Jacobian elliptic function $\cn^{-1}$ in 
equation~(\ref{r2_int}) is well defined.

Therefore, if we can find the four roots of $R = 0$, we can work out the
integral over $r$. The standard procedure for finding the roots of a quartic 
equation can be found in, e.g., \citet{bir65,zwi02}.

Each photon emitted by the disk has three possible fates: escaping to infinity,
returning to the disk, or being captured by the black hole. In the following
we discuss the solutions for the photon orbits for each of those the
possibilities separately.

\subsection{Solutions for photons escaping to infinity}
\label{sol_inf}

We consider an observer at infinity (in practice corresponding to the limit 
that the distance from the black hole to the observer is $\gg$ the radius of
the disk) with a polar angle $\vartheta_{\rm obs}$, and calculate the orbit of 
a photon that reaches him/her. As the photon leaves the disk, we have 
$\vartheta=\vartheta_e = \pi/2$ and $\mu=\mu_e = \cos\vartheta_e = 0\,$. As 
the photon reaches the observer, we have $\vartheta = \vartheta_{\rm obs}$ and 
$\mu = \mu_{\rm obs}\equiv\cos\vartheta_{\rm obs}$. 

The position of a photon on the sky as seen by the observer is specified by a 
pair of impact parameters $(\alpha,\beta)$ (see Figure~\ref{coord1} and the 
second part of Appendix~\ref{const} for details). The coordinate $\alpha$ 
measures the displacement of the photon image perpendicular to the projection 
of the rotation axis of the black hole, and $\beta$ measures the displacement 
parallel to the 
projection of the axis. The line of sight to the black hole center marks the
origin of the coordinates, where $\alpha = \beta = 0$. It can be checked that
as the photon reaches the observer, on the photon orbit we have $d\vartheta/dr
>0$ (i.e., $d\mu/dr <0$) if $\beta >0$, and $d\vartheta/dr<0$ (i.e., $d\mu/dr 
>0$) if $\beta <0\,$. Therefore, when $\beta > 0$, the photon must encounter 
a turning point at $\mu = \mu_+$: $\mu$ starts from $0$, goes up to $\mu_+$,
then goes down to $\mu_{\rm obs}$ (which is $\le \mu_+$ according to the 
earlier argument). When $\beta<0$, the photon must not encounter a turning 
point at $\mu = \mu_+$: $\mu$ starts from $0$ and monotonically increases to 
$\mu_{\rm obs}\,$.

Based on the above arguments, we can calculate the total integration over 
$\mu$ along the path of the photon from the disk to the observer. The
results are
\begin{eqnarray}
    \tau_\mu \equiv \int_{{\cal S}_{\rm ph}}\frac{d\mu}{\sqrt{\Theta_\mu}} 
        = \left\{\begin{array}{ll}\left(\int_0^{\mu_+}+\int_{\mu_{\rm 
        obs}}^{\mu_+}\right)\frac{d\mu}{\sqrt{\Theta_\mu}} \;, 
	   & (\beta>0) \\[2mm]
	   \int_0^{\mu_{\rm obs}}\frac{d\mu}{\sqrt{\Theta_\mu}} = \left(
	   \int_0^{\mu_+}-\int_{\mu_{\rm obs}}^{\mu_+}\right)\frac{d\mu}
	   {\sqrt{\Theta_\mu}} \;, & (\beta<0)
	   \end{array}\right. \;, \label{tau_mu}
\end{eqnarray}
where $\int_{{\cal S}_{\rm ph}}$ stands for the integration along the path of 
the photon. Note, by definition, $\tau_\mu$ is positive. Substituting 
equation~(\ref{mu_int}) into equation~(\ref{tau_mu}), we get
\begin{eqnarray}
    \tau_\mu = \frac{1}{\sqrt{a^2\left(\mu_+^2+\mu_-^2\right)}}\,\left[K(m_\mu) 
	   + {\rm sign}(\beta)\; \cn^{-1} \left(\left.\frac{\mu_{\rm obs}}
	   {\mu_+}\right|m_\mu\right)\right] \;, \label{tau_mu1}
\end{eqnarray}
where ${\rm sign}(\beta) = 1$ if $\beta> 0$, $0$ if $\beta=0$, and $-1$ if 
$\beta < 0$; $K(m)$ is the complete elliptic integral of the first kind.

Now let us consider the integration over $r$. Since the observer is at infinity, 
the photon reaching him/her must have been moving in the allowed region
defined by $r\ge r_1$ when $R=0$ has four real roots (case A), or the allowed 
region defined by $r\ge r_3$ when $R=0$ has two complex roots and two real roots 
(case B). There are then two possibilities for the photon during its trip: it 
has encountered a turning point at $r = r_+$ ($r_+ = r_1$ in case A, 
$r_+ = r_3$ in case B), or it has not encountered any turning point in $r$. 
Define
\begin{eqnarray}
    \tau_\infty\equiv\int_{r_+}^\infty\frac{dr}{\sqrt{R}} \;, \hspace{1cm}
    \tau_e\equiv\int_{r_+}^{r_e}\frac{dr}{\sqrt{R}} \;,  \label{tau_inf}
\end{eqnarray}
where $r_e$ is the radius in the disk where the photon is emitted. Obviously,
according to equation~(\ref{r-mu}), a necessary and sufficient condition for
the occurrence of a turning point in $r$ on the path of the photon is that 
$\tau_\infty< \tau_\mu$. Then, the total integration over $r$ along the path 
of the photon from the disk to the observer is
\begin{eqnarray}
    \tau_r \equiv \int_{{\cal S}_{\rm ph}}\frac{dr}{\sqrt{R}} =
        \left\{\begin{array}{ll}
        \tau_\infty + \tau_e \;, & (\tau_\infty < \tau_\mu) \\[2mm]
	   \int_{r_e}^\infty\frac{dr}{\sqrt{R}} = \tau_\infty - \tau_e \;, 
	     & (\tau_\infty \ge \tau_\mu)
        \end{array}\right. \;. \label{tau_r}
\end{eqnarray}
By definition, $\tau_\infty$, $\tau_e$, and $\tau_r$ are all positive. 
According to equation~(\ref{r-mu}), we must have $\tau_r = \tau_\mu$ for 
the orbit of a photon.

{\em Case A: $R=0$ has four real roots~~} When $r_1\neq r_2$, substitute
equation~(\ref{r4_int1}) into equation~(\ref{tau_inf}) and set $r_+ = r_1$. We
obtain
\begin{eqnarray}
    \tau_\infty &=& \frac{2}{\sqrt{(r_1-r_3)(r_2-r_4)}}\,\sn^{-1}\left(
        \left.\sqrt{\frac{r_2-r_4}{r_1-r_4}}\right| m_4\right) \;, 
	   \label{tau1_r4}\\[1mm]
    \tau_e &=& \frac{2}{\sqrt{(r_1-r_3)(r_2-r_4)}}\,\sn^{-1}\left[
        \left.\sqrt{\frac{(r_2-r_4)(r_e-r_1)}{(r_1-r_4)(r_e-r_2)}}
	   \right|m_4\right] \;. \label{tau2_r4}
\end{eqnarray}
Substituting equations~(\ref{tau1_r4}) and (\ref{tau2_r4}) into 
equation~(\ref{tau_r}) and letting $\tau_r = \tau_\mu\,$, we can solve for 
$r_e$---the radius in the disk where the photon is emitted. The solution is
\begin{eqnarray}
    r_e = \frac{r_1(r_2-r_4)-r_2(r_1-r_4)\,\sn^2(\xi_4|m_4)}{(r_2-r_4)-
	   (r_1-r_4)\,\sn^2(\xi_4|m_4)} \;, \label{sol1}
\end{eqnarray}
where
\begin{eqnarray}
    \xi_4 \equiv \frac{1}{2}(\tau_\mu-\tau_\infty)\sqrt{(r_1-r_3)
	   (r_2-r_4)} \;.
\end{eqnarray}

Since $\sn^2(-\xi_4|m_4)=\sn^2(\xi_4|m_4)$, the solution given by 
equation~(\ref{sol1}) applies whether $\tau_\mu-\tau_\infty$ is positive or 
negative, i.e. no matter whether there is a turning point in $r$ or not along 
the path of the photon. Equation~(\ref{sol1}) is essentially equivalent to
equation~(37) of \citet{cad98}.

When $r_1 = r_2$, the integration from $r = r_1$ to any $r>r_1$ is infinite
(see eq.~[\ref{r4a_int1}]), so $\tau_\infty = \infty$ and there cannot be a 
turning point in $r$. Then, by equation~(\ref{r4a_int1}), we have
\begin{eqnarray}
    \tau_r &=& \int_{r_e}^\infty\frac{dr}{\sqrt{R}} = \frac{1}{\sqrt{(r_1-
	     r_3)(r_1-r_4)}}\left\{- \ln \left[1+\frac{2r_1}{\sqrt{(r_1-r_3)(r_1
	     -r_4)}}\right]\right. \nonumber\\
        &&+\left.\ln\left[\frac{\sqrt{(r_e-r_3)(r_e-r_4)}}{r_e-r_1}+\frac{
	     r_1^2+r_3r_4+2 r_1 r_e}{(r_e-r_1)\sqrt{(r_1-r_3)(r_1-r_4)}}
	     \right]\right\} \;. \label{tau_r4a}
\end{eqnarray}
Setting $\tau_r = \tau_\mu$, we then obtain the solution for $r_e$
\begin{eqnarray}
    r_e = \frac{r_3 r_4 - \left[\gamma r_1+\left(r_1^2+r_3 r_4\right)/
	   \sqrt{(r_1-r_3)(r_1-r_4)}\right]^2}{r_1\left\{1-\left[\gamma-
	   2r_1/\sqrt{(r_1-r_3)(r_1-r_4)}\right]^2\right\}} \;, \label{sol2}
\end{eqnarray}
where the identity $2r_1 + r_3 +r_4 = 0$ has been used, and
\begin{eqnarray}
    \gamma \equiv \left[1+\frac{2r_1}{\sqrt{(r_1-r_3)(r_1-r_4)}}\right]
        \exp\left[{\tau_\mu\sqrt{(r_1-r_3)(r_1-r_4)}}\right] \;.
\end{eqnarray}

The numerical value for $r_e$ obtained from equation~(\ref{sol2}) should be 
substituted back in equation~(\ref{tau_r4a}) (with $\tau_r = \tau_\mu$) to 
check if $r_e$ is a true root, since the square operation in the process of 
solving equation~(\ref{tau_r4a}) might produce a false root.

{\em Case B: $R=0$ has two complex roots and two real roots~~} Substituting 
equation~(\ref{r2_int}) into equation~(\ref{tau_inf}) and setting $r_+=r_3$, 
we obtain
\begin{eqnarray}
    \tau_\infty &=& \frac{1}{\sqrt{AB}}\,\cn^{-1}\left(\left.\frac{A-B}{A+B}
	   \right| m_2\right) \;, \label{tau1_r2}\\[1mm]
    \tau_e &=& \frac{1}{\sqrt{AB}}\,\cn^{-1}\left[\left.\frac{(A-B)r_e+r_3 
	   B -r_4 A}{(A+B)r_e -r_3 B -r_4 A}\right| m_2\right] \;. \label{tau2_r2}
\end{eqnarray}
Substituting equation~(\ref{tau1_r2}) and (\ref{tau2_r2}) into 
equation~(\ref{tau_r}) and letting $\tau_r = \tau_\mu$, we can solve for 
$r_e$. The solution is
\begin{eqnarray}
    r_e = \frac{r_4 A -r_3 B - (r_4 A +r_3 B)\, \cn(\xi_2|m_2)}
        {(A-B) - (A+B)\, \cn(\xi_2|m_2)} \;, \label{sol3}
\end{eqnarray}
where
\begin{eqnarray}
    \xi_2 \equiv (\tau_\mu-\tau_\infty)\sqrt{AB} \;.
\end{eqnarray}

Since $\cn(-\xi_2|m_2)=\cn(\xi_2|m_2)$, the solution given by 
equation~(\ref{sol3}) applies no matter whether $\tau_\mu-\tau_\infty$ is 
positive or negative, i.e. no matter whether there is a turning in $r$ or not 
along the path of the photon. Equation~(\ref{sol3}) is essentially equivalent 
to equation~(39) of \citet{cad98}, although equation~(\ref{sol3}) is much
simpler.

In the above derivation we have implicitly assumed that $r_+ > r_{\rm H}$, 
where $r_{\rm H}$ is the radius of the horizon of the black hole. If $r_+ \le 
r_{\rm H}$, then obviously the orbit of the photon cannot have a turning 
point at $r=r_+$, since if a photon falls into a black hole it cannot get 
out. Therefore, when $r_+ \le r_{\rm H}$, the solutions given by 
equations~(\ref{sol1}) and (\ref{sol3}) hold only if $\tau_\mu<\tau_\infty$, 
i.e. when there is no turning in $r$ along the path of the photon. 
Finally, the solution that we have found above for $r_e$ represents a 
physical solution---i.e., it represents a photon emitted by the disk---only if 
it satisfies $r_{\rm in} \le r_e \le r_{\rm out}$, where $r_{\rm in}$ is the 
radius of the inner boundary of the disk, and $r_{\rm out}$ is the radius 
of the outer boundary of the disk. 

\subsection{Solutions for photons returning to the disk}
\label{sol_ret}

A photon returning to the disk must satisfy $\mu = 0$ (i.e., $\vartheta =
\pi/2$) as it approaches the disk at radius $r_a$. Since as it leaves the disk 
at radius $r_e$ it must also have $\mu = 0$, the photon must have encountered
a turning in $\vartheta$ during its trip. Therefore, along the path of the 
photon, $\mu$ must first increases from $\mu=0$ to $\mu = \mu_+$, then 
decreases from $\mu = \mu_+$ to $\mu=0$.\footnote{Of course this assumes that
the returning photon crosses the equatorial plane only once. This is reasonable 
since we assume that as the photon 
reaches the equatorial plane it will be captured by either the disk or the 
gas in the plunging region.} Then, along the path of the photon the total
integration over $\mu$ is
\begin{eqnarray}
    \tau_\mu = 2\int_0^{\mu_+} \frac{d\mu}{\sqrt{\Theta_\mu}}
	   = \frac{2}{\sqrt{a^2\left(\mu_+^2+\mu_-^2\right)}}\,K(m_\mu) \;. 
    \label{tau_mu2}
\end{eqnarray}

The integration over $r$ depends on the allowed region which the photon moves 
in. In case A ($R=0$ has four real roots), there are two allowed regions: 
region I and region II. If the photon moves in region I, which is bounded at 
$r=r_1$, it may encounter a turning point at $r=r_1$. If the photon moves in 
region II, which is bounded at $r=r_2$ and $r=r_3$, it may encounter many
turning points at $r=r_2$ and $r=r_3$. In case B ($R=0$ has two complex roots and
two real roots), there is only one allowed region that is given by $r\ge r_3$, 
so the photon may encounter a turning point at $r=r_3$. In each of these cases 
the path of the photon also depends on the fact whether $r_{\rm H}$ appears in 
the region that the photon moves in, and on the direction of the photon  as it
approaches $r=r_a$ in the disk, determined by the value of $k_r = k_a(\partial/
\partial r)^a$ at $r = r_a$, where $k^a$ is the four-wavevector of the photon. 
Let us discuss each of these cases separately.

{\em Case A1: $R=0$ has four real roots, region I $(r\ge r_1)$~~} 
If $r_1\neq r_2$, the fundamental integral over $r$ is given by 
equation~(\ref{r4_int1}). Let us define 
\begin{eqnarray}
    \tau_a \equiv \int_{r_1}^{r_a} \frac{dr}{\sqrt{R}} 
        = \frac{2}{\sqrt{(r_1-r_3)(r_2-r_4)}}\,\sn^{-1}\left[
        \left.\sqrt{\frac{(r_2-r_4)(r_a-r_1)}{(r_1-r_4)(r_a-r_2)}}
	   \right|m_4\right] \;. \label{taua_r4}
\end{eqnarray}
Other relevant integrations are $\tau_\infty$ and $\tau_e$, defined by
equations~(\ref{tau1_r4}) and (\ref{tau2_r4}) respectively.

When $k_r <0$ at $r=r_a$, i.e., the photon approaches $r_a$ from the side of 
$r>r_a$, the solution for $r_e$ is determined by the equation $\tau_\mu =
\tau_r$, where the integral over $r$ along the path of the photon is (there
is no turning point in $r$)
\begin{eqnarray}
    \tau_r = \int_{r_a}^{r_e}\frac{dr}{\sqrt{R}} = \left(\int_{r_1}^{r_e}
        - \int_{r_1}^{r_a}\right)\frac{dr}{\sqrt{R}} = \tau_e-\tau_a \;.
\end{eqnarray}
Obviously, if $\tau_\mu \ge \tau_\infty - \tau_a$, then $r_e = \infty$, which
is of course not a solution for a photon returning to the disk. So, the 
solution for $r_e$ exists only if $\tau_\mu < \tau_\infty - \tau_a$, which is 
given by the solution to $\tau_\mu = \tau_e - \tau_a$. The solution is
\begin{eqnarray}
    r_e = \frac{r_1(r_2-r_4)-r_2(r_1-r_4)\,\sn^2(\xi_{4a}|m_4)}{(r_2-r_4)
	   -(r_1-r_4)\,\sn^2(\xi_{4a}|m_4)} \;, \label{sol4}
\end{eqnarray}
where
\begin{eqnarray}
    \xi_{4a} \equiv \frac{1}{2}(\tau_\mu+\tau_a)\sqrt{(r_1-r_3)
	   (r_2-r_4)} \;. \label{xi4a1}
\end{eqnarray}

When $k_r >0$ at $r=r_a$, i.e. the photon approaches $r_a$ from the side of 
$r<r_a$, the integral over $r$ along the path of the photon is
\begin{eqnarray}
    \tau_r = \left\{\begin{array}{ll}
        \int_{r_e}^{r_a}\frac{dr}{\sqrt{R}} = \left(\int_{r_1}^{r_a}
          - \int_{r_1}^{r_e}\right)\frac{dr}{\sqrt{R}} = \tau_a-\tau_e \;,
	     & (\tau_\mu\le \tau_a) \\[2mm]
	   \left(\int_{r_1}^{r_a}+\int_{r_1}^{r_e}\right)\frac{dr}{\sqrt{R}}
	     = \tau_a+\tau_e \;, & (\tau_a < \tau_\mu <\tau_a+\tau_\infty)
	   \end{array}\right. \;.
\end{eqnarray}
When $\tau_\mu\le \tau_a$, there is no turning point in $r$; when $\tau_a < 
\tau_\mu <\tau_\infty$, there is a turning point at $r=r_1$; and when $\tau_\mu 
\ge \tau_a+ \tau_\infty$, a solution for $r_e$ does not exist. Setting $\tau_r = 
\tau_\mu$, we can solve for $r_e$. The solution is given by equation~(\ref{sol4}) 
with $\xi_{4a}$ replaced by
\begin{eqnarray}
    \xi_{4a} \equiv \frac{1}{2}(\tau_\mu-\tau_a)\sqrt{(r_1-r_3)
	   (r_2-r_4)} \;. \label{xi4a2}
\end{eqnarray}
Since $\sn^2(-\xi_{4a}|m_4) = \sn^2(\xi_{4a}|m_4)$, the above solution applies
if $\tau_\mu <\tau_a+\tau_\infty$, no matter whether there is a turning point 
in $r$ or not.

Of course, when $r_{\rm H} \ge r_1$, there cannot be a turning point in $r$
along the path of the photon. Then, when $k_r >0$, the solution is given by
equations~(\ref{sol4}) and (\ref{xi4a2}) if and only if $\tau_\mu<
\tau_a$.\footnote{A more strict condition should be $\tau_\mu < \tau_{\rm H}$,
where $\tau_{\rm H} = \int_{r_1}^{r_{\rm H}}\frac{dr}{\sqrt{R}}$ is given by 
the right hand side of eq.~(\ref{taua_r4}) with $r_a$ replaced by $r_{\rm H}$. 
This condition ensures that $r_e > r_{\rm H}$. However, since in the final 
stage of our computation we select only the solutions satisfying 
$r_{\rm in}\le r\le r_{\rm out}$, $\tau_\mu<\tau_a$ and $\tau_\mu<\tau_{\rm 
H}$ lead to the same final results. (Note, $r_{\rm in} \ge r_{\rm H}$ always.)}

Now let us consider the case $r_1=r_2$. In region I the 
fundamental integral over $r$ is given by equation~(\ref{r4a_int1}). Since
the integration from $r=r_1$ to any $r>r_1$ is infinite, there cannot be
a turning point along the path of the photon. Let us define
\begin{eqnarray}
    \hat{\tau}_\infty &\equiv& \frac{1}{\sqrt{(r_1-r_3)(r_1-r_4)}}\,\ln
        \left[1+\frac{2r_1}{\sqrt{(r_1-r_3)(r_1-r_4)}}\right] \;, \\[2mm]
    \hat{\tau}_a &\equiv& \frac{1}{\sqrt{(r_1-r_3)(r_1-r_4)}}\,\ln
        \left[\frac{\sqrt{(r_a-r_3)(r_a-r_4)}}{r_a-r_1}\right. \nonumber\\
	   &&\left.+\frac{r_1^2+r_3r_4
	   +2r_1r_a}{(r_a-r_1)\sqrt{(r_1-r_3)(r_1-r_4)}}\right] \;, \\[2mm]
    \hat{\tau}_e &\equiv& \frac{1}{\sqrt{(r_1-r_3)(r_1-r_4)}}\,\ln
        \left[\frac{\sqrt{(r_e-r_3)(r_e-r_4)}}{r_e-r_1}\right. \nonumber\\
	   &&\left.+\frac{r_1^2+r_3r_4
	   +2r_1r_e}{(r_e-r_1)\sqrt{(r_1-r_3)(r_1-r_4)}}\right] \;.
\end{eqnarray}
We have
\begin{eqnarray}
    \int_{r_a}^\infty\frac{dr}{\sqrt{R}} = \hat{\tau}_a - 
        \hat{\tau}_\infty \;.
\end{eqnarray}

Then, when $k_r<0$ at $r=r_a$, the integration over $r$ along the path of the 
photon is
\begin{eqnarray}
    \tau_r = \int_{r_a}^{r_e}\frac{dr}{\sqrt{R}} = \hat{\tau}_a -
        \hat{\tau}_e \;.
\end{eqnarray}
The solution for $r_e$ is determined by $\tau_r=\tau_\mu$. Obviously, if
$\tau_\mu\ge \hat{\tau}_a-\hat{\tau}_\infty$, the solution for $r_e$ does
not exist. So, if $\tau_\mu< \hat{\tau}_a-\hat{\tau}_\infty$, we have the
solution for $r_e$
\begin{eqnarray}
    r_e = \frac{r_3 r_4 - \left[\gamma_a r_1+\left(r_1^2+r_3 r_4\right)/
        \sqrt{(r_1-r_3)(r_1-r_4)}\right]^2}{r_1\left\{1-\left[\gamma_a-
        2r_1/\sqrt{(r_1-r_3)(r_1-r_4)}\right]^2\right\}} \;, \label{sol4a}
\end{eqnarray}
where
\begin{eqnarray}
    \gamma_a \equiv \exp\left[{(\hat{\tau}_a-\tau_\mu)\sqrt{(r_1-r_3)
	   (r_1-r_4)}}\right] \;.
\end{eqnarray}

When $k_r>0$ at $r=r_a$, the integration over $r$ along the path of the 
photon is
\begin{eqnarray}
    \tau_r = \int_{r_e}^{r_a}\frac{dr}{\sqrt{R}} = \hat{\tau}_e -
        \hat{\tau}_a \;.
\end{eqnarray}
The solution for $r_e$ is determined by $\tau_r=\tau_\mu$. Since 
$\int_{r_1}^{r_a}\frac{dr}{\sqrt{R}}=\infty$, the solution for $r_e$ always
exists. It is given by equation~(\ref{sol4a}) with $\gamma_a$ replaced by
\begin{eqnarray}
    \gamma_a \equiv \exp\left[{(\hat{\tau}_a+\tau_\mu)\sqrt{(r_1-r_3)
	   (r_1-r_4)}}\right] \;.
\end{eqnarray}

As in Appendix~\ref{sol_inf}, the value of $r_e$ 
obtained above for the case of $r_1=r_2$ should be substituted back in the 
original equation $\tau_r=\tau_\mu$ to check if $r_e$ is a true solution.

{\em Case A2: $R=0$ has four real roots, region II $(r_3\le r \le r_2)$~~}
If $r_1\neq r_2$, the fundamental integral over $r$ is given by 
equation~(\ref{r4_int2}). Let us define the following integrals
\begin{eqnarray}
    \tau_{e2} &\equiv& \int_{r_e}^{r_2}\frac{dr}{\sqrt{R}} = \frac{2}{
	   \sqrt{(r_1-r_3)(r_2-r_4)}}\,\sn^{-1}\left[\left.\sqrt{\frac{(r_1-
	   r_3)(r_2-r_e)}{(r_2-r_3)(r_1-r_e)}}\right|m_4\right] \;, \\[1mm]
    \tau_{a2} &\equiv& \int_{r_a}^{r_2}\frac{dr}{\sqrt{R}} = \frac{2}{
	   \sqrt{(r_1-r_3)(r_2-r_4)}}\,\sn^{-1}\left[\left.\sqrt{\frac{(r_1-
	   r_3)(r_2-r_a)}{(r_2-r_3)(r_1-r_a)}}\right|m_4\right] \;, \\[1mm]
    \tau_{32} &\equiv& \int_{r_3}^{r_2}\frac{dr}{\sqrt{R}} = \frac{2}{
	   \sqrt{(r_1-r_3)(r_2-r_4)}}\,K(m_4) \;, 
\end{eqnarray}
and
\begin{eqnarray}
    \tau_{3a} \equiv \int_{r_3}^{r_a}\frac{dr}{\sqrt{R}} =
        \left(\int_{r_3}^{r_2} -\int_{r_a}^{r_2}\right)\frac{dr}{\sqrt{R}}
        = \tau_{32}- \tau_{a2} \;.
\end{eqnarray}

When $r_{\rm H} \ge r_3$, there cannot be a turning point at $r=r_3$. However, 
there may be a turning point at $r=r_2$, since $r_a$ must be between $r_{\rm H}$ and
$r_2$. Then, when $k_r <0$ at $r=r_a$, the integral over $r$ is
\begin{eqnarray}
    \tau_r = \left\{\begin{array}{ll}
        \int_{r_a}^{r_e}\frac{dr}{\sqrt{R}} = \left(\int_{r_a}^{r_2}
          - \int_{r_e}^{r_2}\right)\frac{dr}{\sqrt{R}} = \tau_{a2}-\tau_{e2}
	     \;, & (\tau_\mu\le \tau_{a2}) \\[2mm]
	   \left(\int_{r_a}^{r_2}+\int_{r_e}^{r_2}\right)\frac{dr}{\sqrt{R}}
	     = \tau_{a2}+\tau_{e2} \;, & (\tau_{a2} < \tau_\mu <\tau_{a2}+
		\tau_{32})
	   \end{array}\right. \;.
\end{eqnarray}
When $\tau_\mu\le \tau_{a2}$, there is no turning point in $r$; when $\tau_{a2} 
<\tau_\mu <\tau_{a2}+\tau_{32}$, there is a turning point at $r=r_2$; and when 
$\tau_\mu\ge \tau_{a2}+\tau_{32}$, a solution for $r_e$ (satisfying $r_e > 
r_{\rm H}$) does not exist since it would have a turning point at $r=r_3$ which
is impossible. Setting $\tau_r = \tau_\mu$, we can then solve for $r_e$. The 
solution is
\begin{eqnarray}
    r_e = \frac{r_2(r_1-r_3)-r_1(r_2-r_3)\,\sn^2(\xi_{4b}|m_4)}{(r_1-r_3)
	   -(r_2-r_3)\,\sn^2(\xi_{4b}|m_4)} \;, \label{sol5}
\end{eqnarray}
where
\begin{eqnarray}
    \xi_{4b} = \frac{1}{2}(\tau_\mu-\tau_{a2})\sqrt{(r_1-r_3)
	   (r_2-r_4)} \;.
\end{eqnarray}
Since $\sn^2(-\xi_{4b}|m_4) = \sn^2(\xi_{4b}|m_4)$, the above solution applies
if $\tau_\mu <\tau_{a2}+\tau_{32}$, no matter whether there is a turning point 
at $r=r_2$ or not.

When $r_{\rm H} \ge r_3$ and $k_r >0$ at $r=r_a$, there cannot be a turning
point in $r$ along the path of the photon. Then, the integral over $r$ is 
\begin{eqnarray}
    \tau_r = \int_{r_e}^{r_a}\frac{dr}{\sqrt{R}} = \left(\int_{r_e}^{r_2}
        - \int_{r_a}^{r_2}\right)\frac{dr}{\sqrt{R}} =\tau_{e2}-\tau_{a2} \;.
\end{eqnarray}
Setting $\tau_r = \tau_\mu$, we then get the solution for $r_e$. The solution
exists only if $\tau_\mu \le \tau_{3a}$, and is given by equation~(\ref{sol5}) 
with $\xi_{4b}$ replaced by
\begin{eqnarray}
    \xi_{4b} = \frac{1}{2}(\tau_\mu+\tau_{a2})\sqrt{(r_1-r_3)
        (r_2-r_4)} \;.
\end{eqnarray}

When $r_{\rm H} < r_3$, the horizon of the black hole does not affect the
motion of the photon, and the photon moves in a region bounded by $r_2$ and
$r_3$. So, the photon can have many turning points at $r=r_2$ and $r= r_3$. 
To properly account for this fact, we define the following two numbers
\begin{eqnarray}
    n \equiv {\rm int}\left(\frac{\tau_\mu}{2\tau_{32}}\right) \;,
    \hspace{1cm} \tau_\mu^{\prime} \equiv \tau_\mu - 2n\tau_{32} \;,
\end{eqnarray}
where $2\tau_{32}$ is a whole period of the integration over $r$, and ${\rm int}
(x)$ means the integer part of a real number $x$. Obviously, we must have 
$0\le \tau_\mu^{\prime} < 2\tau_{32}\,$.

Then, when $r_{\rm H} < r_3$ and $k_r<0$ at $r=r_a$, the integration over $r$ 
is $\tau_r = 2n\tau_{32}+ \tau_r^\prime$, where
\begin{eqnarray}
    \tau_r^\prime = \left\{\begin{array}{ll}
        \left(\int_{r_a}^{r_2}-\int_{r_e}^{r_2}\right)\frac{dr}{\sqrt{R}}
	     = \tau_{a2}-\tau_{e2} \;, & (\tau_\mu^\prime \le \tau_{a2}) \\[2mm]
	   \left(\int_{r_a}^{r_2}+\int_{r_e}^{r_2}\right)\frac{dr}{\sqrt{R}}
          = \tau_{a2}+\tau_{e2} \;, & (\tau_{a2}<\tau_\mu^\prime \le
		\tau_{a2}+\tau_{32}) \\[2mm]
	   \left(\int_{r_a}^{r_2}+\int_{r_3}^{r_2}+\int_{r_3}^{r_e}\right)
	     \frac{dr}{\sqrt{R}}= \tau_{a2}+2\tau_{32}-\tau_{e2} \;, & 
		(\tau_\mu^\prime >\tau_{a2}+\tau_{32})
        \end{array}\right. \;.
\end{eqnarray}
When $\tau_\mu^\prime \le \tau_{a2}$, there are $n$ turning points at $r=r_2$ 
and $n$ turning points at $r=r_3$. When $\tau_{a2}<\tau_\mu^\prime \le\tau_{a2}+
\tau_{32}$, there are $n+1$ turning points at $r=r_2$ and $n$ turning points at 
$r=r_3$. When $\tau_\mu^\prime >\tau_{a2}+\tau_{32}$, there are $n+1$ turning
points at $r=r_2$ and $n+1$ turning points at $r=r_3$. Setting $\tau_r = 
\tau_\mu$, we then obtain the solution for $r_e$. It is given by 
equation~(\ref{sol5}) with $\xi_{4b}$ replaced by
\begin{eqnarray}
    \xi_b = \left\{\begin{array}{ll}
        \frac{1}{2}(\tau_\mu^\prime-\tau_{a2})\sqrt{(r_1-r_3)(r_2-r_4)} \;,
	     & (\tau_\mu^\prime \le \tau_{a2}+\tau_{32}) \\[2mm]
	   \frac{1}{2}(\tau_{a2}+2\tau_{32}-\tau_\mu^\prime)
          \sqrt{(r_1-r_3)(r_2-r_4)} \;, & (\tau_\mu^\prime > \tau_{a2}+
		\tau_{32})
\end{array}\right. \;.
\end{eqnarray}

When $r_{\rm H} \ge r_3$ and $k_r>0$ at $r=r_a$, the integration over $r$ 
is $\tau_r = 2n\tau_{32}+ \tau_r^\prime$, where
\begin{eqnarray}
    \tau_r^\prime = \left\{\begin{array}{ll}
        \left(\int_{r_e}^{r_2}-\int_{r_a}^{r_2}\right)\frac{dr}{\sqrt{R}}
	     = \tau_{e2}-\tau_{a2} \;, & (\tau_\mu^\prime \le \tau_{3a}) \\[2mm]
	   \left(\int_{r_3}^{r_a}+\int_{r_3}^{r_e}\right)\frac{dr}{\sqrt{R}}
          = \tau_{3a}+\tau_{32}-\tau_{e2} \;, & (\tau_{3a}<\tau_\mu^\prime
		\le \tau_{3a}+\tau_{32}) \\[2mm]
	   \left(\int_{r_3}^{r_a}+\int_{r_3}^{r_2}+\int_{r_e}^{r_2}\right)
	     \frac{dr}{\sqrt{R}}= \tau_{3a}+\tau_{32}+\tau_{e2} \;, & 
		(\tau_\mu^\prime >\tau_{3a}+\tau_{32})
        \end{array}\right. \;.
\end{eqnarray}
When $\tau_\mu^\prime \le \tau_{3a}$, there are $n$ turning points at $r=r_2$ 
and $n$ turning points at $r=r_3$. When $\tau_{3a}<\tau_\mu^\prime \le\tau_{3a}+
\tau_{32}$, there are $n$ turning points at $r=r_2$ and $n+1$ turning points at 
$r=r_3$. When $\tau_\mu^\prime >\tau_{3a}+\tau_{32}$, there are $n+1$ turning
points at $r=r_2$ and $n+1$ turning points at $r=r_3$. Setting $\tau_r = 
\tau_\mu$, we then obtain the solution for $r_e$. It is given by 
equation~(\ref{sol5}) with $\xi_{4b}$ replaced by
\begin{eqnarray}
    \xi_b = \left\{\begin{array}{ll}
        \frac{1}{2}(\tau_\mu^\prime+\tau_{a2})\sqrt{(r_1-r_3)(r_2-r_4)} \;,
	     & (\tau_\mu^\prime \le \tau_{3a}) \\[2mm]
	   \frac{1}{2}(\tau_{3a}+\tau_{32}-\tau_\mu^\prime)
          \sqrt{(r_1-r_3)(r_2-r_4)} \;, & (\tau_\mu^\prime > \tau_{3a})
        \end{array}\right. \;.
\end{eqnarray}

Now let us consider the case of $r_1=r_2$. When $r_1=r_2$, in region II the 
fundamental integral over $r$ is given by equation~(\ref{r4a_int2}). Since
the integration from $r=r_1$ to $r_a<r_1=r_2$ is infinite, there cannot 
be a turning point at $r=r_2$. However, there can be a turning point at 
$r=r_3$. Let us define
\begin{eqnarray}
    \hat{\tau}_3 &\equiv& \frac{1}{\sqrt{(r_1-r_3)(r_1-r_4)}}\,\ln
        \left[\frac{r_1+r_3}{\sqrt{(r_1-r_3)(r_1-r_4)}}\right] \;, \\[2mm]
    \hat{\tau}_a &\equiv& \frac{1}{\sqrt{(r_1-r_3)(r_1-r_4)}}\,\ln
        \left[\frac{\sqrt{(r_a-r_3)(r_a-r_4)}}{r_1-r_a}\right. \nonumber\\
	   &&\left.+\frac{r_1^2+r_3r_4
	   +2r_1r_a}{(r_1-r_a)\sqrt{(r_1-r_3)(r_1-r_4)}}\right] \;, \\[2mm]
    \hat{\tau}_e &\equiv& \frac{1}{\sqrt{(r_1-r_3)(r_1-r_4)}}\,\ln
        \left[\frac{\sqrt{(r_e-r_3)(r_e-r_4)}}{r_1-r_e}\right. \nonumber\\
	   &&\left.+\frac{r_1^2+r_3r_4
	   +2r_1r_e}{(r_1-r_e)\sqrt{(r_1-r_3)(r_1-r_4)}}\right] \;.
\end{eqnarray}
We have
\begin{eqnarray}
    \int_{r_3}^{r_a}\frac{dr}{\sqrt{R}} = \hat{\tau}_a -\hat{\tau}_3 \;.
\end{eqnarray}

Then, when $k_r<0$ at $r=r_a$, the integration over $r$ along the path of the 
photon is (there is no turning point)
\begin{eqnarray}
    \tau_r = \int_{r_a}^{r_e}\frac{dr}{\sqrt{R}} = \hat{\tau}_e -
        \hat{\tau}_a \;.
\end{eqnarray}
The solution for $r_e$ is determined by $\tau_r=\tau_\mu$. Since 
$\int_{r_a}^{r_1}\frac{dr}{\sqrt{R}}=\infty$, the solution for $r_e$ 
always exists. It is given equation~(\ref{sol4a}) with $\gamma_a$ replaced by 
\begin{eqnarray}
    \gamma_a \equiv -\exp\left[{(\hat{\tau}_a+\tau_\mu)\sqrt{(r_1-r_3)
	   (r_1-r_4)}}\right] \;.
\end{eqnarray}

When $k_r>0$ at $r=r_a$, the integration over $r$ along the path of the 
photon is
\begin{eqnarray}
    \tau_r = \left\{\begin{array}{ll}
        \int_{r_e}^{r_a}\frac{dr}{\sqrt{R}}=\hat{\tau}_a -\hat{\tau}_e\;,
	     & (\tau_\mu \le \hat{\tau}_a-\hat{\tau}_3) \\[2mm]
	   \left(\int_{r_3}^{r_a}+\int_{r_3}^{r_e}\right)\frac{dr}
		{\sqrt{R}} = \hat{\tau}_a + \hat{\tau}_e -2\hat{\tau}_3 \;,
		& (\tau_\mu> \hat{\tau}_a-\hat{\tau}_3)
	   \end{array}\right. \;.
\end{eqnarray}
When $\tau_\mu \le \hat{\tau}_a-\hat{\tau}_3$, there is no turning point in 
$r$; and when $\tau_\mu> \hat{\tau}_a-\hat{\tau}_3$, there is a turning point
at $r=r_3$. Since $\int_{r_3}^{r_1}\frac{dr}{\sqrt{R}}=\infty$, the solution 
for $r_e$ always exists. It is given by equation~(\ref{sol4a}) with $\gamma_a$  
replaced by
\begin{eqnarray}
    \gamma_a \equiv \left\{\begin{array}{ll}
        -\exp\left[{(\hat{\tau}_a-\tau_\mu)\sqrt{(r_1-r_3)(r_1-r_4)}}
		\right] \;, & (\tau_\mu \le \hat{\tau}_a-\hat{\tau}_3) \\[2mm]
        -\exp\left[{(\tau_\mu-\hat{\tau}_a+2\hat{\tau}_3)\sqrt{(r_1-r_3)
		(r_1-r_4)}}\right] \;, & (\tau_\mu> \hat{\tau}_a-\hat{\tau}_3)
	   \end{array}\right. \;. \label{gamma}
\end{eqnarray}

Of course, when $r_{\rm H} > r_3$, there cannot be a turning point. Then, 
when $k_r>0$ at $r=r_a$, the solution is given by equation~(\ref{sol4a})
and the first line of equation~(\ref{gamma}) (i.e., for the case $\tau_\mu 
\le \hat{\tau}_a-\hat{\tau}_3$ only).

As in Appendix~\ref{sol_inf}, the value of $r_e$ obtained above for the case 
of $r_1=r_2$ should be substituted back in the original equation $\tau_r=
\tau_\mu$ to check if $r_e$ is a true solution.

{\em Case 2: $R=0$ has two complex roots and two real roots, region $r\ge r_3$~~}
The fundamental integral over $r$ is given by equation~(\ref{r2_int}).
Let us define
\begin{eqnarray}
    \tau_a\equiv\frac{1}{\sqrt{AB}}\,\cn^{-1}\left[\left.\frac{(A-B)r_a+r_3 
	   B -r_4 A}{(A+B)r_a -r_3 B -r_4 A}\right| m_2\right] \;. 
    \label{taua_r2}
\end{eqnarray}
Other relevant integrations are $\tau_\infty$ and $\tau_e$, defined by
by equations~(\ref{tau1_r2}) and (\ref{tau2_r2}) respectively.

When $k_r<0$ at $r=r_a$, the integral over $r$ along the path of the photon is
(there is no turning point)
\begin{eqnarray}
    \tau_r = \int_{r_a}^{r_e}\frac{dr}{\sqrt{R}} = \left(\int_{r_3}^{r_e}
        - \int_{r_3}^{r_a}\right)\frac{dr}{\sqrt{R}} = \tau_e-\tau_a \;.
\end{eqnarray}
Obviously, the solution for $r_e$ exists only if $\tau_\mu < \tau_\infty - 
\tau_a$. Then, setting $\tau_r = \tau_\mu$, we get the solution for $r_e$
\begin{eqnarray}
    r_e = \frac{r_4 A -r_3 B - (r_4 A +r_3 B)\, \cn(\xi_{2a}|m_2)}
        {(A-B) - (A+B)\, \cn(\xi_{2a}|m_2)} \;, \label{sol6}
\end{eqnarray}
where
\begin{eqnarray}
    \xi_{2a} \equiv (\tau_\mu+\tau_a)\sqrt{AB} \;.
\end{eqnarray}

When $k_r >0$ at $r=r_a$, the integral over $r$ along the path of the photon is
\begin{eqnarray}
    \tau_r = \left\{\begin{array}{ll}
        \left(\int_{r_3}^{r_a}- \int_{r_3}^{r_e}\right)\frac{dr}{\sqrt{R}} 
	     = \tau_a-\tau_e \;, & (\tau_\mu\le \tau_a) \\[2mm]
        \left(\int_{r_3}^{r_a}+\int_{r_3}^{r_e}\right)\frac{dr}{\sqrt{R}}
          = \tau_a+\tau_e \;, & (\tau_a < \tau_\mu <\tau_a+\tau_\infty) 
        \end{array}\right. \;.
\end{eqnarray}
When $\tau_\mu\le \tau_a$, there is no turning point in $r$; when $\tau_a < 
\tau_\mu <\tau_\infty$, there is a turning point at $r=r_3$; and when $\tau_\mu 
\ge \tau_a+ \tau_\infty$, a solution for $r_e$ does not exist. Setting $\tau_r 
= \tau_\mu$, we get the solution for $r_e$. The solution is given by
equation~(\ref{sol6}) with $\xi_{2a}$ replaced by
\begin{eqnarray}
    \xi_{2a} \equiv (\tau_\mu-\tau_a)\sqrt{AB} \;.
\end{eqnarray}
When $r_{\rm H} < r_3$, the above solution applies if $\tau_\mu <\tau_a+
\tau_\infty$, no matter whether there is a turning point in $r$ or not. When 
$r_{\rm H}\ge r_3$, there cannot be a turning point at $r=r_3$, then the above 
solution applies only if $\tau_\mu <\tau_a$.

\subsection{Conditions for photons to be captured by the black hole}
\label{con_bh}

To derive the conditions for a photon emitted by the disk to be captured by
the black hole, we make the following simplified assumption: Any radiation
that returns to the equatorial plane inside the inner boundary of the disk 
is captured by the black hole. This radiation, if it is not swallowed by the 
black hole directly, will be advected or scattered inward by
the infalling gas, which has a large inward radial velocity \citep{ago00}.

Following \citet{tho74}, we define a photon capture function $C$ as follow: 
$C = 1$ if a photon emitted at $r_e$ in the disk is eventually captured by the 
black hole; $C = 0$ if the photon eventually escapes to infinity or returns 
to the disk. The task in this subsection is to find out under what conditions 
we have $C=1$.

Obviously, a sufficient and necessary condition for $C=1$ is that the photon
neither escapes to infinity nor returns to the disk at a radius beyond the
inner boundary. In the following we
translate this condition to mathematical expressions for each case discussed
in Appendix~\ref{sol_ret}. These results are obtained by carefully following 
the analysis in Appendix~\ref{sol_ret}. However, for our purpose in this 
subsection, we follow the orbit of a photon by starting from the radius $r_e$ 
where the photon is emitted, rather than the radius at the end of the orbit 
as we did in Appendix~\ref{sol_ret} when discussing the photons returning to the 
disk. Then, $k_r>0$ at $r=r_e$ means that the photon moves to the side $r>r_e$, 
and $k_r<0$ at $r=r_e$ means that the photon moves to the side $r<r_e$. Since 
the photon must be emitted by the disk, we must have $r_{\rm in} \le r_e \le 
r_{\rm out}$.

In the analysis presented below, in each case (Case A1, Case A2, and Case B) 
the symbols have the same meanings as those in the corresponding case in 
Appendix~\ref{sol_ret}.

{\em Case A1: $R=0$ has four real roots, region I $(r\ge r_1)$~~} 
When $r_1\neq r_2$ and $r_{\rm in}>r_1$, let us define 
\begin{eqnarray}
    \tau_{\rm in} \equiv \int_{r_1}^{r_{\rm in}} \frac{dr}{\sqrt{R}} 
        = \frac{2}{\sqrt{(r_1-r_3)(r_2-r_4)}}\,\sn^{-1}\left[
        \left.\sqrt{\frac{(r_2-r_4)(r_{\rm in}-r_1)}{(r_1-r_4)(r_{\rm 
	   in}-r_2)}}\right|m_4\right] \;. \label{tauin_r4}
\end{eqnarray}
Then, $C = 1$ if and only if one of the following conditions is satisfied:
\begin{description}
\item[Condition 1:] ($r_1 \le r_{\rm H}<r_{\rm in}$) \& ($k_r<0$ at 
    $r=r_e$) \& ($\tau_\mu>\tau_e-\tau_{\rm in}$)\,;
\item[Condition 2:] ($r_{\rm H}<r_1<r_{\rm in}$) \& ($k_r<0$ at 
    $r=r_e$) \& ($\tau_e-\tau_{\rm in}<\tau_\mu<\tau_e+\tau_{\rm in}$)\,.
\end{description}

When $r_1=r_2$ and $r_{\rm in}>r_1$, we define
\begin{eqnarray}
    \hat{\tau}_{\rm in} &\equiv& \frac{1}{\sqrt{(r_1-r_3)(r_1-r_4)}}\,\ln
        \left[\frac{\sqrt{(r_{\rm in}-r_3)(r_{\rm in}-r_4)}}{r_{\rm in}
	   -r_1}\right. \nonumber\\
	   &&\left.+\frac{r_1^2+r_3r_4+2r_1r_{\rm in}}{(r_{\rm in}-r_1)
	   \sqrt{(r_1-r_3)(r_1-r_4)}}\right] \;. 
\end{eqnarray}
Then, $C = 1$ if and only if the following condition is satisfied:
\begin{description}
\item[Condition 3:] ($r_{\rm in}>r_1$) \& ($k_r<0$ at $r=r_e$) \& 
    ($\tau_\mu>\hat{\tau}_{\rm in}-\hat{\tau}_e$)\,.
\end{description}

{\em Case A2: $R=0$ has four real roots, region II $(r_3\le r \le r_2)$~~}
When $r_1\neq r_2$ and $r_{\rm in}>r_3$, let us define
\begin{eqnarray}
    \tau_{\rm in} &=& \int_{r_{\rm in}}^{r_2}\frac{dr}{\sqrt{R}} = \frac{2}
	   {\sqrt{(r_1-r_3)(r_2-r_4)}}\,\sn^{-1}\left[\left.\sqrt{\frac{(r_1
	   -r_3)(r_2-r_{\rm in})}{(r_2-r_3)(r_1-r_{\rm in})}}
	   \right|m_4\right] \;.
\end{eqnarray}
Then, $C = 1$ if and only if one of the following conditions is satisfied:
\begin{description}
\item[Condition 4:] ($r_3\le r_{\rm H}<r_{\rm in}$) \& ($k_r>0$ at 
    $r=r_e$) \& ($\tau_\mu>\tau_{e2}+\tau_{\rm in}$)\,;
\item[Condition 5:] ($r_3\le r_{\rm H}<r_{\rm in}$) \& ($k_r<0$ at 
    $r=r_e$) \& ($\tau_\mu>\tau_{\rm in}-\tau_{e2}$)\,;
\item[Condition 6:] ($r_{\rm H}<r_3<r_{\rm in}$) \& ($k_r>0$ at 
    $r=r_e$) \& ($\tau_{e2}+\tau_{\rm in}<\tau_\mu^\prime<\tau_{e2}+
    2\tau_{32}-\tau_{\rm in}$)\,;
\item[Condition 7:] ($r_{\rm H}<r_3<r_{\rm in}$) \& ($k_r<0$ at 
    $r=r_e$) \& ($\tau_{\rm in}-\tau_{e2}<\tau_\mu^\prime<2\tau_{32}-
    \tau_{\rm in}-\tau_{e2}$)\,.
\end{description}

When $r_1=r_2$ and $r_{\rm in}>r_3$, we define
\begin{eqnarray}
    \hat{\tau}_{\rm in} &\equiv& \frac{1}{\sqrt{(r_1-r_3)(r_1-r_4)}}\,\ln
        \left[\frac{\sqrt{(r_{\rm in}-r_3)(r_{\rm in}-r_4)}}{r_1-r_{\rm in}
	   }\right. \nonumber\\
	   &&\left.+\frac{r_1^2+r_3r_4+2r_1r_{\rm in}}{(r_1-r_{\rm in})
        \sqrt{(r_1-r_3)(r_1-r_4)}}\right] \;.
\end{eqnarray}
Then, $C = 1$ if and only if one of the following conditions is satisfied:
\begin{description}
\item[Condition 8:] ($r_3\le r_{\rm H} <r_{\rm in}$) \& ($k_r<0$ at 
    $r=r_e$) \& ($\tau_\mu>\hat{\tau}_e-\hat{\tau}_{\rm in}$)\,;
\item[Condition 9:] ($r_{\rm H} <r_3<r_{\rm in}$) \& ($k_r<0$ at 
    $r=r_e$) \& ($\hat{\tau}_e-\hat{\tau}_{\rm in}<\tau_\mu<
    \hat{\tau}_e+\hat{\tau}_{\rm in}-2\hat{\tau}_3$)\,.
\end{description}

{\em Case 2: $R=0$ has two complex roots and two real roots, region $r\ge r_3$~~}
When $r_{\rm in}>r_3$, let us define
\begin{eqnarray}
    \tau_{\rm in}\equiv\frac{1}{\sqrt{AB}}\,\cn^{-1}\left[\left.\frac{(A-B)
	   r_{\rm in}+r_3 B -r_4 A}{(A+B)r_{\rm in} -r_3 B -r_4 A}\right| 
	   m_2\right] \;.
\end{eqnarray}
Then, $C = 1$ if and only if one of the following conditions is satisfied:
\begin{description}
\item[Condition 10:] ($r_3\le r_{\rm H}<r_{\rm in}$) \& ($k_r<0$ at 
    $r=r_e$) \& ($\tau_\mu>\tau_e-\tau_{\rm in}$) \,;
\item[Condition 11:] ($r_{\rm H}<r_3 \le r_{\rm in}$) \& ($k_r<0$ at 
    $r=r_e$) \& ($\tau_e-\tau_{\rm in}<\tau_\mu<\tau_e+\tau_{\rm 
    in}$)\,.
\end{description}

\section{Integration of Photon Orbits in a Kerr Spacetime II: The $E_\infty
\approx 0$ Case}
\label{integral_p}

A Kerr black hole has an ergosphere inside which the energy of particles
and photons can be negative \citep{mis73}. The radius of the ergosphere is
equal to $2M$ on the equatorial plane. When the black hole rotates fast enough, 
the inner boundary of the disk may enter the ergosphere, i.e., $r_{\rm in}< 
2M$. For instance, when $r_{\rm in} = r_{\rm ms}$, the inner boundary of the
disk enters the ergosphere of the black hole when $a>0.9428M$. When this 
happens, for some photons emitted by the inner part of the disk, 
$E_\infty$---the energy measured at infinity---might be close to zero. This 
needs special treatment since when $E_\infty = 0$ the approach described in 
Appendix~\ref{integral} does not apply. Even when $E_\infty$ is nonzero but 
the ratio $|E_\infty|/E_{l0} \ll 1$, where $E_{l0}$ is the energy of the photon 
measured by a local observer at rest with respect to the gas at radius $r_0$ on 
the disk, the numerical approach based on 
the procedure in Appendix~\ref{integral} cannot give solutions with sufficient 
precision since then $|\lambda|/r_0$ and $Q/r_0^2$ are large numbers. The task 
of this Appendix is to develop special relations for the case of $E_\infty
\approx 0$.

Let us define
\begin{eqnarray}
    \varepsilon \equiv \frac{E_\infty}{E_{l0}} \;, \hspace{1cm}
    |\varepsilon|\ll 1 \;,
\end{eqnarray}
and neglect all terms of order $\varepsilon^2$ and higher in our calculations. 
Then we have
\begin{eqnarray}
    R^\prime(r) &\equiv& R(r)\,\frac{E_\infty^2}{{\cal Q}+L_z^2} 
        ~=~ -r^2 +2M\left(1- \frac{2 a \lambda^\prime \varepsilon}{
        Q^\prime+\lambda^{\prime 2}}\right)r - \frac{a^2 Q^\prime}{
	   Q^\prime+\lambda^{\prime 2}}\;, \label{rr_p} \\[2mm]
    \Theta_\mu^\prime(\mu) &\equiv& \Theta_\mu(\mu)\,\frac{E_\infty^2}
	   {{\cal Q}+L_z^2} ~=~ \frac{Q^\prime}{Q^\prime+\lambda^{\prime 
	   2}} - \mu^2\;, \label{thmu_p}
\end{eqnarray}
where $\lambda^\prime \equiv L_z/E_{l0}$\,, $Q^\prime \equiv {\cal Q}/
E_{l0}^2$\,, which are geometric quantities that do not depend on the energy
of photons. The motion of the photon in the $r-\vartheta$ plane is then 
described by
\begin{eqnarray}
    \int_{r_e}^r \frac{dr}{\sqrt{R^\prime(r)}} = \pm 
        \int_{\mu_e}^\mu \frac{d\mu}{\sqrt{\Theta_\mu^\prime(\mu)}} \;.
    \label{r-mu_p}
\end{eqnarray}

The integral over $\mu$ can be worked out by the following indefinite 
integration
\begin{eqnarray}
    \int \frac{d\mu}{\sqrt{\Theta_\mu^\prime(\mu)}}
        = \arctan\left(\frac{\mu}{\sqrt{\mu_m^2 -\mu^2}}\right) \;, 
	   \hspace{1cm}
    \mu_m \equiv \sqrt{\frac{Q^\prime}{Q^\prime +\lambda^{\prime 
	   2}}} \;. \label{mu_int_p}
\end{eqnarray}

The two roots of $R^\prime(r) = 0$ are $r_1^\prime = r_1 (1+\delta)$ and
$r_2^\prime = r_2 (1-\delta)$, where
\begin{eqnarray}
    r_{1,2} \equiv M\pm\sqrt{M^2 - \frac{a^2 Q^\prime}{Q^\prime + 
	   \lambda^{\prime 2}}} \;, \hspace{1cm}
	   \delta \equiv \frac{-2Ma\lambda^\prime\varepsilon}{\sqrt{
	   \left(Q^\prime +\lambda^{\prime 2}\right)\left[M^2 \left(Q^\prime 
	   +\lambda^{\prime 2}\right)-a^2 Q^\prime\right]}} \;.
\end{eqnarray}
The physically allowed region is given by $R^\prime \ge 0$, i.e. $r_- \le r 
\le r_+$, where $r_+ = \max\left(r_1^\prime, r_2^\prime\right)$, $r_- =
\min\left(r_1^\prime, r_2^\prime\right)$. Thus, the integral over $r$ can be 
worked out by
\begin{eqnarray}
    \int\frac{dr}{\sqrt{R^\prime(r)}} = \int\frac{dr}{\sqrt{\left(r_+
	   -r\right)\left(r-r_-\right)}} =\arctan\left[\frac{2r - 
	   r_1^\prime-r_2^\prime}{2\sqrt{\left(r_1^\prime-r\right)\left(r-
	   r_2^\prime\right)}}\right] \;. \label{r_int_p}
\end{eqnarray}

Since the physically allowed region is bounded by $r_+$ and $r_-$, the photon 
must return to the disk or be captured by the black hole. It cannot escape to
infinity. 

First let us consider the case when the photon returns to the disk. Then, we
must have $\mu = \cos(\pi/2) =0$ as the photon approaches the disk at radius 
$r_a$. Since as it leaves the disk at radius $r_e$ we also have $\mu =0$, the 
photon must have encountered a turning point in $\vartheta$ during its trip. 
Thus, by equation~(\ref{mu_int_p}), along the photon orbit the total integral 
over $\mu$ is
\begin{eqnarray}
    \tau_\mu \equiv 2\int_0^{\mu_m} \frac{d\mu}{\sqrt{
	   \Theta_\mu^\prime(\mu)}} = \pi \;.
\end{eqnarray}

The integral over $r$ along the photon orbit depends on the sign of $k_r$ at
$r=r_a$---i.e., it depends on the direction from which the photon approaches the
disk at $r_a$. It can be shown that we always have $r_- < r_{\rm H}$ for $a^2
<M^2$ and $Q^\prime > 0$. Since a photon cannot get out of a black hole after 
it falls in, its orbit cannot have a turning point in $r$ at $r=r_-$. However, 
it can have a turning point at $r=r_+$. Let us define the following integrals
\begin{eqnarray}
    \tau_{e+} &\equiv& \int_{r_e}^{r_+}\frac{dr}{\sqrt{R^\prime}} = 
        \frac{\pi}{2}-\arctan\left[\frac{2r_e-r_1^\prime-r_2^\prime}{2
	   \sqrt{\left(r_1^\prime-r_e\right)\left(r_e-r_2^\prime\right)}}
	   \right] \;, \\[1mm]
    \tau_{a+} &\equiv& \int_{r_a}^{r_+}\frac{dr}{\sqrt{R^\prime}} = 
        \frac{\pi}{2}-\arctan\left[\frac{2r_a-r_1^\prime-r_2^\prime}
	   {2\sqrt{\left(r_1^\prime-r_a\right)\left(r_a-r_2^\prime\right)}}
	   \right] \;, \\[1mm]
    \tau_{-+} &\equiv& \int_{r_-}^{r_+}\frac{dr}{\sqrt{R^\prime}} = 
        \pi \;.
\end{eqnarray}

Since $\tau_{-+} = \tau_\mu$ and $r_-<r_{\rm H}<r_a<r_+$, for a photon emitted 
by the disk and returning to the disk, it must have $k_r <0$ at $r=r_a$
(i.e., the photon must approach the disk at $r_a$ from the side of $r>r_a$) 
and it must have encountered a turning point at $r=r_+$. Then, the total integral 
over $r$ is
\begin{eqnarray}
    \tau_r = \left(\int_{r_a}^{r_+}+\int_{r_e}^{r_+}\right)\frac{dr}
	   {\sqrt{R^\prime}} = \tau_{a+} + \tau_{e+} \;.
\end{eqnarray} 
Setting $\tau_r=\tau_\mu=\pi$, we can solve for $r_e$. The solution is given
by
\begin{eqnarray}
    r_e + r_a = r_1^\prime + r_2^\prime = 2M \left(1- \frac{2a
	   \lambda^\prime}{Q^\prime + \lambda^{\prime 2}}\,
        \varepsilon\right) \;. \label{sol_p}
\end{eqnarray} 
Of course, a physical solution must satisfy $r_e+ r_a \ge 2 r_{\rm in}$. 
According to equation~(\ref{sol_p}), this condition corresponds to
\begin{eqnarray}
    -2a\lambda^\prime\varepsilon \ge \left(\frac{r_{\rm in}}{M}-1\right)
        \left(Q^\prime+\lambda^{\prime 2}\right) \;. \label{ieq}
\end{eqnarray}
Note that the solution in equation~(\ref{sol_p}) and the condition in 
equation~(\ref{ieq}) do not depend on whether we choose $E_{l0} = E_{la}$ or
$E_{l0} = E_{le}$.

From the above results it is easy to obtain the condition for the photon
to be captured by the black hole, which is as follow: either $k_r<0$ at $r=
r_e$, or $k_r>0$ at $r= r_e$ but the inequality~(\ref{ieq}) is violated.

\section{Calculation of the Constants of Motion}
\label{const}

In the first part of this section we express the constants of motion of a 
photon (in particular, $\lambda$ and $Q$) in terms of the direction of the 
velocity of the photon and the radius as the photon crosses the equatorial 
plane. In the second part, we express the constants of motion in terms of the 
impact parameters of the photon at infinity. The former is useful for 
calculating the orbits of the photons that return to the disk or are captured 
by the black hole, while the latter is useful for calculating the 
orbits of photons that escape to infinity. We will also calculate the redshift 
factor of the photon in each case.

{\em 1. Evaluation of the constants of motion in terms of quantities on the 
equatorial plane}
We need to relate the local rest frame attached to the disk fluid to the 
locally nonrotating frame. The relation is given by the Lorentz transformation
\begin{eqnarray}
    e_t^a &=& \Gamma\left[e_{(t)}^a - v_\varphi e_{(\varphi)
        }^a\right] \;, \label{ett}\\
    e_\varphi^a &=& \Gamma\left[-v_\varphi e_{(t)}^a + e_{(\varphi)}^a
        \right] \;, \label{efit}\\
    e_r^a &=& e_{(r)}^a \;, \label{ert}\\
    e_z^a &=& e_{(z)}^a \label{ezt}\;,
\end{eqnarray}
where $\left\{e_t^a, e_r^a, e_\varphi^a, e_z^a\right\}$ are the locally 
nonrotating frame (see, e.g., Bardeen, Press, \& Teukolsky 1972), 
$\left\{e_{(t)}^a, e_{(r)}^a, e_{(\varphi)}^a, e_{(z)}^a\right\}$ are the 
local rest frame of the disk 
[$e_{(t)}^a = u^a$ is the four-velocity of a disk particle], $v_\varphi$ is 
the azimuthal velocity of a disk particle relative to the locally nonrotating 
frame, and $\Gamma = \left(1-v_\varphi^2\right)^{-1/2}$ is the corresponding 
Lorentz factor. 

In the coordinates $(t,r,\varphi,z)$, the locally nonrotating frame are defined 
by
\begin{eqnarray}
    e_t^a &=& \frac{1}{\chi}\left[\left(\frac{\partial}{\partial t}
	   \right)^a + \omega \left(\frac{\partial}{\partial \varphi}
	   \right)^a\right] \;, \label{et0}\\
    e_\varphi^a &=& \left(\frac{r^2}{A}\right)^{1/2} \left(\frac{\partial}
        {\partial \varphi}\right)^a \;, \label{efi0}\\
    e_r^a &=& \left(\frac{\Delta}{r^2}\right)^{1/2} \left(\frac{\partial}
        {\partial r}\right)^a \;, \label{er0}\\
    e_z^a &=& \left(\frac{\partial}{\partial z}\right)^a 
        = -\frac{1}{r}\left(\frac{\partial}{\partial\vartheta}\right)^a  
	   \;, \label{ez0}
\end{eqnarray}
where $\Delta = r^2 -2M r +a^2$, $A = r^4 + a^2 r(r+2M)$, $\chi = (r^2
\Delta/A)^{1/2}$ (the lapse function in the equatorial plane), and $\omega = 
2Mar/A$ (the frame dragging angular velocity in the equatorial plane).

The direction of the velocity of a photon as it crosses the disk plane is
specified by the normalized four-wavevector of the photon, $n^a \equiv k^a/
k^{(t)} =k^a/E_l$, where $k^a$ is the four-wavevector of the photon, $E_l = 
k^{(t)} = -k_a e_{(t)}^a = -k_a u^a$ is the energy (frequency) of the photon 
measured in the local rest frame. The components of $n^a$ in the local
rest frame of the disk are
\begin{eqnarray}
    n^{(t)} = 1 \;, \hspace{1cm} n^{(z)} = \cos\theta \;,
    \hspace{1cm} n^{(r)} = \sin\theta\cos\phi \;,
    \hspace{1cm} n^{(\phi)} = \sin\theta\sin\phi \;, \label{nka}
\end{eqnarray}
where $(\theta,\phi)$ are spherical coordinates in the local rest 
frame, with the polar angle $\theta$ measured from the normal of the disk, and
the azimuthal angle $\phi$ measured from $e_{(r)}^a$ along the disk radial 
direction.

By equations~(\ref{efit}) and (\ref{nka}), we have
\begin{eqnarray}
    k_a e_\varphi^a = \Gamma\left[-v_\varphi k_a e_{(t)}^a + k_a
        e_{(\varphi)}^a\right] = \Gamma E_l \left(v_\varphi + \sin\theta
	   \sin\phi\right) \;. \label{kaefi1}
\end{eqnarray}
By equation~(\ref{efi0}) and $L_z = k_a(\partial/\partial\phi)^a$, we have
\begin{eqnarray}
    k_a e_\varphi^a = \left(\frac{r^2}{A}\right)^{1/2} k_a \left(
        \frac{\partial}{\partial\varphi}\right)^a = \left(\frac{r^2}
        {A}\right)^{1/2} L_z \;. \label{kaefi2}
\end{eqnarray}
Equations~(\ref{kaefi1}) and (\ref{kaefi2}) lead to
\begin{eqnarray}
    L_z = \left(\frac{A}{r^2}\right)^{1/2} \Gamma \left(v_\varphi + 
        \sin\theta\sin\phi\right) E_l \;. \label{lz1}
\end{eqnarray}

Similarly, equations~(\ref{ett}), (\ref{et0}), (\ref{nka}), and $E_\infty = 
-k_a(\partial/\partial t)^a$ lead to
\begin{eqnarray}
    E_\infty &=& \Gamma\chi(1+v_\varphi\sin\theta\sin\phi)E_l + \omega L_z 
          \nonumber\\[1mm]
        &=& \Gamma\chi\left\{1+\frac{\omega}{\chi}\left(\frac{A}{r^2}
		\right)^{1/2}v_\varphi + \left[v_\varphi+\frac{\omega}{\chi}
		\left(\frac{A}{r^2}\right)^{1/2}\right]\sin\theta\sin\phi
		\right\}E_l \;, \label{e_inf1}
\end{eqnarray}
where in the last step equation~(\ref{lz1}) has been used. 

Equations~(\ref{lz1}) and (\ref{e_inf1}) can be simplified by introducing 
$L^\dagger = \Gamma(A/r^2)^{1/2}v_\varphi$, $E^\dagger = \Gamma\chi + \omega 
L^\dagger$, and 
$\Omega = \omega + \chi(r^2/A)^{1/2} v_\varphi$, which are respectively the 
specific angular momentum, the specific energy-at-infinity, and the angular 
velocity of disk particles. The results are
\begin{eqnarray}
    L_z &=& \left[L^\dagger + \left(\frac{A}{r^2}\right)^{1/2} \Gamma \sin\theta
	   \sin\phi\right] E_l \;. \label{lz2} \\[1mm]
    E_\infty &=& \left[E^\dagger + \left(\frac{A}{r^2}\right)^{1/2}\Gamma\Omega
        \sin\theta\sin\phi\right]E_l \;. \label{e_inf2}
\end{eqnarray}
Note that, although $E_l$ is always positive, $E_\infty$ can be negative when 
the black hole rotates so fast that the radius where the photon is emitted is
inside the ergosphere of the black hole (see Appendix~\ref{integral_p}). Let
$\psi$ denote the angle between the photon wavevector and the direction of
$e_{(\phi)}^a$ in the local rest frame. Then $\cos\psi=\sin\theta\sin\phi$,
and equation~(\ref{e_inf2}) implies that $E_\infty<0$ if and only if $\psi_0
<\psi \le \pi$, where
\begin{eqnarray}
    \psi_0 = \arccos\left(-\frac{rE^\dagger}{A^{1/2}\Gamma\Omega}\right) \;.
\end{eqnarray}
The solution for $\psi_0$ exists for $r<2M\,$, i.e. inside the ergosphere of
the black hole. At $r=2M$ (the boundary of the ergosphere), we have $\psi_0 = 
\pi$ for a Keplerian disk.

Equations~(\ref{ezt}), (\ref{ez0}), and (\ref{nka}) lead to
\begin{eqnarray}
    {\cal Q} = \left[k_a\left(\frac{\partial}{\partial\vartheta}
	   \right)^a\right]^2_{\vartheta=\pi/2} = r^2 \left(k_a e_z^a\right)^2 = 
        r^2 E_l^2 \cos^2\theta \;, \label{cq}
\end{eqnarray}
where the first identity comes from the definition of ${\cal Q}$ \citep{bar72}.

From equations~(\ref{lz2}), (\ref{e_inf2}), and (\ref{cq})  we have
\begin{eqnarray}
    \lambda &=& \frac{L_z}{E_\infty} ~=~ \frac{L^\dagger + (A/r^2)^{1/2} \Gamma 
	   \sin\theta\sin\phi}{E^\dagger + (A/r^2)^{1/2}\Gamma\Omega\sin\theta
	   \sin\phi} \;, \label{lam} \\[1mm]
    Q &=& \frac{{\cal Q}}{E_\infty^2} ~=~ \frac{r^2 \cos^2\theta}{\left[
	   E^\dagger + (A/r^2)^{1/2}\Gamma\Omega\sin\theta\sin\phi\right]^2} \;.
        \label{qq}
\end{eqnarray}
These two equations relate $\lambda$ and $Q$ to the direction of the photon
wavevector ($\theta$, $\phi$), the radius ($r$), as well as the quantities 
specifying the disk motion ($L^\dagger$, $E^\dagger$, $\Omega$, and 
$\Gamma$---of course they are not independent) at the point where the orbit of 
the photon crosses the disk plane. For the formulae for calculating $L^\dagger$, 
$E^\dagger$, $\Omega$ ... in the case of a Keplerian disk, see \cite{pag74}. 
The parameters $\lambda$ and $Q$ do not depend on the energy of the photon 
$E_l$, which is expected under the geometric optics approximation.

Equation~(\ref{lam}) can be rewritten as
\begin{eqnarray}
    \Omega\lambda = 1- \frac{E^\dagger-\Omega L^\dagger}{E^\dagger + 
        (A/r^2)^{1/2}\Gamma\Omega\sin\theta\sin\phi} \;, \label{ome_lam}
\end{eqnarray}
which by equation~(\ref{e_inf2}) leads to
\begin{eqnarray}
    \frac{E_\infty}{E_l} = \frac{E^\dagger-\Omega L^\dagger}{1-\Omega\lambda} \;.
    \label{einf_el}
\end{eqnarray} 
Equation~(\ref{einf_el}) is useful for calculating the redshift factor 
of a photon. For example, if a photon is emitted from the disk at $r = r_e$
and then absorbed by the disk at $r = r_a$, the ratio of locally measured 
energy (frequency) of the photon at $r_e$ and $r_a$ is simply
\begin{eqnarray}
    \frac{E_{la}}{E_{le}} = \frac{E_e^\dagger - \Omega_e L_e^\dagger}{E_a^\dagger 
        - \Omega_a L_a^\dagger} \frac{1-\Omega_a\lambda}{1-\Omega_e\lambda} \;,
    \label{el12}
\end{eqnarray}
where label ``$e$'' means evaluation at $r=r_e\,$, and label ``$a$'' means 
evaluation at $r=r_a\,$. Equation~(\ref{el12}) implies that if a photon is 
emitted and returns to the disk at the same radius, then the locally measured 
energy of the photon does not change. 

Given $\lambda$ and $Q$, we can also calculate the value of $\cos\theta$.
From equations~(\ref{qq}) and (\ref{ome_lam}), we have
\begin{eqnarray}
    \cos\theta = \pm \frac{\sqrt{Q}}{r}\frac{E^\dagger-\Omega L^\dagger}{|1-
	   \Omega\lambda|} \;, \label{coth}
\end{eqnarray}
where the sign of $\cos\theta$ should be chosen according to the problem at
hand. If the photon is leaving the disk, then $0\le\theta<\pi/2$, and $\cos
\theta$ is positive. If the photon is approaching the disk, then $\pi/2
<\theta\le\pi$, and $\cos\theta$ is negative. Note, according to 
equation~(\ref{einf_el}), $1-\Omega\lambda$ has 
the same sign as $E_\infty$ since both $E_l$ and $E^\dagger-\Omega L^\dagger$ are 
positive.\footnote{$E^\dagger-\Omega L^\dagger = \chi/\Gamma$ for any particle whose 
velocity has only an azimuthal component.} Since $E_\infty$ can be negative, 
$1-\Omega\lambda$ can also be negative. This is the reason for the absolute 
value signs in equation~(\ref{coth}).

{\em 2. Evaluation of the constants of motion in terms of photon impact 
parameters at infinity~~}
The apparent position of the disk image as seen by an observer is 
conveniently represented by two impact parameters $\alpha$ and $\beta$, measured 
relative to the direction to the center of the black hole. The impact parameters 
$\alpha$ and $\beta$ are, respectively, the displacement of the image perpendicular 
to the projection of the rotation axis of the black hole on the sky and the 
displacement parallel to the projection of the axis. They are related to the 
conserved parameters $\lambda$ and $Q$ by
\citep{cun73,cun75}
\begin{eqnarray}
    \alpha = - \lambda\csc\vartheta_{\rm obs} \;, \hspace{1cm}
    \beta = \pm\left(Q + a^2\cos^2\vartheta_{\rm obs} - \lambda^2\cot^2
        \vartheta_{\rm obs}\right)^{1/2} \;, \label{alp_beta}
\end{eqnarray}
where $\vartheta_{\rm obs}$ is the polar angle of the observer with respect to
the rotation axis of the black hole (i.e., the inclination angle). From 
equation~(\ref{alp_beta}) we can solve for $\lambda$ and $Q$. The solutions are
\begin{eqnarray}
    \lambda = - \alpha \sin\vartheta_{\rm obs} \;, \hspace{1cm}
    Q = \beta^2 + (\alpha^2 - a^2) \cos^2\vartheta_{\rm obs} \;.
    \label{lq_ab}
\end{eqnarray}

For ray-tracing, it is more convenient to use polar coordinates in the plane of 
the disk since disk particles are on circular orbits. Figure~\ref{coord1} shows 
the relation between the impact parameters in the observer's sky (the plane $S$, 
the image plane) and the polar coordinates in the plane of the disk (the 
plane $S^\prime$). For a point $P$ on $S$, labeled by a pair of impact 
parameters $(\alpha,\beta)\,$, there is a corresponding point $P^\prime$ on 
$S^\prime$. $P^\prime$ is obtained from $P$ by drawing a straight line from $P$ 
parallel to the line of sight. The polar coordinates $(r^\prime,\varphi^\prime)$ 
of $P^\prime$ are related to the impact parameters $(\alpha,\beta)$ by
\begin{eqnarray}
    \alpha = r^\prime\cos\varphi^\prime \;, \hspace{1cm}
    \beta = r^\prime\sin\varphi^\prime\cos\vartheta_{\rm obs} \;.
    \label{ab_rfi}
\end{eqnarray}
Therefore, the image of a circular orbit in the disk plane is a circle in 
$S^\prime$, but an ellipse in $S$. 

From equations~(\ref{lq_ab}) and (\ref{ab_rfi}) we obtain
\begin{eqnarray}
    \lambda = - r^\prime\cos\varphi^\prime\sin\vartheta_{\rm obs} \;,
    \hspace{1cm}
    Q = (r^{\prime 2}-a^2)\cos^2\vartheta_{\rm obs} \;.
    \label{lq_rfi}
\end{eqnarray}
These two equations relate the conserved quantities $\lambda$ and $Q$ to the
polar coordinates on the $S^\prime$ plane which is gridded into many surface
elements to make the ray-tracing computation.\footnote{The fictitious 
image plane is gridded as follow: Using the polar coordinates $r^\prime$ and
$\varphi^\prime$ in $S^\prime$, with the origin at the intersection of the
line of sight with $S^\prime$, we divide ${\rm log}\,r^\prime$ uniformly from 
${\rm log}\,r^\prime= {\rm log}\,r_{\rm in}$ to ${\rm log}\,r^\prime = 
{\rm log}\,r_{\rm out}$, and divide $\varphi$ uniformly from $\varphi=0$ to 
$\varphi=2\pi$. Because of the light focusing effect of the
black hole, the image of the inner boundary of the disk on $S^\prime$ has
a somewhat larger radius than $r_{\rm in}$. So, starting the computation
from $r^\prime = r_{\rm in}$ does not miss points in the inner region of
the disk. Since $r_{\rm out} \gg r_{\rm g}$, the image of the outer boundary 
of the disk on $S^\prime$ has a radius that is almost equal to $r_{\rm out}$.} 
The element of solid angle seen by the observer is then
\begin{eqnarray}
    d\Omega_{\rm obs} = \frac{d\alpha d\beta}{D^2} = \frac{
	   \cos\vartheta_{\rm obs}\,r^\prime d r^\prime d\varphi^\prime}
        {D^2} \;, \label{solid_obs}
\end{eqnarray}
where $D$ is the distance from the observer to the black hole.

\section{The Radiation Flux of the Disk and the Balance of Energy}
\label{flux}

The conservation of angular momentum in a geometrically thin accretion disk 
rotating around a Kerr black hole in the equatorial plane is described by 
\citep{nov73,pag74}
\begin{eqnarray}
    \frac{d}{dr}\left(\dot{M}L^\dagger-g\right) = 4\pi r \left.T_\varphi^{\;z}
        \right|_{z=H} \;. \label{cons_l}
\end{eqnarray}
Here, as usual, $\dot{M}$ is the mass accretion rate, $L^\dagger$ is the specific 
angular momentum of disk particles, $g$ is the internal torque of the disk 
which transports angular momentum outward in the radial direction, 
$T_\varphi^{\;z}$ is the $\varphi$-$z$ component of the stress-energy
tensor of the radiation produced by the disk, which describes the flow and 
transport of angular momentum in the vertical direction, and $H$ is the 
half-thickness of the disk.

Similarly, the conservation of energy in the disk is described by
\begin{eqnarray}
    \frac{d}{dr}\left(\dot{M}E^\dagger-g\Omega\right) = -4\pi r \left.T_t^{\;z}
        \right|_{z=H} \;,\label{cons_e}
\end{eqnarray}
where $E^\dagger$ is the specific energy of disk particles, $\Omega$ is the 
angular velocity of the disk, and $T_t^{\;z}$ is the $t$-$z$ component of the 
stress-energy tensor of the radiation, which describes the flow and transport
of energy in the vertical direction.

In a local Lorentz frame, the stress-energy tensor of photons can be 
expressed as an integral \citep{mis73,tho74}
\begin{eqnarray}
    T^{ab} = \int n^a n^b I d\Omega \;,
\end{eqnarray}
where $n^a = k^a/E_l$, $k^a$ is the four-wavevector of the photon, $E_l$ is 
the energy of the photon measured in the local Lorentz frame ($n^a$ does 
not depend on the energy of the photon), $I$ is the intensity of photons which 
is a function of direction, and $d\Omega$ is the element of solid angle defined 
by the direction of the three-wavevector of the photon. 

The evaluation of $T^{ab}$ on the surface of the disk can be written in
spherical coordinates $(\theta,\phi)$ in the local rest frame of the disk 
particles, where $\theta$ is measured from the normal of the disk surface
(see Appendix~\ref{const}). Then we have
\begin{eqnarray}
    \left.T_t^{\;z}\right|_{z=H} = \int n_t \cos\theta\, I(\theta,\phi) 
	   d\Omega\;, \hspace{1cm}
    \left.T_\varphi^{\;z}\right|_{z=H} = \int n_\varphi \cos\theta\,
        I(\theta,\phi) d\Omega \;, \label{tfz}
\end{eqnarray}
where $d\Omega = \sin\theta d\theta d\phi$, and we have used $n^z = n^{(z)} 
= \cos\theta$ (see eq.~[\ref{nka}]). 

From the definition of $n^a$, we have $n_t = k_t/E_l = -E_\infty/E_l$,
$n_\varphi = k_\varphi/E_l = L_z/E_l$, where $E_\infty$ is the 
energy-at-infinity of the photon, and $L_z=\lambda E_\infty$ is the angular 
momentum of the photon about the axis of the black hole. Then, by 
equations~(\ref{lz2}) and (\ref{e_inf2}) we have
\begin{eqnarray}
    n_t = -E^\dagger-\left(\frac{A}{r^2}\right)^{1/2}\Gamma\Omega \cos\psi \;, 
    \hspace{1cm} n_\varphi = L^\dagger+\left(\frac{A}{r^2}\right)^{1/2}\Gamma 
        \cos\psi \;, \label{ntf}
\end{eqnarray}
where $\cos\psi = \sin\theta\sin\phi$, $\psi$ is the angle between the 
wavevector of the photon and the direction of $e_{(\phi)}^a$ in the local
rest frame.

Substituting equations~(\ref{tfz}) and (\ref{ntf}) into equations~(\ref{cons_l})
and (\ref{cons_e}), we get
\begin{eqnarray}
    \frac{d}{dr}\left(\dot{M}L^\dagger-g\right) &=& 4\pi r(L^\dagger F +S) \;, 
        \label{cons_l2} \\
    \frac{d}{dr}\left(\dot{M}E^\dagger-g\Omega\right) &=& 4\pi r(E^\dagger F 
        +S\Omega) \;, \label{cons_e2}
\end{eqnarray}
where
\begin{eqnarray}
    F \equiv \int \cos\theta\, I d\Omega  \label{flux0}
\end{eqnarray}
is the local energy flux density of the radiation,
\begin{eqnarray}
    S \equiv \left(\frac{A}{r^2}\right)^{1/2}\Gamma\int 
        \cos\psi\cos\theta\, I d\Omega  \label{ss}
\end{eqnarray}
is the stress density of the radiation which represents the transport of 
angular momentum in the vertical direction. When $S>0$, the angular momentum 
is transported along the $+z$ direction, which means that the radiation exerts 
a negative torque on the disk.

Equations~(\ref{cons_l2}) and (\ref{cons_e2}) have the formal solution
\citep{cun76,li02a}
\begin{eqnarray}
    F = F_0 + F_S \;, \label{f_sol}
\end{eqnarray}
where
\begin{eqnarray}
    F_0 &\equiv& \frac{\dot{M}}{4\pi r}f + \frac{g_{\rm in}}{4\pi}
        \left(E_{\rm in}^\dagger-\Omega_{\rm in} L_{\rm in}^\dagger\right) 
        \frac{1}{r}\left(-\frac{d\Omega}{dr}\right)(E^\dagger-\Omega 
        L^\dagger)^{-2} \;, \label{f0}\\
    F_S &\equiv& -\frac{1}{r}\left(-\frac{d\Omega}{dr}\right)(E^\dagger-
        \Omega L^\dagger)^{-2} \int_{r_{\rm in}}^r(E^\dagger-\Omega 
        L^\dagger) Sr dr \;, \label{fs}
\end{eqnarray}
the subscript ``in'' denotes evaluation at the inner boundary of the disk, 
and
\begin{eqnarray}
    f \equiv \left(-\frac{d\Omega}{dr}\right)(E^\dagger-\Omega L^\dagger)^{-2} 
        \int_{r_{\rm in}}^r (E^\dagger-\Omega L^\dagger) 
        \frac{dL^\dagger}{dr} dr\;.
\end{eqnarray}
The quantity $g_{\rm in}$ denotes the value of the internal torque of the disk
at the inner boundary. The function $f$ has been worked out by \citet{pag74} 
and is given by their equation~(15n). Note, $F_0$ is just the solution for the 
radiation flux density of the disk when the effect of returning radiation is 
ignored.

We call the solution given by equation~(\ref{f_sol}) a ``formal solution''
since the stress density $S$ is a functional of the flux density of the
returning radiation in general. 

The flux density $F$ defined by equation~(\ref{flux0}) is the ``net''
flux of energy, which usually contains two components: an outgoing component, 
corresponding to the radiation leaving the disk surface; and an ingoing 
component, corresponding to the radiation approaching the disk surface.
To decompose the net flux density into an outgoing component and an ingoing 
component, let us write
\begin{eqnarray}
    I(\theta,\phi)= \left\{\begin{array}{ll}
        I_{\rm out}(\theta,\phi) \;, & 0\le \theta <\pi/2 \\
	   I_{\rm in}(\theta,\phi) \;, & \pi/2< \theta \le\pi
	   \end{array}\right. \;.
\end{eqnarray}
That is, we use $I_{\rm out}$ to denote the radiation intensity for the 
outgoing photons, and $I_{\rm in}$ to denote the radiation intensity for the 
ingoing photons. Then, the outgoing flux density $F_{\rm out}$ defined earlier, 
is given by
\begin{eqnarray}
    F_{\rm out} \equiv \int_{\Omega_+} \cos\theta\, I_{\rm out} d\Omega =
        \int_0^{2\pi}\int_0^{\pi/2} I_{\rm out}(\theta,\phi)\cos\theta 
	   \sin\theta d\theta d\phi\;, \label{flux_o}
\end{eqnarray}
where $\Omega_+$ denotes the whole solid angle above the disk surface, i.e. 
the solid angle defined by $0\le\theta<\pi/2$ and $0\le\phi<2\pi$. The ingoing
flux density $F_{\rm in}$ is similarly given by
\begin{eqnarray}
    F_{\rm in} &\equiv& -\int_{\Omega_-} \cos\theta\, I_{\rm in} d\Omega =
        -\int_0^{2\pi}\int_{\pi/2}^{\pi} I_{\rm in}(\theta,\phi) \cos\theta
	   \sin\theta d\theta d\phi \nonumber\\
	 &=& \int_0^{2\pi}\int_0^{\pi/2} I_{\rm in}(\pi-\theta,\pi+\phi) 
	   \cos\theta\sin\theta d\theta d\phi \;,\label{flux_i}
\end{eqnarray}
where $\Omega_-$ denotes the whole solid angle below the disk surface, i.e. 
the solid angle defined by $\pi/2 <\theta\le\pi$ and $0\le\phi<2\pi$. 

In the last line of equation~(\ref{flux_i}), we have expressed $F_{\rm in}$ as 
an integral over the reverse direction of the wavevector of the incoming 
photons, i.e. an integral over $\Omega_+$, which in practice is easier to 
handle. With the above definitions, we have $F_{\rm out} \ge 0$, $F_{\rm in}\ge 
0$, and $F = F_{\rm out}-F_{\rm in}$. Then, by equation~(\ref{f_sol}),
we have
\begin{eqnarray}
    F_{\rm out} = F_0 + F_{\rm in} + F_S \;. \label{sol_fout}
\end{eqnarray}
Equation~(\ref{sol_fout}) shows that the returning radiation makes two 
distinct contributions to the outgoing flux density, $F_{\rm in}$ and $F_S$. 
$F_{\rm in}$ comes directly from the energy carried by the returning
radiation, while $F_S$ represents the work done by the returning radiation 
on the disk arising from the fact that the returning radiation is
nonaxisymmetric about the normal of the disk surface at the point
where the photon crosses the disk (see below).

The outgoing radiation, which is emitted by the disk locally, is symmetric 
around the normal of the disk surface. However the ingoing radiation, emitted 
by the disk remotely and focused by the gravity of the black 
hole, must be highly nonsymmetric. Thus we have in the frame moving with the 
disk fluid
\begin{eqnarray}
    \partial I_{\rm out}/\partial \phi = 0 \;, \hspace{1cm}
    \partial I_{\rm in}/\partial \phi \neq 0 \;. \label{dI}
\end{eqnarray}
Because of the presence of $\cos\psi\equiv\sin\theta\sin\phi$ in 
equation~(\ref{ss}) that defines $S$, only the part of $I$ that is 
nonsymmetric around the disk normal contributes to $S$. By equation~(\ref{dI}), 
this means that only the ingoing component of the radiation contributes to $S$. 
Thus we have
\begin{eqnarray}
    S &=& \left(\frac{A}{r^2}\right)^{1/2}\Gamma\int_{\Omega_-} 
        I_{\rm in}(\theta,\phi) \cos\psi\cos\theta d\Omega \nonumber\\
      &=& \left(\frac{A}{r^2}\right)^{1/2}\Gamma \int_{\Omega_+} I_{\rm in}
        (\pi-\theta,\pi+\phi) \cos\theta\sin\theta\sin\phi\, d\Omega \;.
    \label{ss2}
\end{eqnarray}

We assume that the radiation emitted by the disk is either isotropic, i.e.,
$I_{\rm out}$ is independent of $\theta$ and $\phi$, or limb-darkened which
corresponds to $I_{\rm out}\propto 2+3\cos\theta$ (see Chandrasekhar
1950; Cunningham 1976). Then, by equation~(\ref{flux_o}), the outgoing 
intensity is related to the outgoing flux by
\begin{eqnarray}
    I_{\rm out} = \frac{1}{\pi} F_{\rm out} \Upsilon \;, \hspace{1cm}
    \Upsilon \equiv\left\{\begin{array}{ll}
        1 \;, & \mbox{for isotropic radiation} \\[1mm]
	   \frac{1}{2}+\frac{3}{4}\cos\theta \;, & 
	   \mbox{for limb-darkened radiation}
	   \end{array}\right. \;. \label{iout}
\end{eqnarray}

The evaluation of $I_{\rm in}$ in a given direction at radius $r=r_a$ in the 
disk can be calculated from the outgoing intensity at radius $r=r_e$ at the
point where the photon comes from. Since $I_{E_l}/E_l^3$ 
is invariant along the orbit of the photon \citep{mis73}, where $I_{E_l}$ is 
the specific intensity that is related to the intensity $I$ by $I = 
\int_0^\infty I_{E_l} d E_l$, we have
\begin{eqnarray}
    I_{\rm in}(r=r_a) ~=~ \hat{g}^4 I_{\rm out}(r=r_e) \;, \label{iin}
\end{eqnarray}
where $\hat{g} \equiv E_l(r=r_a)/E_l(r=r_e)$ is the redshift factor of the
returning photon, given in equation~(\ref{el12}).

Since $F_{\rm in}$ and $F_S$ themselves are determined by $F_{\rm out}$ at the 
radii where the photons contributing to $F_{\rm in}$ and $F_S$ come from---as 
equation~(\ref{iin}) shows---equation~(\ref{sol_fout}) must be solved 
iteratively. First, we ignore the effect of returning radiation and take 
$F_{\rm out} = F_0$. With this outgoing flux as an input, we calculate $I_{\rm 
in}$ by equations~(\ref{iout}) and (\ref{iin}), then $F_{\rm in}$ and $S$ 
by equations~(\ref{flux_i}) and (\ref{ss2}), and $F_S$ by equation~(\ref{fs}).
Next, we add $F_{\rm in}$ and $F_S$ to $F_0$ to obtain a new outgoing flux 
density $F_{\rm out}$, and repeat the calculations to obtain a new $F_{\rm in}$ 
and a new $F_S$. This process is repeated until the solution 
converges. In practice, five iterations are sufficient for convergence.

The integrals in equations~(\ref{flux_i}) and (\ref{ss2}) can be computed with 
the ray-tracing technique. We divide the solid angle $\Omega_+$ at radius 
$r_a$ into a number of small elements, and for each element in a direction 
defined by $(\theta,\phi)$ we trace the orbit of the photon---by using the
formulae provided in Appendixes~\ref{sol_ret} and \ref{const}---until the orbit 
crosses the disk at radius $r_e$. Then, we calculate $I_{\rm out}$ at $r_e$ 
through equation~(\ref{iout}), the redshift factor $\hat{g}$ through 
equation~(\ref{el12}), and the corresponding $I_{\rm in}$ at $r_a$ through 
equation~(\ref{iin}). Thus we obtain the integrand in equations~(\ref{flux_i}) 
and (\ref{ss2}) for each element of solid angle. Summing the contributions 
gives rise to $F_{\rm in}$ and $S$ at $r_a$. We remark that since most of the 
returning radiation comes from the general direction of the black hole, to 
obtain results with high precision it is necessary to divide the solid angle 
$\Omega_+$ nonuniformly, with many more resolution elements in the direction 
of the black hole. This is most critical when $r_a \gg M$.

The radiation returning to the disk is, as we have assumed in this paper,
absorbed and reprocessed by the disk, and then reradiated away. Then,
the radiation that 
permanently leaves the disk consists of only two parts: one part escapes to 
infinity, the other is captured by the black hole. The total power of the disk, 
i.e. the total energy carried by the radiation that permanently leaves the 
disk per unit time as measured by an observer at infinity, is \citep{tho74}
\begin{eqnarray}
    {\cal L}_{\rm total} = -4\pi \int_{r_{\rm in}}^{r_{\rm out}} 
        \left.T_t^{\;z}\right|_{z=H} r dr \;. \label{power}
\end{eqnarray}
Substituting equation~(\ref{cons_e}) into equation~(\ref{power}), and using
suitable boundary conditions at $r_{\rm out}$ (i.e., $E^\dagger \approx 1$ and 
$g \Omega \approx 0$ at $r_{\rm out} \gg r_{\rm in}$), we obtain ${\cal 
L}_{\rm total}$ in terms of $\dot{M}$ and $g_{\rm in}$, which is just 
equation~(\ref{power_t}) in \S\ref{assump}. Equation~(\ref{power_t}) clearly 
shows that the power of the disk comes from the gravitational binding energy 
of disk particles and a contribution from a torque at the inner boundary 
of the disk whenever the torque is nonzero.

\section{The Disk Spectrum as Observed by a Distant Observer}
\label{spectra_obs}

The specific flux density of the disk radiation as observed by a remote 
observer, whose distance to the black hole is much larger than the size of 
the disk, is given by
\begin{eqnarray}
    F_{E_{\rm obs}} = \int I_{E_{\rm obs}} d\Omega_{\rm obs} \;, 
    \label{feo}
\end{eqnarray}
where $I_{E_{\rm obs}}$ is the specific intensity of the radiation, $E_{\rm 
obs}$ is the photon energy, both measured by the remote observer, 
and $d\Omega_{\rm obs}$ is the element of the solid angle subtended by the
image of the disk on the observer's sky.

Using the fact that $I_{E_l}/E_l^3$ is invariant along the path of a photon, 
where $E_l$ is the energy measured by any local observer on the path 
\citep{mis73}, equation~(\ref{feo}) can be rewritten as
\begin{eqnarray}
    F_{E_{\rm obs}} = \int g^3 I_{E_{\rm em}} d\Omega_{\rm obs} \;.
    \label{feo2}
\end{eqnarray}
$E_{\rm em}$ is the energy of the photon at its point of emission on the
disk as measured by an observer located at that point who is corotating
with the disk. $I_{E_{\rm em}}$ is the specific intensity measured by that 
observer, and
\begin{eqnarray}
    g\equiv \frac{E_{\rm obs}}{E_{\rm em}}  \label{red_shift}
\end{eqnarray}
is the redshift of the photon (see eq.~[\ref{einf_el}], where $E_\infty =
E_{\rm obs}$, $E_l = E_{\rm em}$).

Suppose the disk radiates like a blackbody. Then the effective temperature of 
the disk measured by a locally corotating observer is simply 
\begin{eqnarray}
    T_{\rm eff}(r) = \left[\frac{F_{\rm out}(r)}{\sigma_{\rm SB}}
	   \right]^{1/4} \;, \label{teff}
\end{eqnarray}
where $\sigma_{\rm SB}$ is the Stefan-Boltzmann constant, and $F_{\rm out}$ is the 
outgoing energy flux of the disk measured by the locally corotating observer
(Appendix~\ref{flux}). Suppose the color temperature $T_{\rm col}$ is related to 
the effective temperature by
\begin{eqnarray}
    T_{\rm col}(r) = f_{\rm col} T_{\rm eff}(r) \;, \label{f_col}
\end{eqnarray}
where $f_{\rm col}$ is a constant. Then, the local specific intensity of the 
radiation emitted by the disk is
\begin{eqnarray}
    I_{E_{\rm em}} = \frac{2 f_{\rm col}^{-4}E_{\rm em}^3}{\exp(E_{\rm em}/
	   k_{\rm B}T_{\rm col}) -1}\Upsilon \;, \label{iem}
\end{eqnarray}
where $k_{\rm B}$ is the Boltzmann constant, $\Upsilon$ is a function of $\theta$
(the angle between the wavevector of the photon emitted by the disk and the normal 
of the disk surface) and is given by equation~(\ref{iout}) when the disk
emission is isotropic or limb-darkened.

Substituting equations~(\ref{red_shift}), (\ref{f_col}), and (\ref{iem}) into
equation~(\ref{feo2}), we have
\begin{eqnarray}
    F_{E_{\rm obs}} = 2f_{\rm col}^{-4} E_{\rm obs}^3 \int\frac{\Upsilon 
	   d\Omega_{\rm obs}}{\exp\left[E_{\rm obs}/\left(gf_{\rm col}
	   k_{\rm B}T_{\rm eff}\right)\right]-1} \;. 
    \label{feo3}
\end{eqnarray}
The integral in equation~(\ref{feo3}) has a form that is suitable for 
computations with the ray-tracing technique (see the second part in
App.~\ref{const}).

\section{Comparison of KERRBB with GRAD}
\label{grad}
We have compared KERRBB with the model GRAD in XSPEC, which considers
a relativistic disk around a Schwarzschild black hole
\citep{han89,ebi91,ebi03}.  For this comparison, we set $a=0$ in
KERRBB and ignored the returning radiation and limb-darkening since
these are the assumptions made by GRAD.  (Note that, for the case of a
Schwarzschild black hole, the effect of returning radiation is not
important, see \S\ref{return}.) We find that, at low inclination
angles KERRBB gives results that are consistent with those of GRAD.
However, at high inclination angles, the results differ.

We find that the inconsistency between KERRBB and GRAD is caused by
the following two problems in GRAD: (1) GRAD uses an incorrect formula
to calculate the redshift factor of a photon approaching the observer
(see below). (2) GRAD takes the 
resolution in the azimuthal direction to be independent of the 
inclination angle.  This leads to large errors for high inclination 
angles.  Our ray-tracing code uses a resolution proportional to $1/\cos
\vartheta_{\rm obs}$, which gives better numerical accuracy. 

The correct formula for calculating the redshift factor of a photon
emitted by a Keplerian disk around a Schwarzschild black hole, adapted
to the form used in GRAD, is
\begin{eqnarray}
    g = \sqrt{1-\frac{3M}{r}} \left[1-\left(1-\frac{2M}{r}\right)^{-1}
        \sqrt{Mr} \left(\frac{d\varphi}{dt}\right)_{\rm ph}\right]^{-1} \;,
        \label{red_sch}
\end{eqnarray}
where $(d\varphi/dt)_{\rm ph}$ is the angular velocity of the photon
evaluated at the point on the disk where it is emitted. 
Equation~(\ref{red_sch}) can easily be obtained from 
equation~(\ref{einf_el}) by using $(d\varphi/dt)_{\rm ph} = 
\lambda\Delta/r^4$ when $a=0$ \citep{bar72,cha83}.

Equation~(\ref{red_sch}) differs from equation~(B2) in
\citet{ebi03}---which is used in GRAD---in the following respects: the
power index of $(1-2M/r)$ in equation~(\ref{red_sch}) is $-1$, whereas in
equation~(B2) in \citet{ebi03} it is taken to be $-0.5$. [Note that
equation~(\ref{red_sch}) agrees with equation~(A15) in \citet{ebi91},
except that $d\varphi_{\rm ph}/dt_{\rm ph}$ in their (A15) should be
$d\varphi_{\rm ph}^\prime/dt_{\rm ph}$ according to their notation.]

%\citet{ebi03} claimed that their equation~(B2) is the correct formula
%for calculating the photon redshift factor. However, we find that
%their equation~(B2) is wrong since it was derived from a set of
%equations in \citet{ebi91} and one of those equations was wrong.

Equation~(B2) in \citet{ebi03} was derived from equations~(A6)--(A14)
in \citet{ebi91}. However, equation~(A7) in \citet{ebi91} is wrong;
it should be
\begin{eqnarray}
    \left(\begin{array}{c}
	\epsilon\\p_r\\p_{\theta^\prime}\\p_{\varphi^\prime}
	\end{array}\right)_{\rm LNRO} = \epsilon_{\rm LNRO}
	\left[\begin{array}{c}
	1\\(1-2M/r)^{-1} (dr_{\rm ph}/d\theta_{\rm ph})/
		(dt_{\rm ph}/d\theta_{\rm ph})\\
	(1-2M/r)^{-1/2} r (d\theta_{\rm ph}^\prime/
		d\theta_{\rm ph})/(dt_{\rm ph}/d\theta_{\rm ph})\\
	(1-2M/r)^{-1/2} r (d\varphi_{\rm ph}^\prime/
		d\theta_{\rm ph})/(dt_{\rm ph}/d\theta_{\rm ph})
	\end{array}\right] \;, \label{e_lnro}
\end{eqnarray}
where we have used the same notation as in \citet{ebi91} and we have
set $G=c=1$. Since on a photon orbit in the Schwarzschild spacetime we
have $ds^2 = 0 = -(1-2M/r) dt^2 + (1-2M/r)^{-1} dr^2 + r^2
d\theta^{\prime 2} + r^2 \sin^2\theta^{\prime} d\varphi^{\prime 2}$,
it is easy to check that equation~(\ref{e_lnro}) satisfies the null
condition $\epsilon^2-p_r^2-p_{\theta^\prime}^2 -p_{\varphi^\prime}^2
= 0$ (note, $\vartheta^\prime = \pi/2$ on the equatorial plane), but
equation~(A7) in \citet{ebi91} does not. In addition, there are a few
other errors and typos in the Appendix of \citet{ebi91}: their
equation~(A8) is incorrect (subsequently corrected in Ebisawa et
al. 2003), and in the $4\times 4$ matrix in their equation~(A11), the
element on the top-right corner should be $-\beta\gamma$, not
$\beta\gamma$ (this is obvious since the Lorentz transformation matrix
should be symmetric).

If the above correct equations are used, then one can derive a
redshift factor that is exactly the same as our
equation~(\ref{red_sch}).

Unfortunately, equation~(B2) of \citet{ebi03} has been used in the
current version of GRAD in XSPEC (both XSPEC11 and XSPEC12) at the
time the present paper was written. In addition, as Ebisawa et
al. applied their equation in GRAD, they replaced the ``$-$'' sign in
front of $(1-2M/r)$ by a ``$+$'' sign. Although the correct formula
was shown in \citet{ebi91}---i.e., their equation~(A15), except that
$d\varphi_{\rm ph}/dt_{\rm ph}$ should be $d\varphi_{\rm
ph}^\prime/dt_{\rm ph}$---lately it was replaced by an incorrect
equation---equation~(B2) in \citet{ebi03}.

In Figure~\ref{grad_kerrbb} we show some examples that compare the
results given by KERRBB, those given by GRAD, and those given by a 
modified GRAD where we replaced
the following statement in GRAD,

Red = 1.0D0/(1.0D0+SQRT(R0/2.0D0)/(1.0D0$-$1.0D0/R0)**0.5D0*dphdx/dtdx),

\noindent
by 

Red = 1.0D0/(1.0D0$-$SQRT(R0/2.0D0)/(1.0D0$-$1.0D0/R0)*dphdx/dtdx) ,

\noindent
(i.e., changing the ``+'' in front of ``SQRT'' to a ``$-$'', and
deleting ``**0.5''). The parameters for the calculated models are:
$M=10M_\odot$, $D=10$kpc, $\dot{M} = 10^{19} {\rm g~sec}^{-1}$,
$f_{\rm col} = 1$, and $\vartheta_{\rm obs} = 20^\circ$, $60^\circ$,
and $85^\circ$ in each panel. (As already mentioned, for KERRBB we set
$a=0$, $\eta=0$, and turned off returning radiation and
limb-darkening.)  The horizontal axis shows the observed photon energy
in units of keV.  The vertical axis shows the specific photon number
density calculated by KERRBB (solid lines), GRAD (dashed lines), and
the modified GRAD (dashed lines), in units of photons keV$^{-1}$
cm$^{-2}$ sec$^{-1}$.

Figure~\ref{grad_kerrbb} shows that, after the incorrect formula in
GRAD is replaced with the correct one, GRAD agrees quite well with
KERRBB. However, for the case of $\vartheta_{\rm obs} = 85^\circ$,
even the modified GRAD differs from KERRBB significantly. This is
caused by insufficient azimuthal resolution in GRAD when the disk is
highly inclined.

\clearpage

\begin{deluxetable}{cccccc}
\tablecolumns{6}
\tablewidth{0pt}
\tabletypesize{\small}
\tablecaption{Average Ratios of Fitted Parameters for U1543}
\tablehead{
	\colhead{Parameter} &
	\colhead{KERRBB} &
	\colhead{GRAD} &
	\colhead{DISKPN} &
	\colhead{EZDISKBB} &
	\colhead{DISKBB}
	}
\startdata

$\dot{M}$	& 1	& 1.03	& 0.61	& 0.25	& 0.017	\\
$D$		& 1	& 1.01	& 0.82	& 0.67	& 0.31	\\

\enddata

\end{deluxetable}

\begin{deluxetable}{cccccc}
\tablecolumns{6}
\tablewidth{0pt}
\tabletypesize{\small}
\tablecaption{Average Ratios of Fitted Parameters for J1550}
\tablehead{
	\colhead{Parameter} &
	\colhead{KERRBB} &
	\colhead{GRAD} &
	\colhead{DISKPN} &
	\colhead{EZDISKBB} &
	\colhead{DISKBB}
	}
\startdata

$\dot{M}$	& 1	& 1.25	& 1.54	& 0.76	& 0.052	\\
$D$		& 1	& 1.11	& 1.47	& 1.21	& 0.55	\\

\enddata

\end{deluxetable}

\begin{deluxetable}{cccc}
\tablecolumns{4} \tablewidth{0pt} \tabletypesize{\small}
\tablecaption{Sensitivity of Results on U1543 to Input
Parameters\tablenotemark{a}} \tablehead{ \colhead{Adjusted Parameter
and Value} & \colhead{$\dot{M}$ (10$^{18}$ g/s)} & \colhead{$a_*$} &
\colhead{$\chi^2$ per dof} } \startdata

$\vartheta_{\rm obs}$ = 22.2$\degr$	& 2.27	& 0.58	& 0.87	\\
$\vartheta_{\rm obs}$ = 19.2$\degr$	& 2.14	& 0.63	& 0.88	\\
$M = 10.4 M_\odot$	& 2.00	& 0.71	& 0.87	\\
$M = 8.4 M_\odot$	& 2.43	& 0.48	& 0.89	\\
$D$ = 8.0 kpc		& 2.65	& 0.54	& 0.88	\\
$D$ = 7.0 kpc		& 1.81	& 0.68	& 0.88	\\
$f_{\rm col}$ = 1.9		& 2.66	& 0.32	& 0.90	\\
$f_{\rm col}$ = 1.5		& 1.76	& 0.83	& 0.87	\\
~~                      & ~~    & ~~    & ~~    \\
Nominal values\tablenotemark{b} & 2.18 & 0.61 & 0.88 \\

\enddata

\tablenotetext{a}{The fits were done with KERRBB using the spectral
data obtained on MJD 52,467.20 (see Figs.~\ref{U1543_norm} and 
\ref{BH_spin}).  For
each fit, one of the following four parameters was varied from its
``correct" value by either adding or subtracting 1$\sigma$:
inclination (20.7 $\pm$ 1.5 degrees), mass (9.4 $\pm$ 1.0 $M_\odot$),
distance (7.5 $\pm$ 0.5 kpc), and spectral hardening factor (1.7 $\pm$
0.2).  The resulting $\dot{M}$, $a_*$ and $\chi^2$ for the fit are
listed.}

\tablenotetext{b}{Values obtained by fixing the parameters at their
central values: $\vartheta_{\rm obs} = 20.7\degr$, $M=9.4M_\odot$,
$D=7.5$ kpc, $f_{\rm col}=1.7$.}

\end{deluxetable}

\clearpage

\begin{figure}
\epsscale{0.75}
\plotone{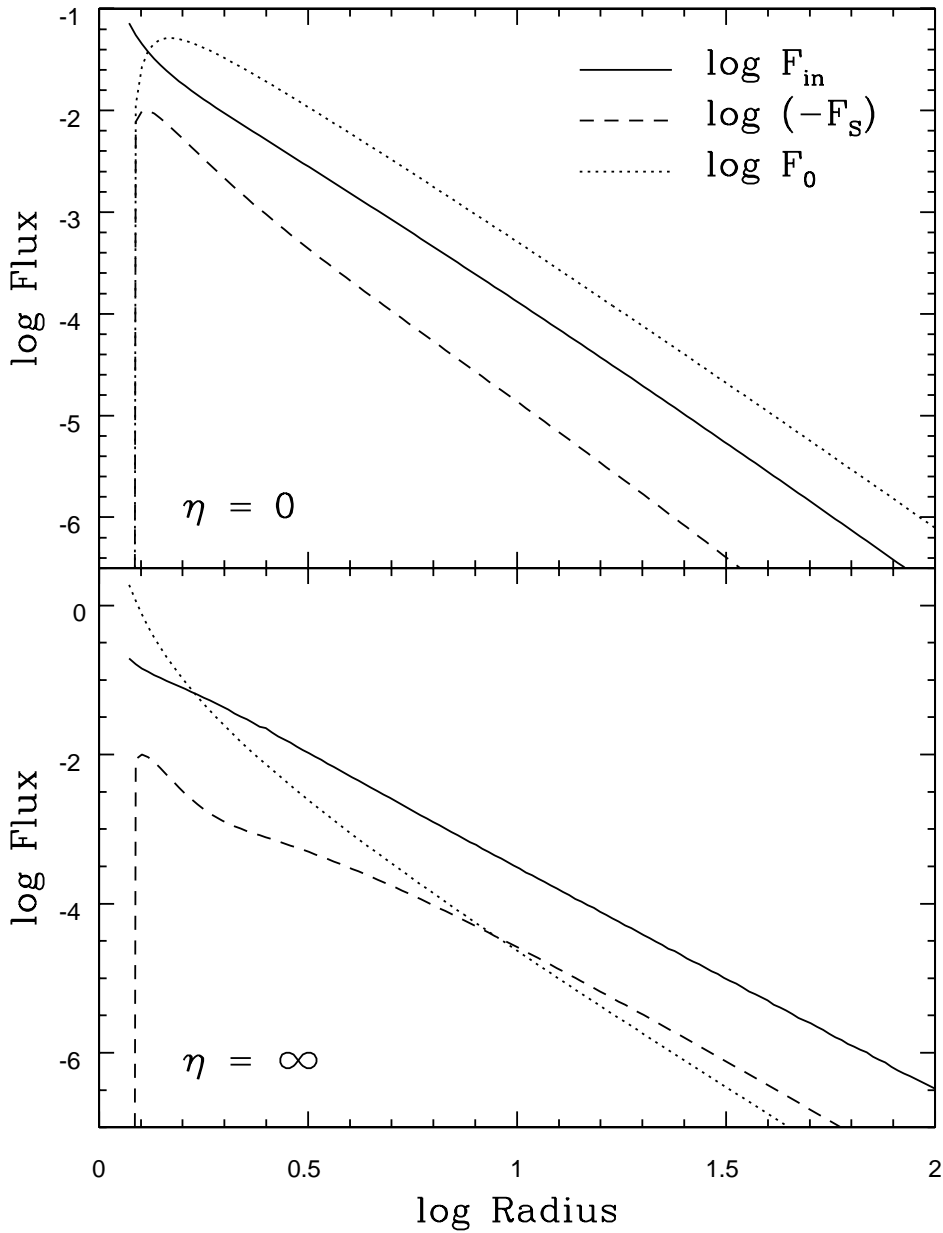}
\caption{The flux density of the radiation returning to the disk for
$\eta=0$ (upper panel) and $\eta=\infty$ (lower panel) for a black
hole with $a=0.999M$.  $F_{\rm in}$ is the flux density of the
returning (ingoing) radiation (eq.~[\ref{flux_i}]), and $F_S$ is the
flux density arising from the angular momentum of the returning
radiation (eq.~[\ref{fs}]). For comparison, the outgoing flux density
when the returning radiation is ignored ($F_0$, eq.~[\ref{f0}]) is
also shown. The disk radius is in units of $r_g =M$ and the fluxes are
in units of $3\dot{M}_{\rm eff}/8\pi r_g^2$. Note that $F_S$ is
negative.
\label{ret_flux_0999}}
\end{figure}

\clearpage

\begin{figure}
\epsscale{1.0}
\plotone{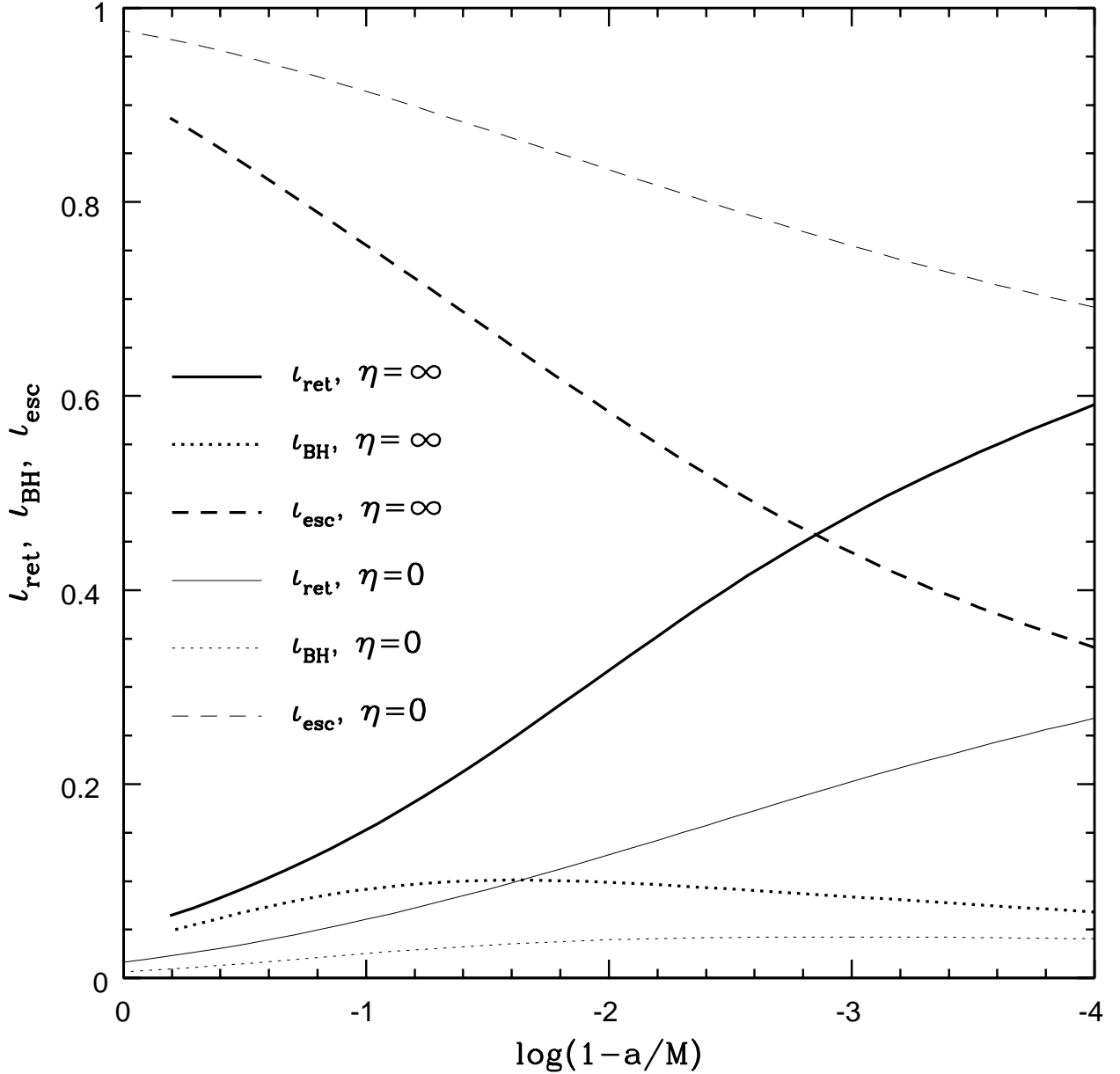}
\caption{Photons emitted by the disk consist of three components: some
photons return to the disk, some go into the black hole, and the rest
escape to infinity. The fraction of each component in the ``total''
energy radiated by the disk is shown as a function of the spin of the
black hole. The definitions of these fractions, which satisfy
$\iota_{\rm ret}+\iota_{\rm BH}+\iota_{\rm esc} =1$, are given in
eq.~(\ref{frac1}).  Thin lines correspond to the case of a standard
Keplerian disk ($\eta=0$, $g_{\rm in} = 0$) and thick lines correspond
to the case of a nonaccreting disk ($\eta = \infty$, $\dot{M} = 0$,
but $g_{\rm in}\neq 0$).
\label{fraction}}
\end{figure}

\begin{figure}
\epsscale{0.75}
\plotone{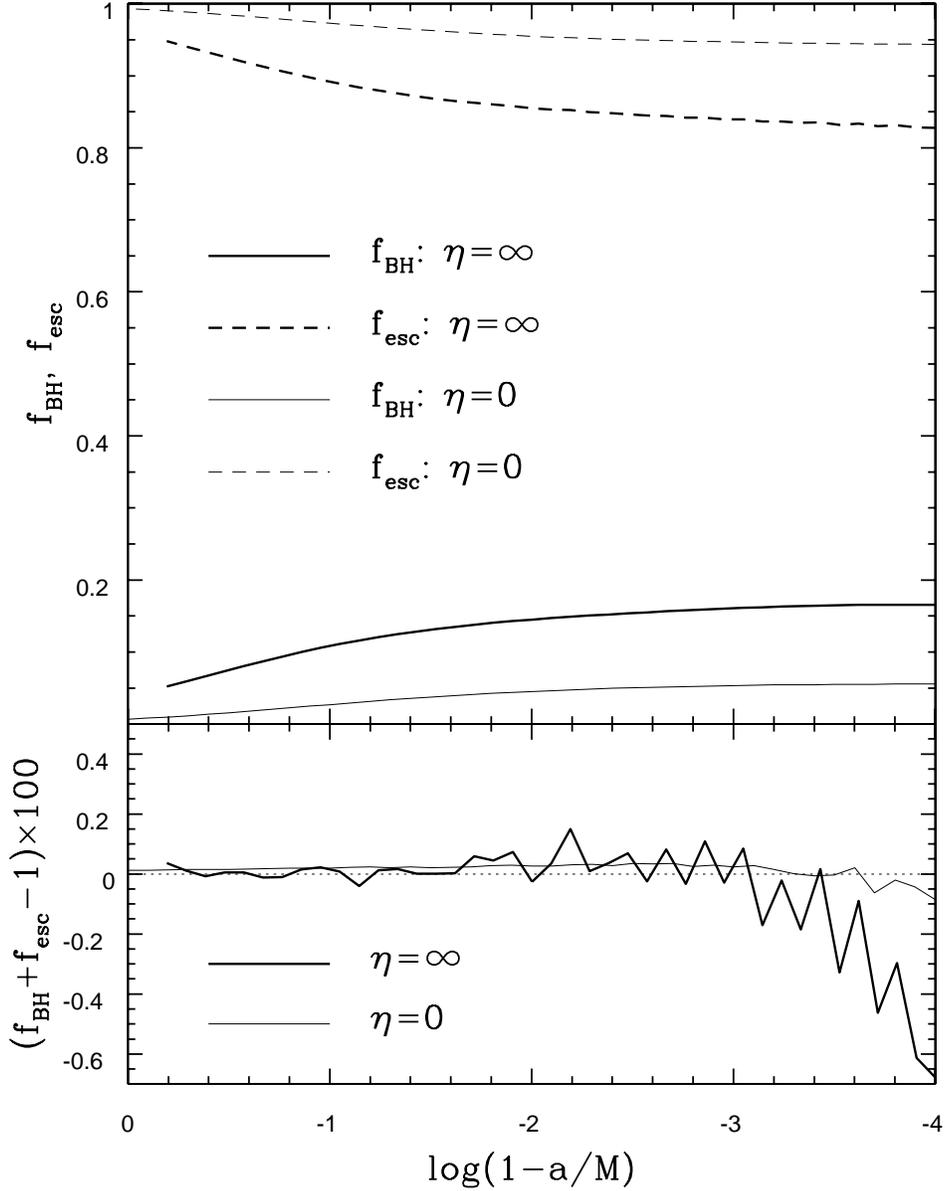}
\caption{The ``net'' radiation from the disk---i.e., the radiation
that permanently leaves the disk---consists of two parts: one falls
into the black hole, the other escapes to infinity. The upper panel
shows the fractions of each as a function of the spin of the black
hole. The definitions of the fractions are given in
eq.~(\ref{frac2}). The lower panel tests the conservation of energy,
$f_{\rm BH} + f_{\rm esc} = 1$, in the calculations.  It is seen to be
satisfied within the errors of the code. Thin lines correspond to the
case of a standard Keplerian disk ($\eta =0$) and thick lines
correspond to the case of a nonaccreting disk ($\eta =\infty$).
\label{fraction2}}
\end{figure}

\clearpage

\begin{figure}
\epsscale{0.75}
\plotone{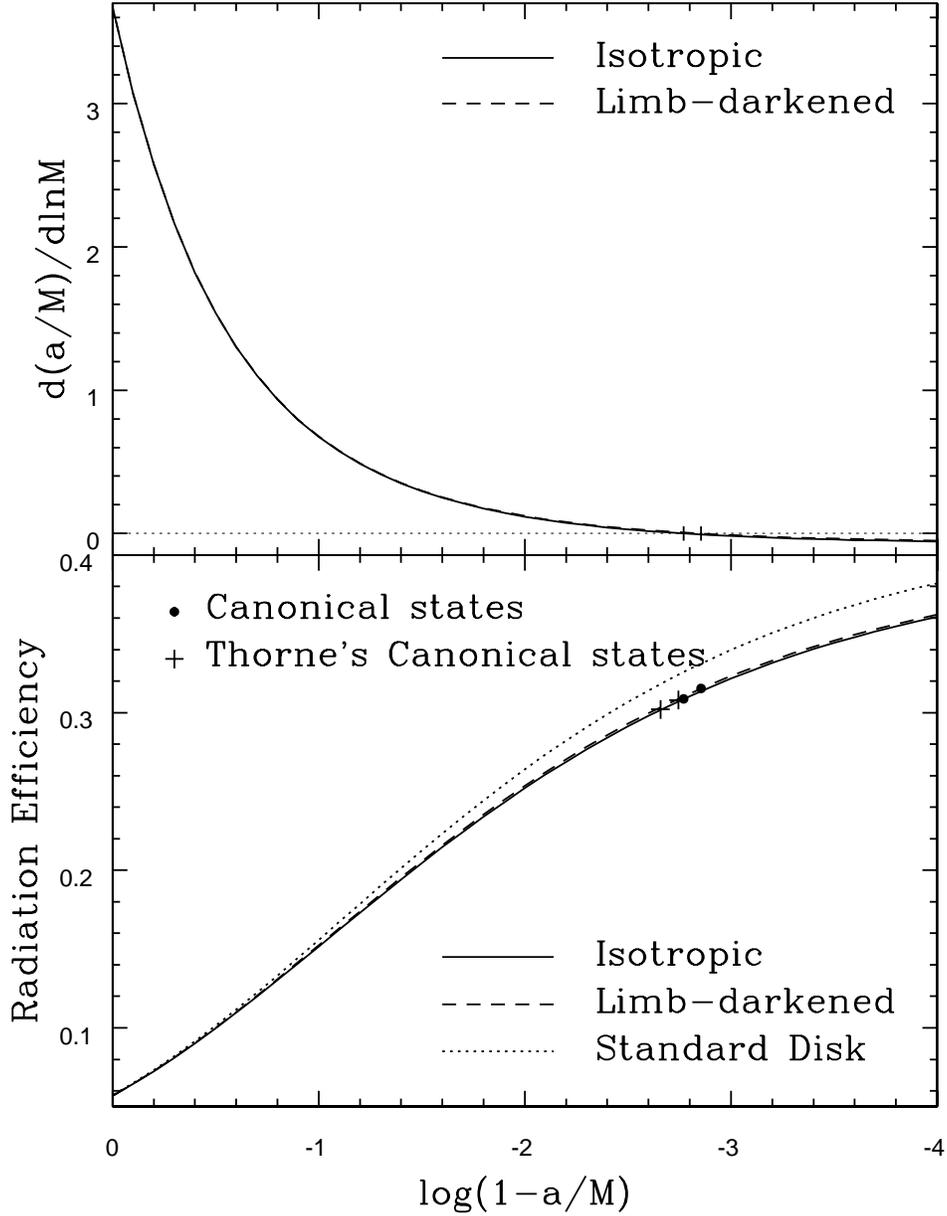}
\caption{Upper panel: The spin-up function defined in
equation~(\ref{daM}).  The equilibrium spin of the black hole [or, the
``canonical'' state, \citet{tho74}] corresponds to the condition
$d(a/M)/d\ln M =0$, which gives $a = 0.9983 M$ when the disk radiation
is isotropic, and $a = 0.9986 M$ when the disk radiation is
limb-darkened (indicated by the two vertical lines).  Lower panel: The
efficiency of a standard Keplerian disk in converting rest mass into 
outgoing radiation (see the text for definition) as a function of the 
spin of the black hole. The corresponding efficiency when the effect of 
returning radiation is ignored is shown by the dotted line.
\label{spin_func}}
\end{figure}

\clearpage

\begin{figure}
\epsscale{0.78}
\plotone{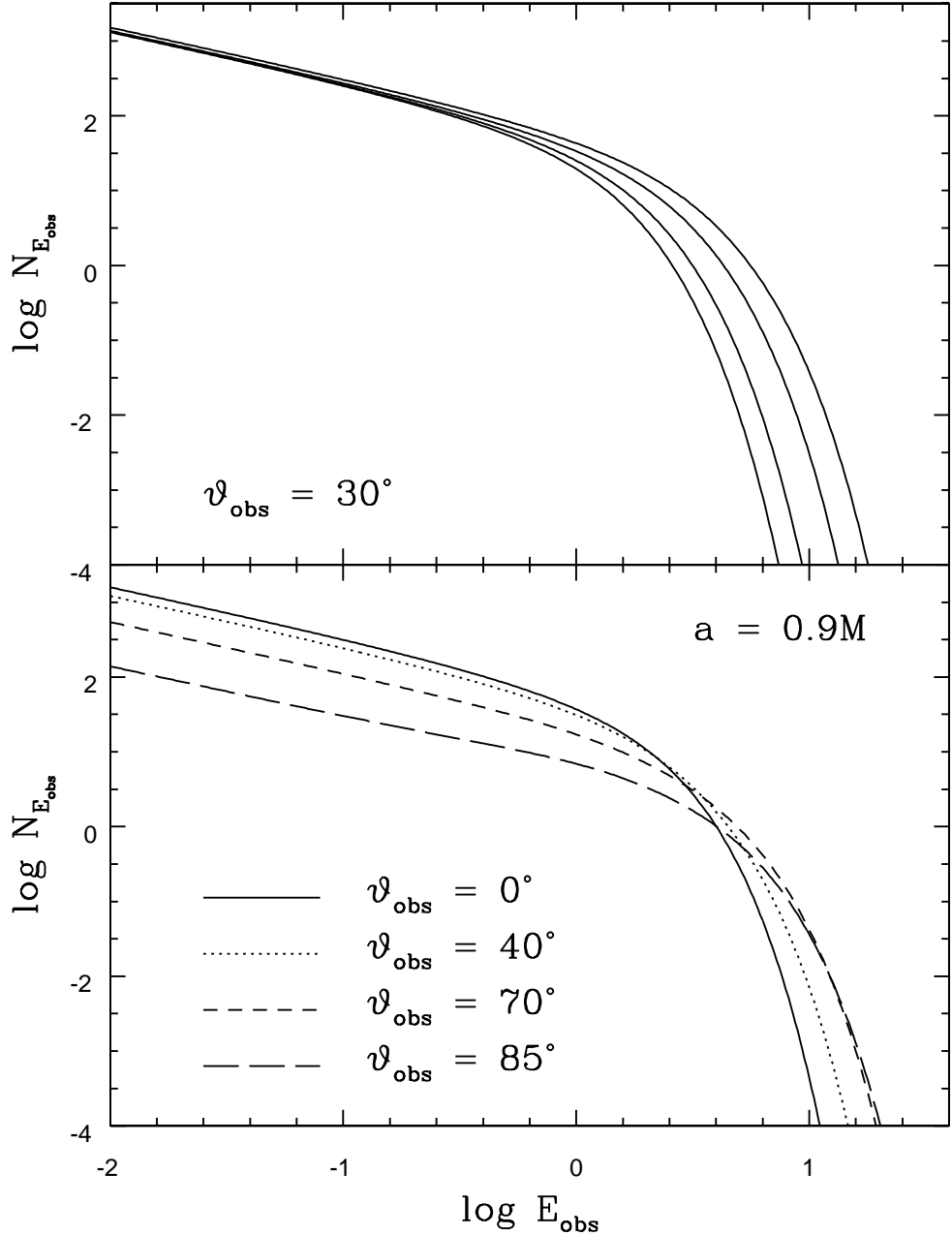}
\caption{Upper panel: Effect of the spin of the black hole on the
observed spectrum of the disk. From left to right: $a/M = 0, 0.5,
0.9$, and $0.999$.  Other parameters are: $\eta = 0$, $\vartheta_{\rm
obs} = 30^\circ$, $M = 10 M_\odot$, $D = 10 {\rm kpc}$, $\dot{M} =
10^{19}{\rm g\,sec}^{-1}$, and $f_{\rm col} = 1$.  Lower panel: Effect
of the inclination angle of the disk on the observed spectrum. The
inclination angles are: $\vartheta_{\rm obs} = 0^\circ, 40^\circ,
70^\circ$, and $85^\circ$, as indicated. Other parameters are: $\eta =
0$, $a = 0.9M$, $M = 10 M_\odot$, $D = 10 {\rm kpc}$, $\dot{M} =
10^{19}{\rm g\,sec}^{-1}$, and $f_{\rm col} = 1$. The energy $E_{\rm
obs}$ is in keV, and the flux density $N_{E_{\rm obs}}$ is in units of
${\rm photons~keV}^{-1}~{\rm cm}^{-2}~{\rm sec}^{-1}$.
\label{spec_spin_incl}}
\end{figure}

\clearpage

\begin{figure}
\epsscale{0.72}
\plotone{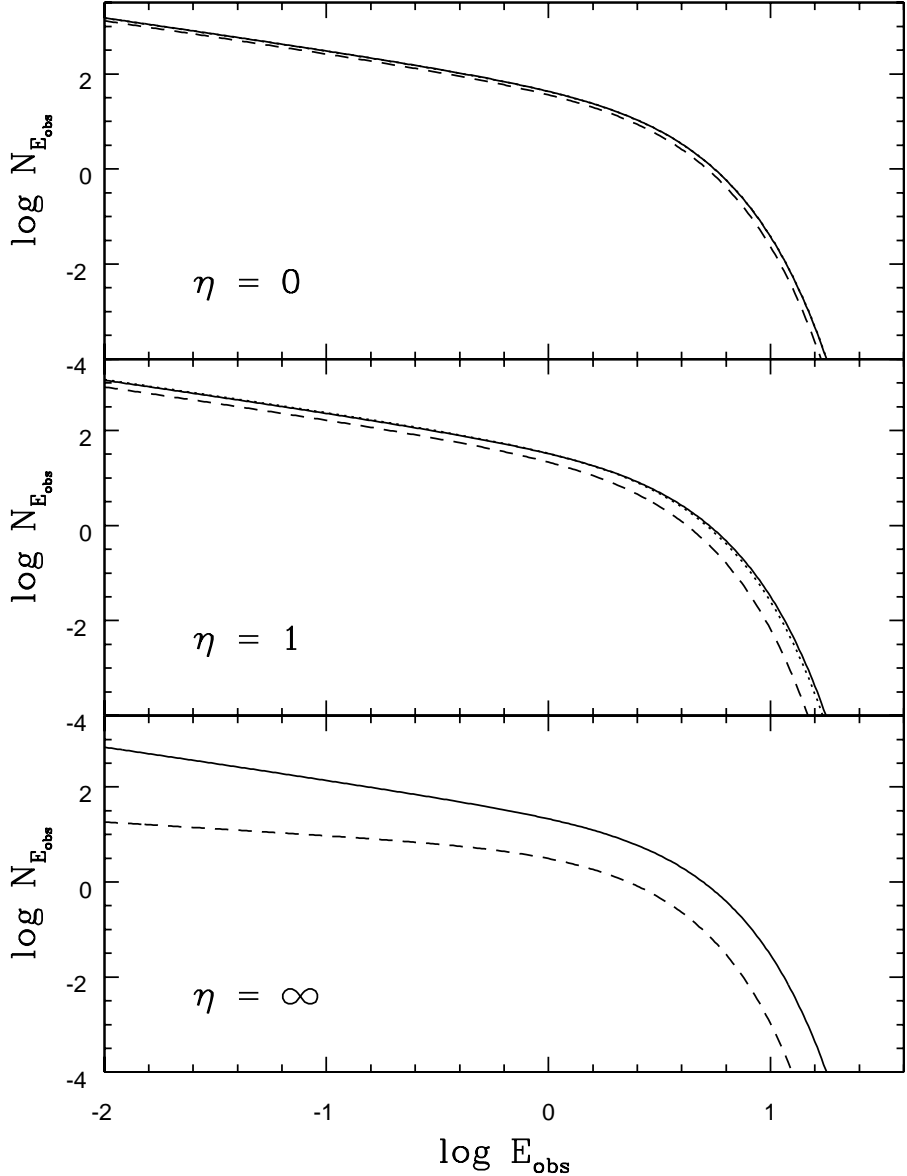}
\caption{Effect of the returning radiation on the observed spectrum of
an accretion disk. The three cases correspond to $\eta = 0$ (upper
panel), $\eta = 1$ (middle panel), and $\eta = \infty$ (lower
panel). The solid line is the spectrum when the returning radiation is
included, and the dashed line is the spectrum when the returning
radiation is ignored. Parameters are: $a = 0.999M$, $\vartheta_{\rm
obs} = 30^\circ$, $M=10M_\odot$, $D=10 {\rm kpc}$, $\dot{M}_{\rm eff}
= 10^{19}{\rm g\,sec}^{-1}$, and $f_{\rm col} = 1$. The dotted line in
the upper panel (almost coincident with the solid line) is the
spectrum when the returning radiation is ignored, but $\dot M$ is
increased to $1.23\times 10^{19}{\rm g\,sec}^{-1}$. The dotted line in
the middle panel is the spectrum when the returning radiation is
ignored, and $\dot{M}_{\rm eff}$ is increased to $1.7\times
10^{19}{\rm g\,sec}^{-1}$.
\label{spectra_0999}}
\end{figure}

\clearpage

\begin{figure}
\epsscale{0.92}
\plotone{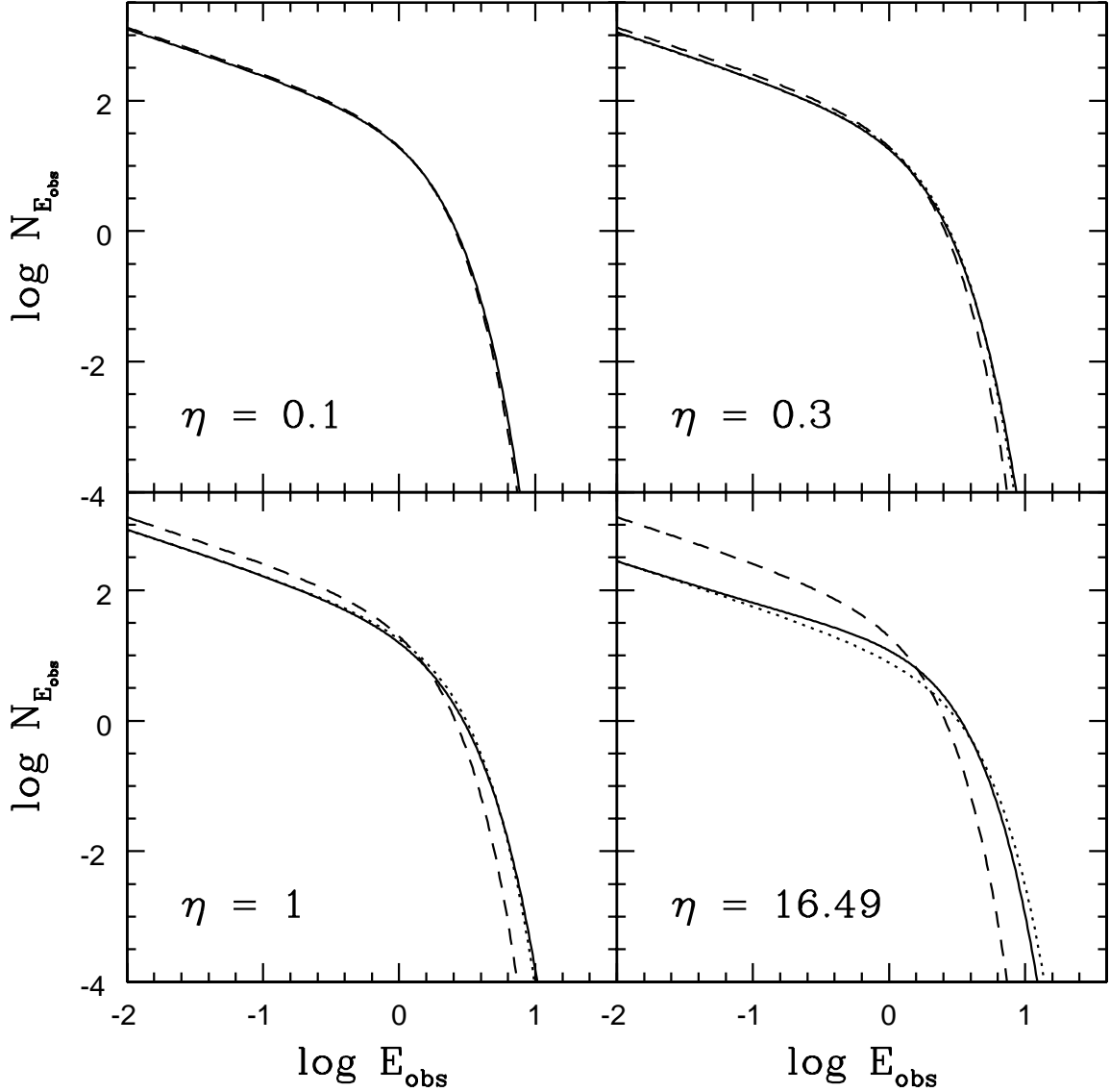}
\caption{Effect of the torque at the inner boundary of the disk.  In
each panel, the solid line is the spectrum of the disk with a torque
corresponding to the indicated value of $\eta$ (see eq.~[\ref{eta}])
and the dashed line is the spectrum of a disk with the same value of
$\dot M_{\rm eff}$ but with the torque at the inner boundary set to
zero. Parameters are: $a = 0$ (Schwarzschild black hole),
$\vartheta_{\rm obs} = 30^\circ$, $M=10M_\odot$, $D=10 {\rm kpc}$,
$\dot{M}_{\rm eff} = 10^{19}{\rm g\,sec}^{-1}$, and $f_{\rm col} =
1$. The dotted lines in the last three panels represent the spectra of
zero-torque models in which the values of $\dot{M}_{\rm eff}$ and
$f_{\rm col}$ have been adjusted for the best fit of the corresponding
solid lines.  The parameters of the dotted line models are:
$\dot{M}_{\rm eff} = 10^{19}{\rm g\,sec}^{-1}$ and $f_{\rm col}=1.15$
for $\eta=0.3$; $\dot{M}_{\rm eff} = 10^{19}{\rm g\,sec}^{-1}$ and
$f_{\rm col}=1.4$ for $\eta =1$; $\dot{M}_{\rm eff} = 6\times
10^{18}{\rm g\,sec}^{-1}$ and $f_{\rm col}=2.5$ for $\eta=16.49$.
\label{spectra_torque_0000}}
\end{figure}

\clearpage

\begin{figure}
\epsscale{1}
\plotone{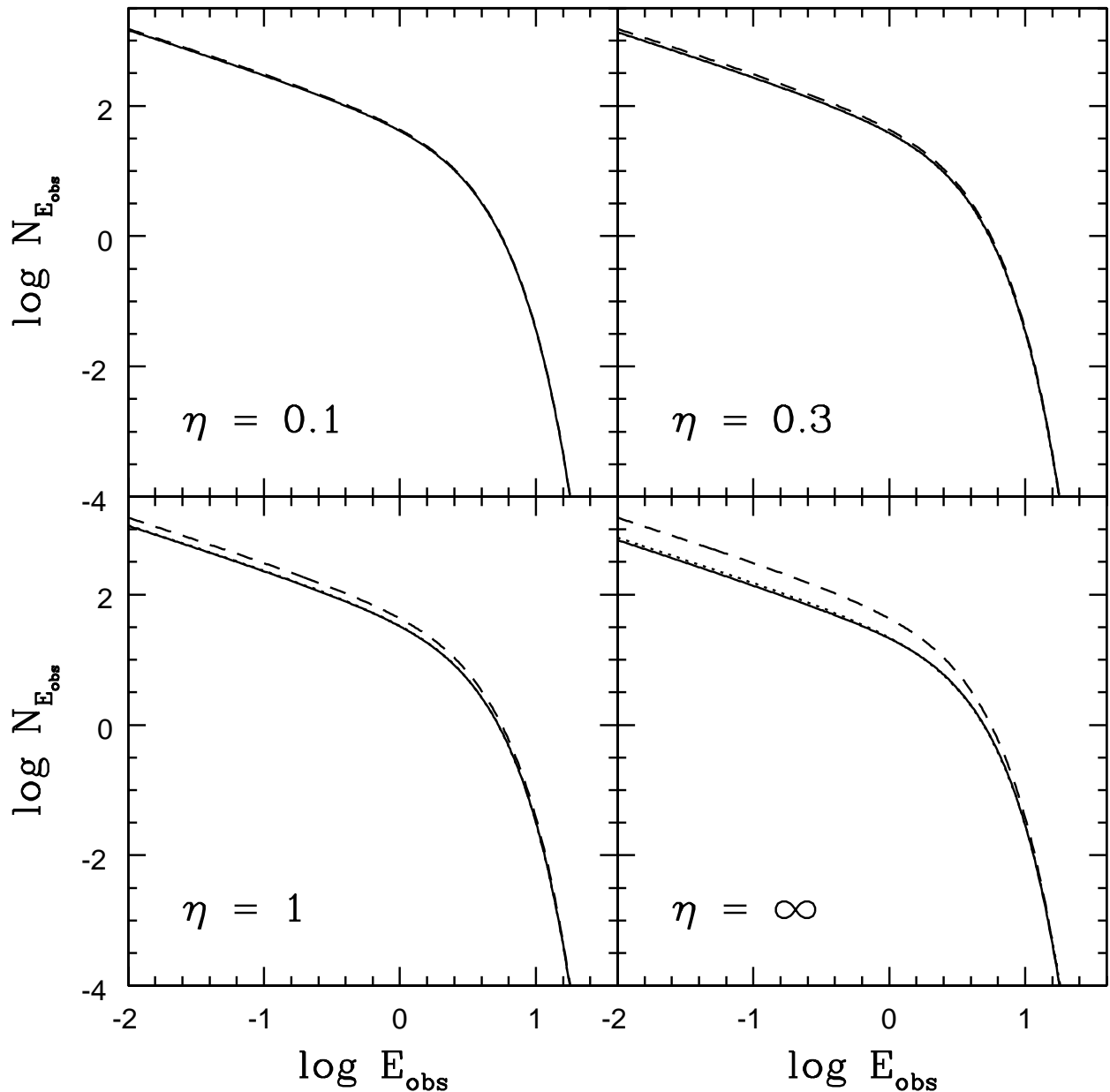}
\caption{Similar to Fig.~\ref{spectra_torque_0000} except that $a/M =
0.999$.  The parameters of the dotted line models are: $\dot{M}_{\rm
eff} = 8.9\times 10^{18}{\rm g\,sec}^{-1}$ and $f_{\rm col}=1.03$ for
$\eta=0.3$; $\dot{M}_{\rm eff} = 7.8\times 10^{18}{\rm g\,sec}^{-1}$
and $f_{\rm col}=1.08$ for $\eta =1$; $\dot{M}_{\rm eff} = 5.4\times
10^{18}{\rm g\,sec}^{-1}$ and $f_{\rm col}=1.25$ for $\eta=\infty$.
\label{spectra_torque_0999}}
\end{figure}

\clearpage

\begin{figure}
\epsscale{0.72}
\plotone{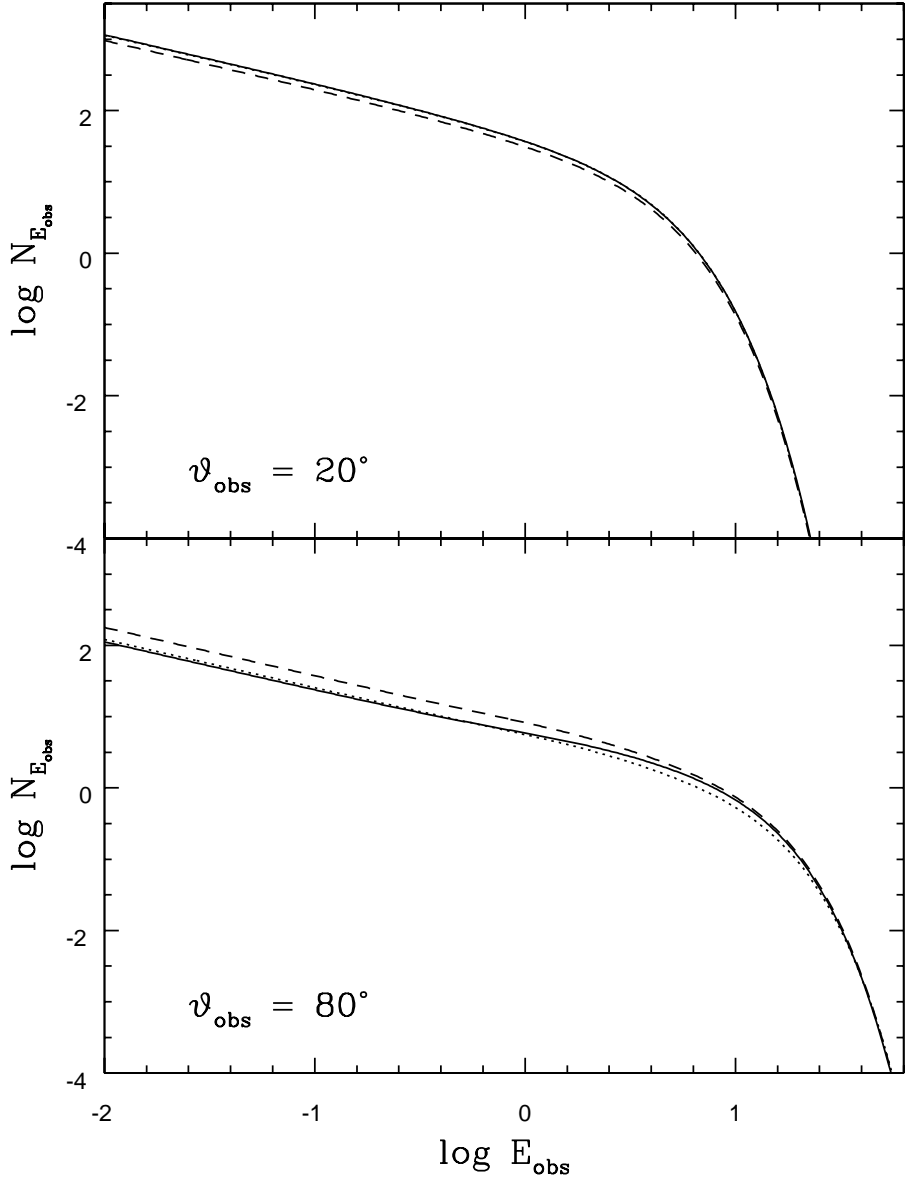}
\caption{Effect of limb-darkening on the observed spectrum. The two
cases correspond to two different disk inclination angles:
$\vartheta_{\rm obs} = 20^\circ$ (upper panel) and $\vartheta_{\rm
obs} = 80^\circ$ (lower panel). The solid line is the spectrum when
the disk emission is limb-darkened, and the dashed line is the
spectrum when the disk emission is isotropic. Other model parameters
are: $\eta =0$, $a = 0.999M$, $M= 10M_\odot$, $D=10 {\rm kpc}$,
$\dot{M} = 10^{19}{\rm g\,sec}^{-1}$, and $f_{\rm col} = 1.5$. The
dotted line in each panel (hardly visible in the upper panel)
represents the best fit of the solid line with a disk with isotropic
emission by adjusting the values of $\dot{M}$ and $f_{\rm col}$. The
parameters of the dotted lines are: $\dot{M} = 1.17\times 10^{19}{\rm
g\,sec}^{-1}$ and $f_{\rm col}=1.43$ in the upper panel, $\dot{M} =
0.72\times 10^{19}{\rm g\,sec}^{-1}$ and $f_{\rm col}=1.7$ in the
lower panel.
\label{spec_limb}}
\end{figure}

\clearpage

\begin{figure}
\epsscale{0.8}
\plotone{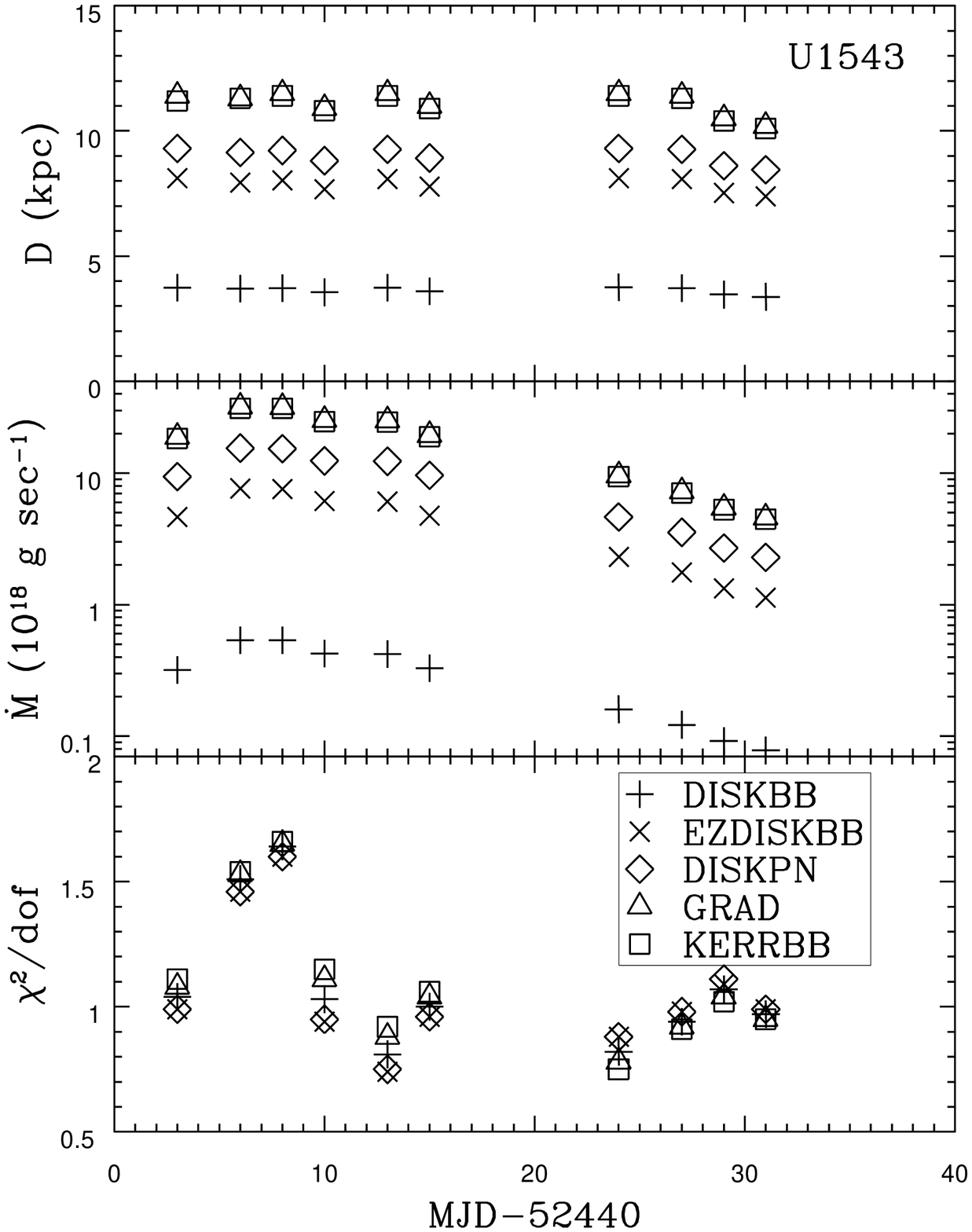}
\caption{Fits of 10 observations of the source U1543 during the
high/soft state.  The data are from \citet{par04}.  Each
observation was fitted separately using KERRBB, GRAD, DISKPN,
EZDISKBB and DISKBB with $f_{\rm col}=1.7$.  The resulting estimates of
the distance $D$ and the mass accretion rate $\dot M$, and the $\chi^2$ 
of the fit, are shown in the three panels.
\label{U1543_norm}}
\end{figure}

\clearpage

\begin{figure}
\epsscale{0.8}
\plotone{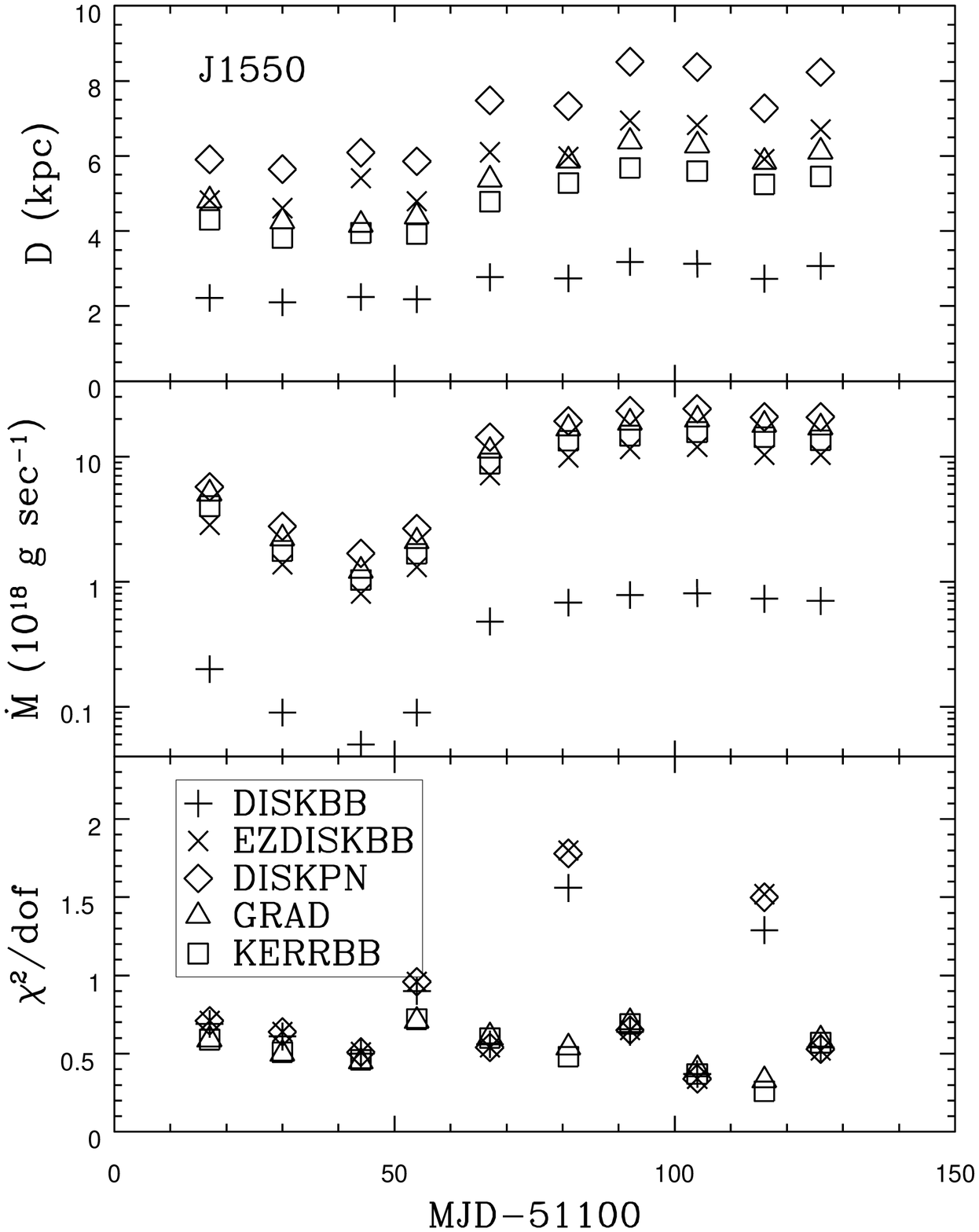}
\caption{Fits of 10 observations of the source J1550 during the
high/soft state.  The data are from \citet{sob00}.  Each
observation was fitted separately using KERRBB, GRAD, DISKPN, EZDISKBB 
and DISKBB with $f_{\rm col}=1.7$.  The resulting estimates of
the distance $D$ and the mass accretion rate $\dot M$, and the $\chi^2$ 
of the fit, are shown in the three panels.
\label{J1550_norm}}
\end{figure}

\clearpage

\begin{figure}
\epsscale{0.8}
\plotone{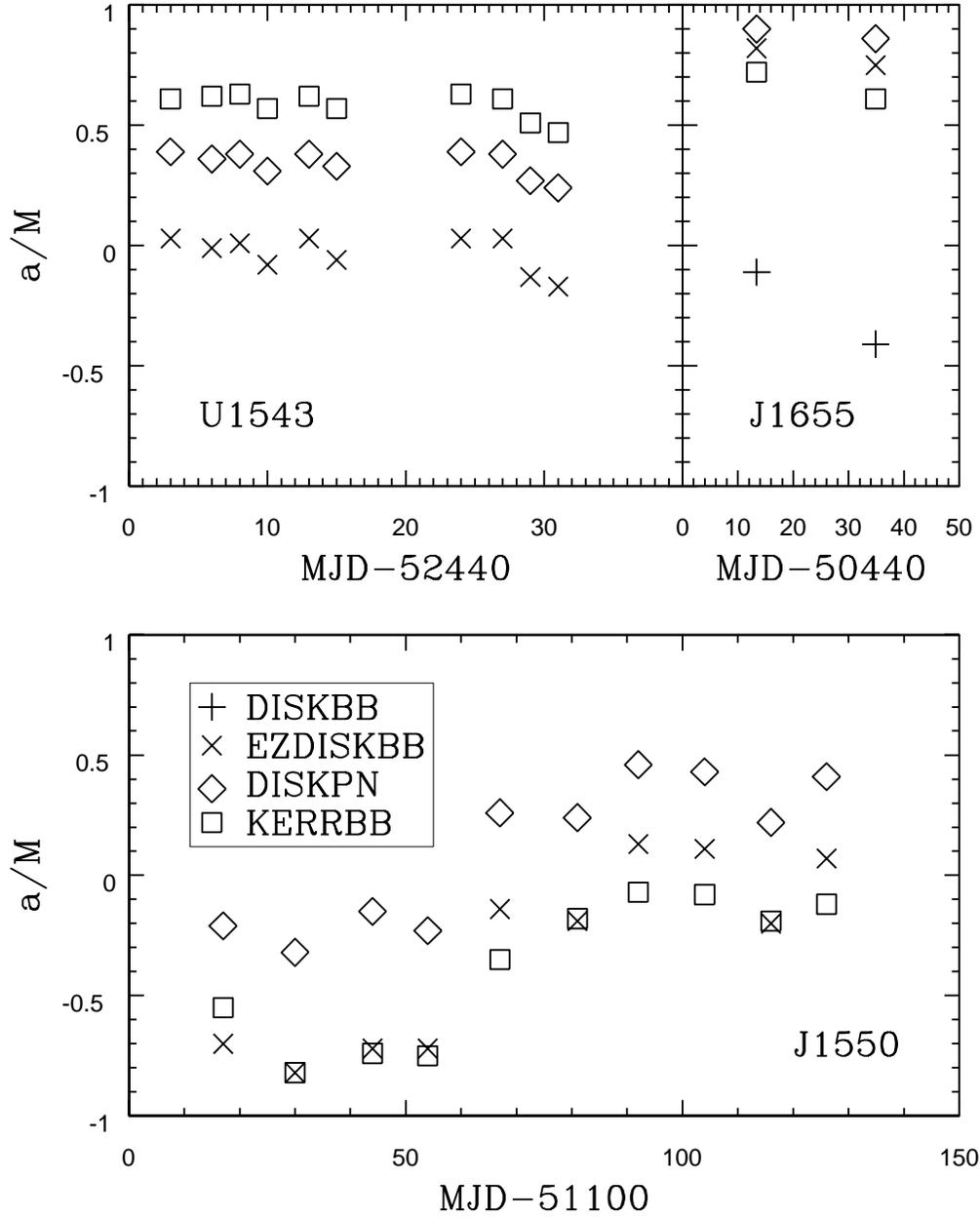}
\caption{Estimates of the black hole spin parameter $a/M$ in U1543,
J1550 and J1655.  The same 10 observations shown in 
Figs.~\ref{U1543_norm} and \ref{J1550_norm} were used for the first
two sources, and two observations from \citet{sob99} were used
for J1655. For each observation, the black hole spin was estimated
separately using KERRBB, DISKPN, EZDISKBB, and DISKBB with $f_{\rm col}=
1.7$.  No consistent solution was obtained with DISKBB for any of the 
observations of U1543 and J1550.
\label{BH_spin}}
\end{figure}

\clearpage

\begin{figure}
\epsscale{.90}
\plotone{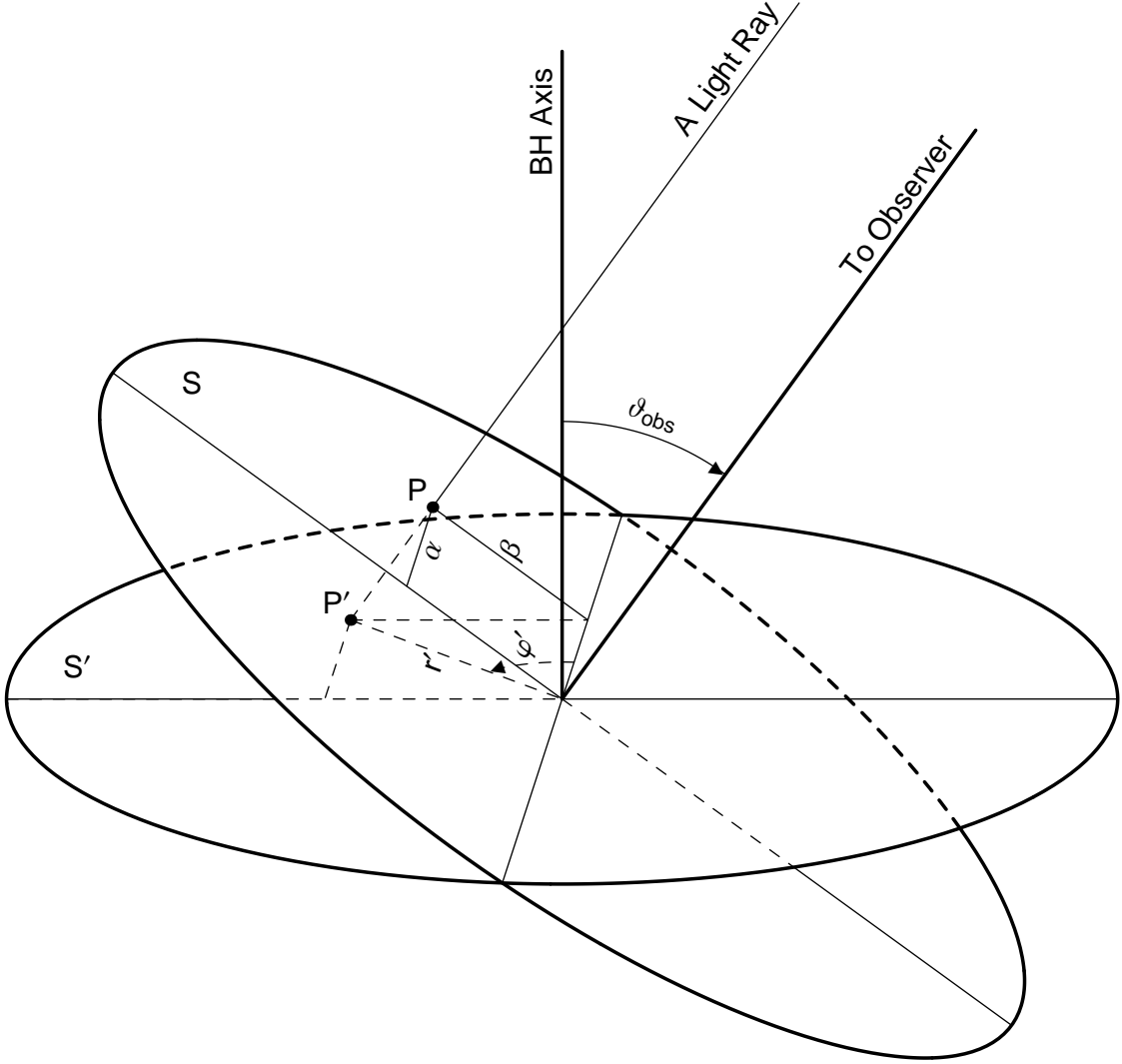}
\caption{Relation between the impact parameters of a photon in the
image plane $S$ of an observer and polar coordinates in the equatorial
plane $S^\prime$ of the black hole. On the image plane $S$, which is
perpendicular to the line of sight, each point $P$ is specified by a
pair of impact parameters $(\alpha,\beta)$.  In the equatorial plane
$S^\prime$, which is perpendicular to the rotation axis of the black
hole, each point $P^\prime$ is specified by a pair of polar
coordinates $(r^\prime,\varphi^\prime)$.  $P$ and $P^\prime$ are
connected by a straight line drawn parallel to the direction of the
line-of-sight.  The coordinates $(\alpha,\beta)$ of $P$ are related to
the coordinates $(r^\prime, \varphi^\prime)$ of $P^\prime$ by
equation~(\ref{ab_rfi}). The polar angle of the observer, or
equivalently the inclination angle of the disk, is $\vartheta_{\rm
obs}$. \label{coord1}}
\end{figure}

\clearpage

\begin{figure}
\epsscale{0.83}
\plotone{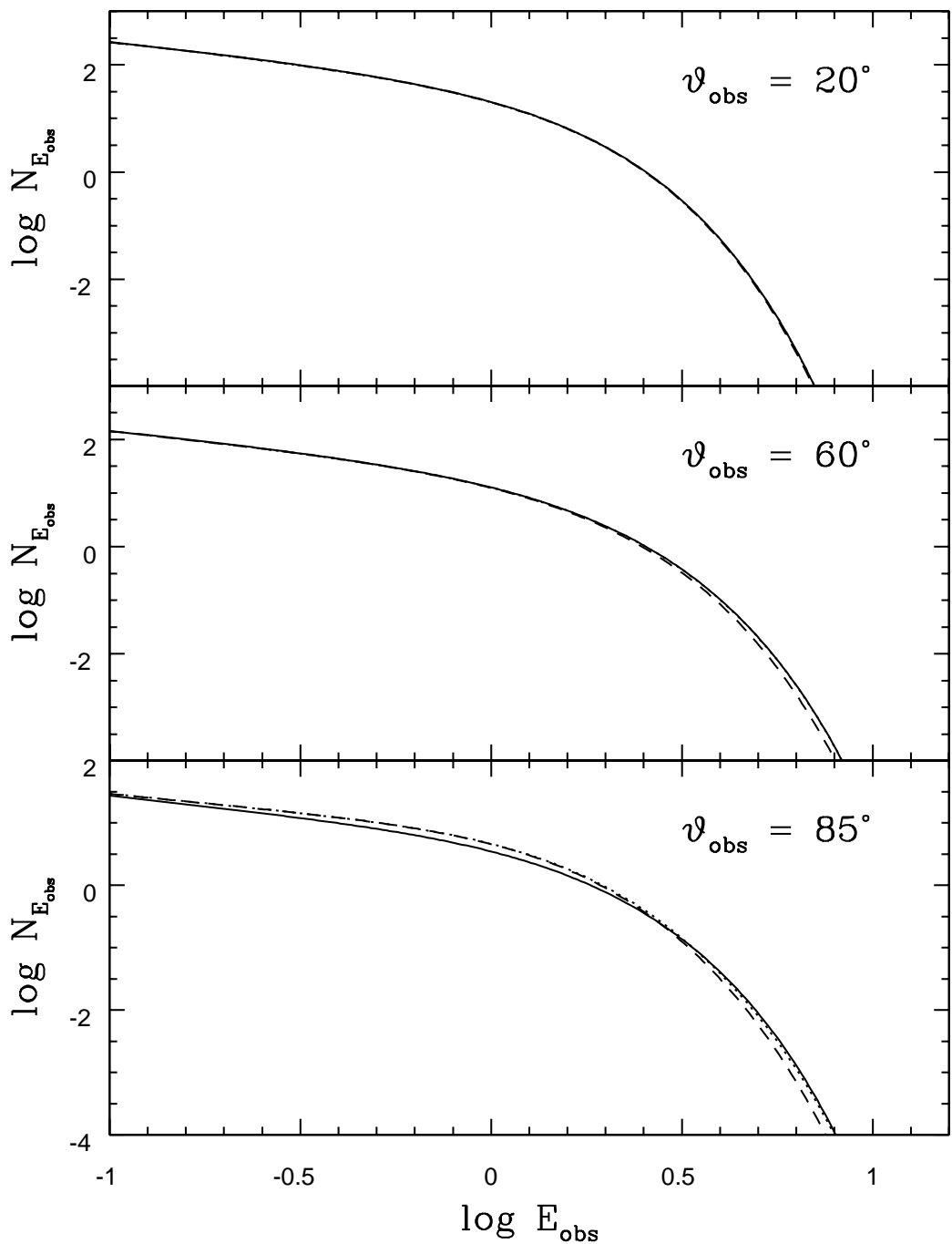}
\caption{Comparison of KERRBB with GRAD. The specific photon number 
density calculated by KERRBB is shown with solid lines, calculated
with GRAD is shown with dashed lines, and that calculated with
modified GRAD is shown with dashed lines. Three panels correspond to 
three different disk inclination angles as labeled. In the upper 
and middle panels, the dashed lines are almost coincident with the 
solid lines.
\label{grad_kerrbb}}
\end{figure}

\end{document}